\documentstyle[12pt]{article}

\begin{document} 

\begin{titlepage}

\hrule 
\leftline{}
\leftline{Preprint
          \hfill   \hbox{\bf CHIBA-EP-103-REV}}
\leftline{\hfill   \hbox{hep-th/9801024(revised)}}
\leftline{\hfill   \hbox{June 1998}}
\vskip 5pt
%\leftline{}
\hrule 
\vskip 1.0cm

\centerline{\large\bf 
Yang-Mills Theory as a Deformation of 
} 
\vskip 0.3cm
\centerline{\large\bf  
Topological Field Theory, Dimensional Reduction 
}%$^*$
\vskip 0.3cm
\centerline{\large\bf  
and Quark Confinement $^*$
}

\vskip 1cm

\centerline{{\bf 
Kei-Ichi Kondo$^{1}{}^{\dagger}$
%and $^2{}^{\ddagger}$
}}  
%\vskip 4mm
\begin{description}
\item[]{\it  %$^1$ 
$^1$ Department of Physics, Faculty of Science,
  Chiba University, Chiba 263, Japan
  %$^\ddagger$
  }
%\item[]{\it 
%$^2$ Graduate School of Science and Technology,
%  Chiba University, Chiba 263, Japan
%  %$^\ddagger$
%  }
\item[]{$^\dagger$ 
  E-mail:  kondo@cuphd.nd.chiba-u.ac.jp 
  }
%\item[]{$^\ddagger$ 
%  E-mail:   
%  }
\end{description}
%\vskip 0.5cm

\centerline{{\bf Abstract}} \vskip .5cm
We propose a reformulation of Yang-Mills theory as a
perturbative deformation of a novel topological (quantum) field
theory. We prove that this reformulation of the four-dimensional QCD
leads to quark confinement in the sense of area law of the Wilson
loop.  First, Yang-Mills theory with a non-Abelian gauge group G is
reformulated as a deformation of a novel topological field
theory.  Next, a special class of topological field theories is
defined by both BRST and anti-BRST exact action corresponding to the
maximal Abelian gauge leaving the maximal torus group H of G
invariant.  Then we find the topological field theory ($D>2$) has a
hidden supersymmetry for a choice of maximal Abelian gauge.  As a
result, the D-dimensional topological field theory is equivalent to
the (D-2)-dimensional coset G/H non-linear sigma model in the sense
of Parisi and Sourlas  dimensional reduction.  After maximal Abelian
gauge fixing, the topological property of magnetic monopole and
anti-monopole of four-dimensional Yang-Mills theory is translated
into that of instanton and anti-instanton in two-dimensional
equivalent model.  It is shown that the linear static potential in
four-dimensions follows from the instanton--anti-instanton gas in
the equivalent two-dimensional non-linear sigma model obtained from
the four-dimensional topological field theory by dimensional
reduction, while the remaining Coulomb potential comes from the
perturbative part in four-dimensional Yang-Mills theory.  The
dimensional reduction opens a path for applying various exact
methods developed in two-dimensional quantum field theory to study
the non-perturbative problem in low-energy physics of
four-dimensional quantum field theories.


%\vskip 0.2cm
\hrule  
%%%%%%%%%%%%%%%%%%%%%%%%%%%%%%
%\vskip 2cm  
%\hrule  
%\bigskip  
%\centerline
%{\bf CHIBA UNIVERSITY}  
%\vfill 

\vskip 0.2cm  
%\hrule  

%\begin{description}
%\item[]{
%$^\ddagger$
%Address from March 1996 to December 1996.
%  On leave of absence from: \\
%  Department of Physics, Faculty of Science,
%  Chiba University, Chiba 263, Japan.
%  }
%\item[]{
$^*$ To be published in Phys. Rev. D.
% Submitted to .
% }  
%\end{description}

\end{titlepage}

%\newpage
%%%%% Table of Contents %%%%%
%\pagenumbering{roman}
%\tableofcontents
%%%%% Table of Contents %%%%%
\pagenumbering{arabic}
%%%%%

\newpage
\section{Introduction and main results}
\setcounter{equation}{0}

In particle physics, perturbation theory is applicable if the
coupling constant as an expansion parameter is small in the energy
region considered.  This is assured in high energy ultraviolet 
region of quantum chromodynamics (QCD) where the effective coupling
constant is small due to asymptotic freedom \cite{GWP73}.   On the
other hand, in the infrared regime of QCD where the effective
coupling is expected to be large, the perturbation theory loses its
validity.  The quark confinement is regarded as a typical example of
indicating the difficulty of treating strongly coupled gauge
theories.  The conventional perturbation theory deals with the small
deviation from the trivial gauge field configuration
${\cal A}_\mu=0$ which is a minimum of the action $S$.  
\par
In the last decade,  various evidences about Abelian dominance and
magnetic monopole dominance in the low energy physics of QCD have
been accumulated based on Monte Carlo simulation of lattice QCD
initiated by the work \cite{KSW87}, see e.g.
\cite{review} for a review.  This urges us to reconsider if there
may exist  any perturbation theory appropriate for QCD with the
expansion parameter being small even in the infrared region.   There
the expansion must be performed about the non-trivial gauge field
configuration ${\cal A}_\mu\not=0$ other than the trivial one
${\cal A}_\mu=0$. 
In gauge field theories, we know that there are soliton solutions
called vortex
\cite{NO73}, magnetic monopole \cite{tHooft74} and instanton
\cite{BPST75,Shifman94}. 
They are candidates for such a non-trivial field configuration.
\par
We know a few examples that such expansions around non-trivial field
configuration successfully have led to the resolution of strong
coupling problem. An example is a proof of quark confinement by
Polyakov 
\cite{Polyakov77} in three-dimensional compact U(1) gauge theory and
three-dimensional compact quantum electrodynamics (QED) in
Georgi-Glashow model with a gauge group $SU(2)$.
He considered the non-trivial minimum $\Omega_\mu$ of the
action given by the instanton (pseudoparticle).  The field 
${\cal A}_\mu$ is decomposed into $\Omega_\mu+Q_\mu$ and
$Q_\mu$ is considered as quantum fluctuation around $\Omega_\mu$. 
The integral over $Q_\mu$ is Gaussian and is exactly integrated
out.  The result is written as the sum over all possible
configurations of instanton and anti-instantons.  In
three-dimensional case, instanton (resp. anti-instanton) is given by
the magnetic monopole (resp. anti-monopole).  
Moreover, Seiberg and Witten \cite{SW94} have shown that in the
four-dimensional
$N=2$ supersymmetric gauge theories,  the non-perturbative
contributions come only from magnetic monopole or instanton in the
prepotential which exactly determines the low-energy effective
Abelian gauge theory.  These  examples show that the quark
confinement is caused by the condensation of magnetic monopoles.
\par
Recently, it has been tried to reformulate the Yang-Mills
(YM) theory as a deformation of a topological (quantum) field theory
\cite{Witten,Schwarz,TQFT}, abbreviated T(Q)FT hereafter. 
The BF theory \cite{TQFT} as a TFT can be regarded
as a zero-coupling limit of
 YM theory
\cite{AN93,Izawa93,BFYM}.  A similar idea was proposed recently by
Abe and Nakanishi
\cite{AN93} where two-dimensional BF theory is essentially equivalent
to the zeroth-order approximation to YM theory in their framework of
the newly proposed method of solving quantum field theory.
In higher dimension, however, the limit is singular due to the fact
that the gauge symmetry in BF theory is larger than that in YM
theory.   In a couple of years, considerable progress has been made
to assure that the YM theory can be obtained as a deformation
(perturbation) of the topological BF theory  by  Fucito, Martellini
and Zeni \cite{BFYM}. This reformulation is the first order
formulation of YM theory, called the BFYM theory
\cite{QR97,KondoI}.  They have checked
the area law behaviour for the Wilson loop average and computed the
string tension.  In this formalism area law arises in a very
simple geometrical fashion, as an higher linking number between
loop and surface. 
\par
In this paper, we reconsider the YM theory from a topological point
of view. First we reformulate the YM theory as a deformation of a
novel TFT.  This is equivalent to
say that the YM theory is described as a perturbation around the
non-trivial field configuration $\Omega_\mu$ given by the TFT. This
formulation of YM theory will be suitable for describing the low
energy region of YM theory, because   the topological property does
not depend on the details of the short distance behavior of the
theory and depends only on the global structure of the theory.  In
order for such a description to be successful, the TFT must include
the most essential or dominant degrees of freedom for describing the
low energy physics in question. The monopole dominance is a hint for
searching an appropriate TFT.  
The TFT we propose in this paper is different from the
conventional TFT's of Witten type
\cite{Witten} or Schwarz type \cite{Schwarz}.
Witten type TFT starts from the gauge fixing condition of
self-duality, 
\begin{eqnarray}
{\cal F}_{\mu\nu} = \pm \tilde {\cal F}_{\mu\nu},
\quad \tilde {\cal F}_{\mu\nu}:= {1 \over 2}
\epsilon_{\mu\nu\rho\sigma}  {\cal F}_{\rho\sigma} ,
\end{eqnarray}
corresponding to the instanton configuration in
four-dimensional YM theory
\cite{BPST75}.
The total action can be written as Becchi-Rouet-Stora-Tyupin (BRST)
transformation $\delta_B$ of some functional $V$ composed of the
fields and their ghosts,
\begin{eqnarray}
 S_{tot} = [ Q_B, V \} = \delta_B V .
\end{eqnarray}
On the other hand, Schwarz type TFT has a non-trivial classical
action
$S_{cl}$ which is metric independent (hence topological) with
non-trivial gauge fixing.  
For example, BF theory and Chern-Simons theory belong to this
type,
\begin{eqnarray}
 S_{tot} = S_{cl} + [ Q_B, V' \} = S_{cl} + \delta_B V'.
\end{eqnarray}
Our
TFT tries to incorporate the magnetic monopole degrees of freedom as
an essential degrees of freedom for low-energy physics. For this, we
use the the maximal Abelian gauge (MAG).   In MAG, we find that the
action is written in the form,
\begin{eqnarray}
 S_{tot} =  \delta_B \bar \delta_B {\cal O} ,
 \label{myTFT}
\end{eqnarray}
using the anti-BRST transformation 
$\bar \delta_B$ \cite{antiBRST}.
\par
In a previous paper \cite{KondoI}, we have proved that the dual
superconductor picture of quark confinement in QCD (proposed by
Nambu, 'tHooft and Mandelstam \cite{Nambu74,tHooft81,Mandelstam76})
can be derived from QCD without any specific assumption. In order to
realize the dual superconductor vacuum of QCD, we need to take the 
MAG.   MAG is an example of Abelian projection proposed by 'tHooft
\cite{tHooft81}.  The basic idea of Abelian projection is that the
off-diagonal non-Abelian parts are made as small as possible. 
Imposing MAG, the gauge degrees of freedom corresponding to
$G/H$ is fixed and the residual gauge invariance for the maximal
torus group
$H$ of the gauge group $G$ remains unbroken.  Under MAG, it is
expected that the off-diagonal gluons (belonging to $G/H$) become
massive and the low energy physics of QCD is described by the
diagonal Abelian part (belonging to $H$) alone. 
All the off-diagonal fields transform as charged
fields under the residual Abelian gauge symmetry $H$ and are
expected to be massive.  It is shown
\cite{KondoI} that an Abelian-projected effective gauge theory
(APEGT) of QCD is obtained by integrating out all the massive
degrees of freedom in the sense of Wilsonian renormalization group
(RG) \cite{Wilson}.   Therefore the resulting APEGT for $G=SU(2)$ is
written in terms of the Abelian field variables only.   In fact, the
APEGT obtained in the previous paper is written in terms of the
maximal Abelian U(1) gauge field
$a_\mu$, the dual Abelian gauge field $b_\mu$ and the magnetic
monopole current
$k_\mu$ which couples to $b_\mu$. This theory is an interpolating
theory in the sense that it gives two dual descriptions of the same
physics, say, quark confinement.   APEGT tells us that the dual
theory which is more suitable in the strong coupling region is given
by the dual Ginzburg-Landau (GL) theory, i.e., dual Abelian gauge
Higgs model \cite{NO73}.    That is to say, monopole condensation
provides the mass
$m_b$ for the dual gauge field and leads to the linear or confining
static potential between quarks and the non-zero string tension
$\sigma$ is given by $\sigma \sim m_b^2$.
APEGT is regarded as a low-energy effective theory of QCD in the
distance scale $R>m_A^{-1}$ with $m_A$ being the non-zero mass of the
off-diagonal gluons. Consequently, the Abelian
dominance
\cite{EI82,SY90} in the physics in the long distance
$R>R_c:=m_A^{-1}$ will be realized in APEGT.   Quite recent
simulation by Amemiya and Suganuma
\cite{ASu97} shows that the propagator of the off-diagonal charged
gluon behaves as the massive gauge boson and provides the
short-range interaction, while the diagonal gluon propagates long
distance.  For SU(2) YM theory, they obtain
$m_A \cong 0.9$ GeV corresponding to the $R_c=4.5$fm. 
In fact, the massiveness of off-diagonal gluons is analytically
derived as a byproduct in this paper.
\par
In our formulation of YM theory, the non-perturbative treatment of YM
theory in the low-energy region can be reduced to that of the TFT in
the sense that any perturbation from the TFT does not change
essentially the result on low-energy physics obtained from the TFT.
So we can hope that the essential contribution for quark confinement
is derived from the TFT alone. In light of monopole dominance, the
TFT should be constructed such that the monopole degrees of freedom
are included as the most dominant topological configuration in
the TFT. If  quark confinement is proved based on the TFT, the
monopole dominance will be naturally understood  by this construction
of the TFT.  Furthermore, this will shed light on a possible
connection with the instanton configuration which is the only
possible topological nontrivial configuration in four-dimensional
Euclidean YM theory without partial gauge fixing.
\par
The purpose of this paper is to prove quark confinement within the
reformulation of the four-dimensional QCD based on the
criterion of area law for the Wilson loop \cite{Wilson74} 
(see section 6).  Here the Wilson loop is taken to be planar and
diagonal
\footnote{
The full non-Abelian Wilson loop will be treated in
a subsequent paper
\cite{KondoIV}, see Discussion.
}
in the maximal torus group $H$ (as taken by Polyakov
\cite{Polyakov77}). Although actual calculations are presented only
for the SU(2) case,
 out strategy of proving quark confinement is also applicable to
SU(N) case and more generally to arbitrary compact Lie group.
\par
This paper is organized as follows.
In section 2, 
the TFT is constructed from a gauge-fixing and Faddeev-Popov term. 
The action is written as a BRST exact form
according to the standard procedure of BRST formalism.  
In other words, the TFT is written as a BRST transformation of
a functional of the field variables including ghosts. 
Here we take the MAG as a gauge fixing condition.  Then the MAG
fixes the coset $G/H$ of the gauge group $G$ and leaves the maximal
torus subgroup $H$ unbroken.
Consequently, YM theory is reformulated as a  (perturbative)
fluctuation around the non-trivial topological configuration given
by TFT.
\par
In section 3, it is shown that a version of MAG allows us to
write the TFT in the form (\ref{myTFT}) which is both BRST and
anti-BRST exact. This version of TFT is called the MAG TFT
hereafter.  We find that the MAG TFT has a hidden supersymmetry
(SUSY) based on the superspace formulation 
\cite{PS79,MNTW83,MNT82,Cardy83,KLF84} of BRST invariant theories
\cite{BT81,Baulieu85}. 
The hidden SUSY plays the quite remarkable role in the next section.
\par
In section 4, it turns out that this choice of MAG leads to the
dimensional reduction in the sense of Parisi and Sourlas (PS)
\cite{PS79}. Consequently
the $D$-dimensional MAG TFT is reduced to the equivalent
$(D-2)$-dimensional coset $G/H$ nonlinear sigma model (NLSM). 
This means the equivalence  of the partition function in two theories. 
Furthermore, the PS dimensional reduction tells us that the
calculation of correlation functions 
in $D$-dimensional TFT can be performed in the equivalent
$(D-2)$-dimensional model if the arguments $x_i$ lie on a certain
$(D-2)$-dimensional subspace, because the correlation function
coincides with the same correlation function calculated in the
$(D-2)$-dimensional equivalent model defined on the subspace on which
$x_i$ lies,
\begin{eqnarray}
  \langle \prod_i {\cal F}_i(x_i) \rangle_{MAGTFT_D}
  = \langle \prod_i {\cal F}_i(x_i) \rangle_{G/H NLSM_{D-2}} .
\end{eqnarray}

\par
In section 5, we study concretely the case of $G=SU(2)$ YM theory in
4 dimensions.  In this case,
$H=U(1)$ and the equivalent dimensionally reduced model is given
by the two-dimensional O(3) NLSM.  The two-dimensional NLSM on group
manifolds or the principal chiral model is exactly solvable 
\cite{Luscher78,BZL76,ZZ79,AAR91,FZ91,FOZ93,BLZ94,HMN90,Wiegmann85,
Tsvelik88,NT91,PW83}.
Therefore, the four-dimensional MAG TFT defined in this paper is
exactly solvable.  It is known that the two-dimensional O(3) NLSM is
renormalizable and asymptotic free
\cite{Polyakov75b,BHZ80}.  Moreover, it has instanton solution as a
topological soliton 
\cite{BP75,Perelomov87,Woo77,GP78,Perelomov78}.
 Instanton is a finite action solution of the field
equation and obtained as a solution of self-duality equation.  The
instanton (resp. anti-instanton) solution is given by the holomorphic
(resp. anti-holomorphic) function. 
\par
We show that the instanton (resp. anti-instanton) configuration in
two-dimensional O(3) NLSM can be identified with the magnetic
monopole (resp. anti-monopole) configuration in higher-dimensions. 
Furthermore, the instanton (resp. anti-instanton) configuration in
two dimensions is considered as the projection of instanton
(resp. anti-instanton) solution of four-dimensional YM theory on the
two-dimensional plane through the dimensional reduction.  From this
observation, we can see intimate connection between magnetic monopole
and instanton.  
In principle, the gluon propagator is calculable according to the
exact treatment of O(3) NLSM.  In the O(3) NLSM, dynamical mass
generation occurs and the correlation length becomes finite and all
the excitations are massive
\cite{PW83}.  This shows that the off-diagonal gluons are massive,
$m_A \not= 0$.  The mass is non-perturbatively generated and
behaves as 
$m_A \sim \exp(-4\pi^2/g^2)$.
\par
In section 6, the planar diagonal Wilson loop in $4$-dimensional
SU(2) MAG TFT is calculated in the two-dimensional equivalent model
by making use of the dimensional reduction.
Actual calculation is done in the dilute-instanton-gas
approximation  \cite{CDG77,RU78,CDG78} in two dimensions.
This is very similar to the calculation
of the Wilson loop in the Abelian Higgs model in two dimensions
\cite{Coleman85,Rajaraman89}.   We can pursue this analogy
further using the
$CP^1$ formulation of the O(3) NLSM.  In $CP^1$ formulation, the
residual U(1) symmetry is manifest and we can introduce the U(1)
gauge field coupled to two complex scalar fields, whereas in the
NLSM, the U(1) gauge invariance is hidden, since the field variable
${\bf n}(z)$ is gauge invariant.  The $CP^1$ formulation indicates
the correspondence of the TFT to the GL theory. As a result,
existence of topological non-trivial configuration corresponding
to the magnetic monopole and anti-monopole in YM theory in MAG is
sufficient to prove quark confinement in the sense of area law of
the diagonal Wilson loop. 
\par
In the end of 1970's, two-dimensional NLSMs were extensively studied
motivated by their similarity with the four-dimensional YM theory. 
Some of the NLSMs exhibit renormalizability, asymptotic freedom,
$\theta$ vacua and instanton solution. 
These analogies are not accidental in our view.  
Now this is understood
as a consequence of dimensional reduction.
The beta function in the two-dimensional O(3) NLSM has been calculated
by Polyakov \cite{Polyakov75b}. This should coincide with the beta
function of four-dimensional SU(2)/U(1) MAG TFT.  
Now we will be able to understand why Migdal and Kadanoff approximate
RG scheme \cite{MK75} yields reasonably good results.
\par
It should be remarked that the dimensional reduction is possible
also for  the gauge fixing other than MAG. Such an example was
proposed by Hata and Kugo \cite{HK85} which is called the pure gauge
model (PGM). However, the choice of MAG as gauge-fixing condition is
essential to prove quark confinement based on the non-trivial
topological configuration, because MAG leads to the G/H NLSM by the
dimensional reduction. The two-dimensional coset $SU(N)/U(1)^{N-1}$
NLSM can have soliton solution as suggested by
\begin{eqnarray}
 \Pi_2(SU(N)/U(1)^{N-1}) = Z^{N-1} .
 \label{homotopy}
\end{eqnarray}
However, the two-dimensional NLSM obtained  from PGM by
dimensional reduction does not have any instanton solution, since
\begin{eqnarray}
 \Pi_2(SU(N)) = 0 .
\end{eqnarray}
Therefore the PGM loses a chance of proving quark confinement based
on non-trivial topological configuration and more endeavor is needed
to prove quark confinement based on the perturbative or
non-perturbative treatment around the topologically trivial
configuration
\cite{Hata84,HK85,HN93,HT94,HT95,Kugo95}. Moreover, MAG has clear
physical meaning which leads to the dual superconductor picture of
QCD vacuum as shown in \cite{KondoI}. This is not the case in PGM.  
In fact, there is a claim \cite{Suzuki83} that the criterion of Kugo
and Ojima  for color confinement \cite{KO79,Nishijima94} is
different from the Wilson criterion.

\par
It is possible to extend our treatment to arbitrary
compact Lie group $G$ along the same line as above, as long as
the existence of instanton solution is guaranteed by the
non-trivial homotopy group, $\Pi_2(G/H) \not=0$. 
Though the dilute-gas approximation is sufficient to deduce the
linear potential,  it is better to compare this result with those
obtained by other methods.  
For this purpose, it is worth performing the $1/N$ expansion to know
the result especially for $N>2$.  
The $O(N)$ and $CP^{N-1}$ models have been extensively studied so far
\cite{BLS76,DLD78,Eichenherr78,Macfarlane79,DDL79,Witten79,Jevicki79,
Affleck80,Iwasaki81}.
However,  
$SU(N)/U(1)^{N-1}$ is isomorphic to $O(N+1)$ or $CP^{N-1}$ only when
$N=2$, and the two-dimensional $O(N)$ NLSM has no instanton solution
for $N>3$. 
To author's knowledge, the $1/N$ analysis of two-dimensional coset
$SU(N)/U(1)^{N-1}$ NLSM has not been worked out, probably due to the
reason that $SU(N)/U(1)^{N-1}$ is not a symmetric space as a
Riemannian manifold
\cite{AJ78}.    
\par
It should be remarked that the resolution of quark confinement is not
simply to show that the full gluon propagator behaves as $1/k^4$ in
the infrared region as
$k \rightarrow 0$.  The correct picture of quark confinement must be
able to explain the anisotropy (or directional dependence) caused by
the existence of widely separated quark--anti-quark pair if we stand
on the dual superconductivity scenario.  This is necessary to deduce
the QCD (hadron) string picture.  Our proof of quark
confinement is possible only when the two-dimensional plane on which
a pair of quark and anti-quark exists is selected as a subspace of
dimensional reduction.  Hence this feature is desirable from the
viewpoint of string picture.  In fact, the effective Abelian gluon
propagator obtained from the dual description in APEGT
shows such an anisotropy
\cite{KondoI}.  
\par
The dimensional reduction of TFT opens a path for analyzing 
non-perturbative problems in four-dimensional YM theory based on  
various technologies developed for two-dimensional field theories,
such as Bethe ansatz \cite{PW83}, conformal field theory (CFT)
\cite{BPZ84,KZ84}.
They are intimately connected to the Wess-Zumino-Novikov-Witten model
\cite{Witten84}, non-Abelian bosonization \cite{Witten84,BQ94},
quantum spin model \cite{Affleck89}, Chern-Simons theory
\cite{Witten89}, induced potential in the path integral
\cite{MT97,KTT97} and so on.  The exact solubility is pulled up at
the level of correlation function, not the field equation.  This
should be compared with the Hamiltonian reduction of YM self-duality
equation
\cite{MW96}.
Furthermore, the APEGT obtained in
MAG can have the same meaning as the low-energy effective theory of
N=2 supersymmetric YM theory and QCD obtained by Seiberg and Witten
\cite{SW94}. This issue will be discussed in subsequent papers.

%\newpage
\section{Yang-Mills theory as a deformation of topological
field theory}
\setcounter{equation}{0}

First, we summarize the BRST formulation of YM theory in the
manifestly covariant gauge and subsequently introduce the MAG.  Next,
we derive the TFT describing the magnetic monopole from the YM theory
in MAG.  The TFT is obtained from the gauge fixing part of the YM
theory.  Finally, the YM theory in MAG is reformulated as a
(perturbative) deformation of the TFT.

\subsection{Yang-Mills theory and gauge fixing}

We consider the Yang-Mills (YM) theory with a
gauge group
$G=SU(N)$ on the $D$ dimensional space-time described by the
action $(D>2)$,
\begin{eqnarray}
 S_{tot}  &=& \int d^Dx ({\cal L}_{QCD}[{\cal A},\psi]
 + {\cal L}_{GF}),
 \\
 {\cal L}_{QCD}[{\cal A},\psi] &:=& -{1 \over 2g^2} 
 {\rm tr}_G({\cal F}_{\mu\nu}{\cal F}_{\mu\nu})
 + \bar \psi (i \gamma^\mu {\cal D}_\mu[{\cal A}] - m) \psi ,
\end{eqnarray}
where 
\begin{eqnarray}
 {\cal F}_{\mu\nu}(x) 
&:=& \sum_{A=1}^{N^2-1} {\cal F}_{\mu\nu}^A(x) T^A
:=   \partial_\mu {\cal A}_\nu(x) 
 -   \partial_\nu {\cal A}_\mu(x)
 - i [{\cal A}_\mu(x), {\cal A}_\nu(x)],
  \\
  {\cal D}_\mu[{\cal A}] &:=& \partial_\mu - i {\cal A}_\mu .
\end{eqnarray}
The gauge fixing term ${\cal L}_{GF}$ is specified below.
We adopt the following convention.
The generators $T^A(A=1, \cdots,
N^2-1)$  of the Lie algebra ${\cal G}$ of the gauge group $G=SU(N)$
are taken to be hermitian satisfying 
$
 [T^A, T^B] = i f^{ABC} T^C
$
and normalized as
$
 {\rm tr}(T^A T^B) =  {1 \over 2} \delta^{AB}.
$
The generators in the adjoint representation are given by
$
 [T^A]_{BC} = -i f_{ABC} .
$
We define the quadratic Casimir operator by
$
 C_2(G) \delta^{AB} = f^{ACD}f^{BCD}.
$
Let $H$ be the maximal torus group of $G$ and $T^a$ be the
generators in the Lie algebra ${\cal G}\setminus{\cal H}$ of the
coset
$G/H$ where ${\cal H}$  is the Lie algebra of $H$.
\par
For $G=SU(2)$,
$T^A = (1/2)
\sigma^A (A=1,2,3)$ with Pauli matrices $\sigma^A$ and the structure
constant is
$f^{ABC} = \epsilon^{ABC}$.
The indices $a, b, \cdots$ denote the off-diagonal parts of the
matrix representation.
The Cartan decomposition is given by
\begin{eqnarray}
 {\cal A}_\mu(x) = \sum_{A=1}^3  {\cal A}_\mu^A(x) T^A
 :=  a_\mu(x)  T^3 
 + \sum_{a=1}^{2}  A_\mu^a(x) T^a .
\end{eqnarray}
\par
Under the gauge transformation, the gauge field ${\cal A}_\mu(x)$
transforms as
\begin{eqnarray}
{\cal A}_\mu(x) \rightarrow  
{\cal A}_\mu^U(x) &:=& 
 U(x) {\cal A}_\mu(x) U^\dagger(x) + i
U(x) \partial_\mu U^\dagger(x) . 
\label{GT0}
\end{eqnarray}
This gauge degrees of freedom is fixed by the procedure of gauge
fixing.  A covariant choice is given by the Lorentz gauge, 
\begin{eqnarray}
   F[{\cal A}] := \partial_\mu {\cal A}^\mu = 0 .
\end{eqnarray}
The procedure of gauge fixing must be done in such a way that the
gauge fixing condition is preserved also for the gauge rotated field
${\cal A}_\mu^U$, i.e., $F[{\cal A}^U]=0$. This is guaranteed by the
Faddeev-Popov (FP) ghost term. 
\par
We formulate the theory based on the BRST formalism.   In the BRST
formalism, the gauge-fixing and FP part ${\cal
L}_{GF}$ is specified by a functional $G_{gf}$ of the field
variables through the relation, 
\begin{eqnarray}
 {\cal L}_{GF} := - i \delta_B G_{gf}[{\cal A}_\mu, {\cal C}, \bar
{\cal C},
\phi] ,
 \label{GF}
\end{eqnarray}
where ${\cal C}, \bar {\cal C}$ are ghost, anti-ghost fields and
$\phi$ is the Lagrange multiplier field for incorporating the gauge
fixing condition. Here $\delta_B$ denotes the nilpotent BRST
transformation,
\begin{eqnarray}
   \delta_B {\cal A}_\mu(x)  &=&  {\cal D}_\mu {\cal C}(x)
   := \partial_\mu {\cal C}(x) - i [{\cal A}_\mu(x), {\cal C}(x)],
    \nonumber\\
   \delta_B {\cal C}(x)  &=& i{1 \over 2}[{\cal C}(x), {\cal C}(x)],
    \nonumber\\
   \delta_B \bar {\cal C}(x)  &=&   i \phi(x)  ,
    \nonumber\\
   \delta_B \phi(x)  &=&  0 ,
    \nonumber\\
   \delta_B \psi(x)  &=&   i {\cal C}(x) \psi(x) .
    \label{BRST0}
\end{eqnarray}
The partition function of QCD is given by
\begin{eqnarray}
 Z_{QCD}[J] := \int [d{\cal A}_\mu][d{\cal C}][d\bar {\cal C}]
 [d\phi][d\psi][d\bar \psi] \exp \left\{ i (S_{tot} + S_J) \right\},
\end{eqnarray}
where the source term is introduced as
\begin{eqnarray}
   S_J := \int d^Dx \{ {\rm tr}[J^\mu {\cal A}_\mu + J_c {\cal C} 
   + J_{\bar c} \bar {\cal C} + J_\phi \phi] +
\bar \eta \psi + \eta \bar \psi \} .
\end{eqnarray}
\par
In the BRST formalism, both the gauge-fixing and the FP terms are
automatically produced according to (\ref{GF}).  The most familiar
choice of $G$ is
\begin{eqnarray}
  G_{gf} = {\rm tr}_{\cal G}[\bar {\cal C}(\partial_\mu {\cal A}^\mu 
+ {\alpha \over 2}\phi)] .
\end{eqnarray}
This yields
\begin{eqnarray}
 {\cal L}_{GF} &:=& - i \delta_B G_{gf}[{\cal A}_\mu, {\cal C}, \bar
{\cal C},
\phi] 
 =  {\rm tr}_{{\cal G}}[\phi \partial_\mu {\cal A}^\mu + i \bar {\cal C}
\partial^\mu {\cal D}_\mu[{\cal A}] {\cal C} + {\alpha \over 2}
\phi^2] .
\end{eqnarray}

\subsection{MAG and singular configuration}
\par
In the previous paper \cite{KondoI}, we examined the maximal
Abelian gauge (MAG) as an example of Abelian
projection \cite{tHooft81}.  For G=SU(2),  MAG is given by
\begin{eqnarray}
 F^{\pm}[A,a] &:=& (\partial^\mu \pm i  a^\mu) A_\mu^{\pm} = 0,
\label{dMAG}
\end{eqnarray}
using the $(\pm, 3)$ basis,
\begin{eqnarray}
{\cal O}^{\pm} := ({\cal O}^1 \pm i{\cal O}^2)/\sqrt{2} .
\end{eqnarray}
The simplest choice of $G_{gf}$ for MAG in $(\pm, 3)$ basis
is given by
\begin{eqnarray}
  G_{gf} = \sum_{\pm}
  \bar C^{\mp} (F^{\pm}[A,a] + {\alpha \over 2} \phi^{\pm}) ,
\label{dMAG0}
\end{eqnarray}
which is equivalently rewritten in the usual basis as
\begin{eqnarray}
  G_{gf} &=& \sum_{a=1,2}
  \bar C^{a} (F^{a}[A,a] + {\alpha \over 2} \phi^{a}) ,
\\
 F^{a}[A,a] &:=& (\partial^\mu \delta^{ab} 
   - \epsilon^{ab3} a^\mu) A_\mu^b 
   := D^\mu{}^{ab}{}[a] A_\mu^b .
   \label{dMAG1}
\end{eqnarray}

\par
The basic idea of Abelian projection proposed by 't Hooft
\cite{tHooft81} is to remove as many non-Abelian degrees of freedom
as possible, by partially fixing the gauge in such a way that the
maximal torus group $H$ of the gauge group $G$ remains unbroken.
Under the Abelian projection, $G=SU(N)$ gauge theory reduces to the
$H=U(1)^{N-1}$ Abelian gauge theory plus magnetic monopoles.
Actually, the choice (\ref{dMAG}) for $G=SU(2)$ is nothing but the
condition of  minimizing the functional ${\cal R}[A]$ for the gauge
rotated off-diagonal gluon fields
$A$, i.e.,
 ${\rm min}_{U} {\cal R}[A^U]$,
\begin{eqnarray}
  {\cal R}[A] := {1 \over 2}\int d^Dx [(A_\mu^1(x))^2 +
(A_\mu^2(x))^2 ]
  = \int d^Dx A_\mu^{+}(x) A_\mu^{-}(x) .
  \label{MAG}
\end{eqnarray}
We can generalize the MAG to arbitrary group $G$ as
\begin{eqnarray}
  {\cal R}[A] := \int d^Dx  {\rm tr}_{{\cal G}\setminus{\cal H}}[ {1
\over 2} {\cal A}_\mu(x) {\cal A}_\mu(x)] ,
  \label{MAG2}
\end{eqnarray}
where the trace is taken over the Lie algebra ${\cal G}\setminus{\cal
H}$. Under the MAG, it is shown \cite{KondoI} that  integration of
the off-diagonal gluon fields
$A_\mu^a \in {\cal G}\setminus{\cal H}$ in the SU(2) YM theory leads
to the Abelian-projected effective gauge theory (APEGT)  
written in terms of the maximal Abelian U(1) gauge field $a_\mu$,
the dual U(1) gauge field
$b_\mu$ and the magnetic (monopole) current $k_\mu$.
\par
In the gauge transformation (\ref{GT0}), the local gauge rotation
$U(x)$ is performed in such a way that the gauge rotated field
${\cal A}_\mu^U(x)$  minimize the functional ${\cal R}[{\cal A}^U]$
and hence satisfies the gauge-fixing condition (\ref{dMAG}). 
We define the magnetic current by
\begin{eqnarray}
 k_\mu(x) &:=& \epsilon_{\mu\nu\rho\sigma} 
 \partial^\mu f^{\rho\sigma}(x),
 \\
 f_{\rho\sigma}(x) &:=& \partial_\rho a_\sigma^U(x) - \partial_\sigma
a_\rho^U(x) ,
\end{eqnarray}
using the Abelian part
(diagonal part)  extracted as
\begin{eqnarray}
 a_\mu^U(x) := {\rm tr}[T^3 {\cal A}_\mu^U(x)] .
\end{eqnarray}
If the gauge field ${\cal A}_\mu(x)$ is not singular, the first piece
$U(x) {\cal A}_\mu(x) U^\dagger(x)$ of ${\cal A}_\mu^U(x)$  is
non-singular and does not give rise to magnetic current.
On the contrary, the second piece $\Omega_\mu(x)$,
\begin{eqnarray}
 \Omega_\mu(x) :=  i U(x) \partial_\mu U^\dagger(x) 
\end{eqnarray}
does give the non-vanishing magnetic monopole current 
(see e.g. \cite{KondoI}) for 
\begin{eqnarray}
 a_\mu^\Omega(x) := \Omega_\mu^3(x) :={\rm tr}[T^3 \Omega_\mu(x)] .
\end{eqnarray}
 According to Monte Carlo simulation on the lattice
\cite{review}, the magnetic monopole part gives the most
dominant contribution in various quantities characterizing the
low-energy physics of QCD, e.g., string tension, chiral condensate,
topological charge, etc.
\par
Therefore, it is expected that the most important degrees of
freedom for the low-energy physics comes from the second piece
$\Omega_\mu(x)$ of ${\cal A}_\mu^U(x)$.
Therefore, we decompose the YM theory into two
pieces, i.e., the contribution from  the part
$\Omega_\mu(x)$ and the remaining part.

\subsection{Magnetic monopole in Non-Abelian gauge theory}

\par
First we recall the calculation of the Abelian (diagonal) field
strength in four-dimensional YM theory.
We introduce three local field variables corresponding to the
Euler angles
\begin{eqnarray}
(\theta(x), \varphi(x), \chi(x)) , ( \theta \in [0,\pi], 
\varphi \in [0,2\pi], \chi \in [0, 2\pi] ) 
\end{eqnarray}
to write an element $U(x) \in SU(2)$ as
\begin{eqnarray}
 U(x)&=& e^{i \chi(x) \sigma_3/2} 
   e^{i \theta(x) \sigma_2/2} 
   e^{i \varphi(x) \sigma_3/2}  
   \nonumber\\
 &=& \pmatrix{
 e^{{i \over 2}(\varphi(x)+\chi(x))} \cos {\theta(x) \over 2} &
 e^{-{i \over 2}(\varphi(x)-\chi(x))} \sin {\theta(x) \over 2} \cr
 -e^{{i \over 2}(\varphi(x)-\chi(x))} \sin {\theta(x) \over 2} &
 e^{-{i \over 2}(\varphi(x)+\chi(x))} \cos {\theta(x) \over 2} } .
\end{eqnarray}
In the usual convention of perturbation theory, we take
\begin{eqnarray}
 \Omega_\mu(x) := {i \over g}
U(x) \partial_\mu U^\dagger(x) .
\end{eqnarray}
Note that the following identity \cite{KondoI} holds
for $\Omega_\mu$,
\begin{eqnarray}
 \partial_\mu \Omega_\nu(x)  - \partial_\nu \Omega_\mu(x)
 = ig[\Omega_\mu(x), \Omega_\nu(x)] + {i \over g}
U(x)[\partial_\mu, \partial_\nu]U^\dagger(x) .
\label{ident}
\end{eqnarray}
Then the diagonal part reads
\begin{eqnarray}
 f_{\mu\nu}^\Omega(x) := 
 \partial_\mu \Omega_\nu^3(x)  - \partial_\nu \Omega_\mu^3(x)
 =   C_{\mu\nu}^{[\Omega]}(x) + {i \over g}
(U(x)[\partial_\mu,
\partial_\nu]U^\dagger(x))^{(3)} ,
\label{dec}
\end{eqnarray}
where $C_{\mu\nu}$ is introduced in the previous paper
\cite{KondoI}, 
\begin{eqnarray}
 C_{\mu\nu}^{[\Omega]} := (ig[\Omega_\mu, \Omega_\nu])^{(3)}
 = g \epsilon^{ab3} \Omega_\mu^a \Omega_\nu^b 
 = ig (\Omega_\mu^+ \Omega_\nu^- - \Omega_\mu^- \Omega_\nu^+) .
\end{eqnarray}
Using Euler angle expression for
$U$, we obtain
\begin{eqnarray}
 i(\Omega_\mu^+ \Omega_\nu^- - \Omega_\mu^- \Omega_\nu^+)
 = {1 \over g^2} 
 \sin \theta (\partial_\mu \theta \partial_\nu \varphi
 - \partial_\mu \varphi \partial_\nu \theta) ,
\end{eqnarray}
which implies
\begin{eqnarray}
 C_{\mu\nu}[\Omega]
= {1 \over g} \sin \theta (\partial_\mu \theta \partial_\nu \varphi
 - \partial_\mu \varphi \partial_\nu \theta) .
\label{corr}
\end{eqnarray}
Now we show that  $C_{\mu\nu}^{[\Omega]}$ denotes the monopole
contribution to the diagonal field strength
$f_{\mu\nu}$.
Note that $C_{\mu\nu}^{[\Omega]}$ is generated from the off-diagonal
gluon fields, $\Omega_\mu^1, \Omega_\mu^2$.
\par
In four dimensions, the magnetic monopole charge is calculated 
from the magnetic current,
\begin{eqnarray}
  k_\mu  =   \partial_\nu \tilde f_{\mu\nu}^\Omega ,
\quad
\tilde f_{\mu\nu} := {1 \over 2}
\epsilon_{\mu\nu\rho\sigma} f_{\rho\sigma} ,
\end{eqnarray}
as
\begin{eqnarray}
 g_m(V^{(3)}) &=& \int_{V^{(3)}} d^3 \sigma_\mu k_\mu 
 =   \int_{V^{(3)}} d^3 \sigma_\mu
 \partial_\nu \tilde f_{\mu\nu}^\Omega
 \nonumber\\
&=&   \int_{S^{(2)}=\partial V^{(3)}} d^2
\sigma_{\mu\nu} \tilde f_{\mu\nu}^\Omega .
\end{eqnarray}
We can identity the first and second parts of right-hand-side (RHS)
of (\ref{dec})  with the  the magnetic monopole and the Dirac
string part respectively contained in the TFT$_4$ theory and hence
the YM$_4$ theory \cite{KondoI}. 
This is clearly seen by the explicit calculation using Euler
angles, since we can rewrite (\ref{dec}) as
\begin{eqnarray}
 f_{\mu\nu}^\Omega 
 = - {1 \over g}\sin \theta (\partial_\mu \theta \partial_\nu \varphi
 - \partial_\mu \varphi \partial_\nu \theta)
 + {1 \over g}([\partial_\mu, \partial_\nu] \chi 
 + \cos \theta  [\partial_\mu, \partial_\nu] \varphi) .
\end{eqnarray}
The magnetic monopole part is given by
\begin{eqnarray}
g_m(V^{(3)}) &=&  
{1 \over 2g}\int_{S^{(2)}} d^2
\sigma_{\rho\sigma} \epsilon_{\mu\nu\rho\sigma} 
 \sin \theta (\partial_\mu \theta \partial_\nu \varphi
 - \partial_\mu \varphi \partial_\nu \theta) ,
\label{magc1}
\end{eqnarray}
while the Dirac string part is
\begin{eqnarray}
 g_{DS}(V^{(3)}) &=&  
{1 \over 2 g}\int_{S^{(2)}} d^2
\sigma_{\rho\sigma} \epsilon_{\mu\nu\rho\sigma} 
 ([\partial_\mu, \partial_\nu] \chi 
 + \cos \theta  [\partial_\mu, \partial_\nu] \varphi) .
 \label{defmc}
\end{eqnarray}
The first definition (\ref{magc1}) of $g_{m}$ gives the quantized
magnetic charge \cite{KondoI}.
The integrand is the Jacobian from
$S^2$ to $S^2$ as will be shown in section 5 and
\begin{eqnarray}
 \Pi_2(SU(2)/U(1)) = \Pi_2(S^2) = Z .
\end{eqnarray}
Then (\ref{magc1}) gives the magnetic charge $g_m$ satisfying the
Dirac quantization condition,
\begin{eqnarray}
 g_m ={2\pi n \over g} , \quad g g_m = 2\pi n \  (n \in Z) .
 \label{DQ}
\end{eqnarray}
In the second definition (\ref{defmc}) of $g_{m}$, if we choose
$\chi=-\varphi$ using residual U(1) gauge invariance, then the
Dirac string appears on the negative $Z$ axis, i.e., 
$\theta=\pi$.  In this case, the surface integral reduces to the line
integral around the string,
\begin{eqnarray}
 g_{DS}(V^{(3)}) &=&  
{1 \over 2 g} \int_{S^{(2)}} d  \sigma_{\mu\nu} 
\epsilon_{\mu\nu\rho\sigma}
 [\partial_\rho, \partial_\sigma] \varphi(x) 
\nonumber\\
&=& - {1 \over 2 g} \int_{S^{(1)}} d  \sigma_{\mu\nu\rho} 
\epsilon_{\mu\nu\rho\sigma} \partial_\rho \varphi(x) .
\end{eqnarray}
This gives the same result (\ref{defmc}) but with the minus sign,
in consistent with 
\begin{eqnarray}
  \Pi_1(U(1)) = Z .
\end{eqnarray}
Actually, two description are equivalent, as can be seen from
\begin{eqnarray}
 \Pi_2(SU(2)/U(1)) = \Pi_1(U(1)) .
\end{eqnarray}

\par
If  the contribution from 
$U(x) {\cal A}_\mu(x) U^\dagger(x) $ 
is completely neglected, i.e., 
${\cal A}_\mu^U(x) \equiv \Omega_\mu(x)
= i U(x) \partial_\mu U^\dagger(x)$, (\ref{ident}) implies
\begin{eqnarray}
 {\cal F}_{\mu\nu}^U(x) \equiv {i \over g}
 U(x) [\partial_\mu, \partial_\nu ]U^\dagger(x) ,
\label{fst}
\end{eqnarray}
where the RHS is identified with the contribution from the Dirac
string, see \cite{KondoI}. 
Note that the original YM theory does not have magnetic monopole
solution.  However, if we partially fix the gauge $G/H=SU(2)/U(1)$
and retain the residual $H=U(1)$ gauge, the theory can have singular
configuration.  This is a reason why the magnetic monopole appears
in the YM theory which does not have Higgs field.  The existence of
Dirac string in the RHS of (\ref{fst}) reflects the fact that the
field strength ${\cal F}_{\mu\nu}^U(x)$ does contain the magnetic
monopole contribution.  We have obtained a gauge theory with magnetic
monopole starting from the YM theory.  Therefore, MAG enables us to
deduce the magnetic monopole without introducing the scalar field, in
contrast to the 't Hooft-Polyakov monopole.
See \cite{KondoI} for more details.

\subsection{TFT and its deformation}

\par
Consequently, since the Dirac string does not contribute
to the action, the topological non-trivial sector with magnetic
monopole in YM theory is described by the gauge-fixing and FP ghost
term alone (we forget the matter field for a while),
\begin{eqnarray}
 S_{TFT}[\Omega_\mu, {\cal C}, \bar {\cal C}, \phi] 
 = \int d^Dx {\cal L}_{TFT},
\quad
 {\cal L}_{TFT} := - i \delta_B G_{gf}[\Omega_\mu, {\cal C}, \bar
{\cal C}, \phi] .
 \label{TFT}
\end{eqnarray}
This theory describes the topological field theory for the magnetic
monopole, which is called MAG TFT hereafter. If we restrict the
gauge rotation $U(x)$ to the regular one,
$\Omega_\mu(x)$ reduces to a pure gauge field 
${\cal F}_{\mu\nu}^U(x) \equiv 0$
and hence the TFT is reduced to the topological trivial theory.  
This
model is called the pure gauge model (PGM) which has been 
studied by Hata, Kugo, Niigata and Taniguchi
\cite{Hata84,HN93,HT94,HT95}.  
However, PGM has only unphysical gauge modes and does not have
physical modes.  We consider that the
topological objects must give the main contribution to the
low-energy physics. From this viewpoint, the PGM is not interesting
in our view, since PGM can not contain the topological non-trivial
configuration as will be shown in the following.
\par
In this paper, we  take into account the topological non-trivial
configuration involved in the theory (\ref{TFT}) and extract the most
important contribution in low-energy physics. 
We consider that the $\Omega_\mu(x)$ gives the most important
dominant contribution and remaining contributions are treated as a
perturbation around it. Whether this is efficient or not crucially
depends on the choice of
$G_{gf}$.   For this purpose,  MAG is most appropriate as will be
shown later.   
\par
Our reformulation of YM theory proceeds as follows.
First of all, we decompose the gauge field
${\cal A}_\mu(x)$ into the non-perturbative piece $\Omega_\mu(x)$
(including topological non-trivial configuration) and the
perturbative piece
$U(x) {\cal A}_\mu(x) U^\dagger(x)$
(including only the topological trivial configuration).
Next, we treat the original YM theory
as a perturbative deformation of the TFT written in terms of 
$\Omega_\mu(x)$ alone. 
Using the normalization of the field in perturbation theory, the TFT
is obtained from the YM theory in the limit of vanishing coupling
constant, $g \rightarrow 0$. 
If we absorb the coupling constant $g$ into the gauge field, TFT
does not have the coupling constant apparently. 
\par
We expect that the TFT of describing the magnetic monopole gives the
most dominant non-perturbative contributions in low-energy physics. 
In fact, the monopole dominance in low energy physics of QCD been
confirmed by Monte Carlo simulations
\cite{review}.  A similar attempt to formulate the YM theory as a
deformation of the topological BF theory was done by Martellini et
al. \cite{BFYM}.  The model is called the BF-YM theory.   A similar
attempt was also made by Izawa
\cite{Izawa93} for the PGM using the BF formulation in 3 dimensions.
The topological BF theory includes the topological non-trivial
configuration.
The APEGT for BF-YM theory can be constructed, see
\cite{KondoI}.
\par
First, we regard the field ${\cal A}_\mu$ and $\psi$ as the gauge
transformation of the fields ${\cal V}_\mu$ and $\Psi$ (we use
different character to avoid confusions),
\begin{eqnarray}
{\cal A}_\mu(x) &:=& 
 U(x) {\cal V}_\mu(x) U^\dagger(x) + \Omega_\mu(x),
 \quad \Omega_\mu(x) := {i \over g}
U(x) \partial_\mu U^\dagger(x) 
\\
  \psi(x) &:=& U(x) \Psi(x), 
\label{GT3}
\end{eqnarray}
where ${\cal V}_\mu$ and $\Psi$ are identified with the perturbative
parts in the topological trivial sector.
\par
Let $[dU]$ be the invariant Haar measure on the group $G$.
Using the gauge invariance of the FP determinant 
$\Delta[{\cal A}]$ given by
\begin{eqnarray}
 \Delta[{\cal A}]^{-1}  := \int [dU] \prod_{x} \delta(\partial^\mu
{\cal A}_\mu^{U^{-1}}(x)) ,
\quad 
\Delta[{\cal A}] =  \Delta[{\cal A}^{U^{-1}}] ,
\end{eqnarray}
we can rewrite
\begin{eqnarray}
 1 &=& \Delta[{\cal A}] \int [dU] \prod_{x} \delta(\partial^\mu {\cal
A}_\mu^{U^{-1}}(x))
\nonumber\\
&=& \Delta[{\cal A}^{U^{-1}}] \int [dU] \prod_{x} \delta(\partial^\mu
{\cal A}_\mu^{U^{-1}}(x))
\nonumber\\
&=& \Delta[{\cal V}] \int [dU] \prod_{x} \delta(\partial^\mu
{\cal V}_\mu(x)) 
\nonumber\\
&=& \int [d\gamma][d\bar \gamma][d\beta] \exp \left\{ i \int d^Dx (
{\rm tr}_{\cal G}(\beta \partial^\mu {\cal V}_\mu + i \bar \gamma
\partial^\mu {\cal D}_\mu[{\cal V}] \gamma ) \right\} 
\nonumber\\
&=& \int [d\gamma][d\bar \gamma][d\beta] \exp \left\{ i \int d^Dx 
[-i \tilde \delta_B \tilde G_{gf}({\cal V}_\mu, \gamma, \bar \gamma,
\beta)] \right\} ,
\end{eqnarray}
where
\begin{eqnarray}
 \tilde G_{gf}({\cal V}_\mu, \gamma, \bar \gamma, \beta) 
 := {\rm tr}_{\cal G}(\bar \gamma \partial^\mu {\cal V}_\mu) .
\end{eqnarray}
Here we have introduced new ghost field $\gamma$, anti-ghost
field $\bar \gamma$ and the multiplier field $\beta$ which are
subject to a new BRST transformation 
$\tilde \delta_B$,
\begin{eqnarray}
   \tilde \delta_B {\cal V}_\mu(x)  &=&  {\cal D}_\mu[{\cal V}]
\gamma(x) 
   := \partial_\mu \gamma(x)  - i [{\cal V}_\mu(x), \gamma(x) ],
    \nonumber\\
   \tilde \delta_B \gamma(x)  &=&  i{1 \over 2} [\gamma(x) ,
\gamma(x) ] ,
    \nonumber\\
   \tilde \delta_B \bar \gamma(x)  &=&   i \beta(x)  ,
    \nonumber\\
   \tilde \delta_B \beta(x)  &=&  0 ,
    \nonumber\\
   \tilde \delta_B \Psi(x)  &=&    \gamma(x) \Psi(x) .
    \label{BRST1}
\end{eqnarray}
Then the partition function is rewritten as
\begin{eqnarray}
 Z_{QCD}[J] &=& \int [dU][d{\cal C}][d\bar {\cal C}]
 [d\phi]
 \int [d{\cal V}_\mu][d\gamma][d\bar \gamma][d\beta]
 [d\Psi][d\bar \Psi] 
 \nonumber\\&&
\times \exp \Big\{ i \int d^Dx \Big[
 -i \delta_B G_{gf}[\Omega_\mu +  U{\cal V}_\mu U^\dagger, {\cal C}, 
 \bar {\cal C}, \phi] 
 \nonumber\\&& \quad \quad \quad \quad
 + {\cal L}_{QCD}[{\cal V},\Psi]  
-i \tilde \delta_B \tilde G_{gf}({\cal V}_\mu, \gamma, \bar \gamma,
\beta) \Big] + i S_{J}
\Big\} ,
\label{formula}
\end{eqnarray}
where
\begin{eqnarray}
   S_J = \int d^Dx \{ {\rm tr}_{{\cal G}}[J^\mu (\Omega_\mu +  U{\cal
V}_\mu U^\dagger) + J_c {\cal C} 
   + J_{\bar c} \bar {\cal C} + J_\phi \phi] +
\bar \eta U \Psi + \eta \bar \Psi U^\dagger \} .
\end{eqnarray}
The correlation functions of the original fundamental field
${\cal A}_\mu, \psi, \bar \psi$ is
obtained by differentiating $Z[J]$ with respect to the source 
$J_\mu, \bar \eta, \eta$.  
The integration over the fields 
$(U, {\cal C}, \bar {\cal C},\phi)$ should be treated
non-perturbatively.  The perturbative expansion around the TFT means
performing the integration over the new fields 
$({\cal V}_\mu, \gamma, \bar \gamma, \beta)$ after power-series
expansions in the coupling constant $g$.
\par
Assume that a choice of $G_{gf}$ allows the separation of the
variable in such a way
\begin{eqnarray}
&& -i \delta_B G_{gf}[\Omega_\mu +  U{\cal V}_\mu U^\dagger, {\cal
C},  \bar {\cal C}, \phi]
 \nonumber\\&&
 = -i \delta_B G_{gf}[\Omega_\mu, {\cal C}, \bar {\cal C}, \phi]
 +  i {\cal V}_\mu^A {\cal M}_\mu^{A}[U] 
 + {i \over 2} {\cal V}_\mu^A {\cal V}_\mu^B {\cal K}^{AB}[U] .
 \label{cond}
\end{eqnarray}
In the next section, we show that the MAG satisfies the condition
(\ref{cond}) and obtain the explicit form for 
${\cal M}_\mu, {\cal K}$.
Then, under the condition (\ref{cond}), the partition function is
rewritten as
\begin{eqnarray}
 Z_{QCD}[J] &:=& \int [dU][d{\cal C}][d\bar {\cal C}]
 [d\phi]
\exp \Big\{ i S_{TFT}[\Omega_\mu, {\cal C}, \bar {\cal C}, \phi] 
 \nonumber\\&&   
  +  i \int d^Dx {\rm tr}_{\cal G}[J^\mu \Omega_\mu + J_c {\cal C} 
   + J_{\bar c} \bar {\cal C} + J_\phi \phi]  
+ i W[U; J^\mu, \bar \eta, \eta]  \Big\} ,
\end{eqnarray}
where $W[U; J^\mu, \bar \eta, \eta]$ is the generating functional of
the connected correlation function of ${\cal V}_\mu$ in the
perturbative sector given by
\begin{eqnarray}
 e^{i W[U; J^\mu, \bar \eta, \eta]}
&:=& 
  \int [d{\cal V}_\mu][d\gamma][d\bar \gamma][d\beta]
 [d\Psi][d\bar \Psi] 
 \exp \Big\{ 
 i S_{pQCD}[{\cal V}_\mu,\Psi,\gamma, \bar \gamma, \beta]
\nonumber\\&&  
 +  i \int d^Dx \Big[  {\cal V}_\mu^A {\cal J}_\mu^A
 + {i \over 2} {\cal V}_\mu^A {\cal V}_\mu^B {\cal K}^{AB}[U]
 + {\rm tr}_{{\cal G}}(\bar \eta U \Psi + \eta \bar \Psi U^\dagger)
\Big] 
\Big\} ,
\end{eqnarray}
where
\begin{eqnarray}
 {\cal J}_\mu^A &:=& (U^\dagger J^\mu  U)^A  + i {\cal M}_\mu^{A}[U]
.
\end{eqnarray}
Here pQCD denotes the perturbative QCD (topological trivial
sector) defined by the action 
$S_{pQCD}$,
\begin{eqnarray}
  S_{pQCD}[{\cal V}_\mu,\Psi,\gamma, \bar \gamma, \beta] &:=& \int
d^Dx
\Big\{ {\cal L}_{QCD}[{\cal V}_\mu,\Psi]   -i \tilde \delta_B \tilde
G_{gf}[{\cal V}_\mu, 
\gamma, \bar \gamma, \beta] \Big\}  .
\end{eqnarray}
The deformation  $W[U; J^\mu, \bar \eta, \eta]$ should be calculated
according to the ordinary perturbation theory in the coupling
constant $g$.  When there is no external source for quarks, we have
\begin{eqnarray}
  i W[U; J^\mu, 0, 0]
&:=& 
\ln \langle 
 \exp \Big\{ i \int d^Dx \Big[
 {\cal J}_\mu^A(x) {\cal V}_\mu^A(x) 
 + {i \over 2} {\cal V}_\mu^A(x) {\cal V}_\mu^B(x) {\cal
K}^{AB}(x) \Big] \Big\}
\rangle_{pQCD} 
\nonumber\\&=& 
{1 \over 2} g^2 \int d^Dx  \int d^Dy 
\langle {\cal V}_\mu^A(x) {\cal V}_\nu^B(y) \rangle^c_{pQCD} 
\nonumber\\&& \times
\{ {\cal J}_\mu^A(x) {\cal J}_\nu^B(y) - \delta^D(x-y)
\delta_{\mu\nu} {\cal K}^{AB}[U](x) \}
 + O(g^4{\cal J}^4) .
 \label{pert}
\end{eqnarray}
Therefore,  
 $W[U; J^\mu, 0, 0]$ is expressed as a power series in the
coupling constant $g$ and goes to zero as $g \rightarrow 0$.
It turns out that the QCD is reduced to the TFT in the vanishing
limit of coupling constant.
Thus QCD has been reformulated as a deformation of TFT.
\par
In a similar way, we can reformulate QED as a deformation of a TFT,
see \cite{KondoIII}.

%\newpage
\section{Maximal Abelian gauge and hidden supersymmetry}
\setcounter{equation}{0}

The purpose of this section is to give some prerequisite which are
necessary to understand the dimensional reduction 
discussed in the next section. 
\par
First of all, we give a special version of the MAG which leads to the
dimensional reduction  of the TFT part obtained from the YM theory
in MAG. Using the BRST $\delta_B$ and anti-BRST
$\bar \delta_B$ transformations, the action of the MAG TFT is
written in the form,
\begin{eqnarray}
   S_{TFT} = \int d^D x \delta_B \bar \delta_B {\cal O}(x) .
\end{eqnarray}

\par
Second, we introduce the superfield formalism. 
The $(D+2)$-dimensional super space 
$X=(x^\mu, \theta, \bar \theta)$ is
defined by introducing two Grassmannian coordinates 
$\theta, \bar \theta$ in addition to the ordinary
(bosonic) $D$-dimensional coordinates $x^\mu (\mu = 1, \cdots, D)$.
We define the supersymmetry transformations and study the property
of the superfield which is invariant under the
supersymmetry transformation.
\par
Third, we give a geometrical meaning of the BRST and anti-BRST
transformations in the superspace.
The gauge field 
${\cal A}_\mu(x)$ is extended into a superfield 
${\cal A}(X)$ as the connection one-form in the superspace.
A merit of this formalism lies in a fact that we can give a
geometrical meaning also to the FP ghost and anti-ghost fields;
actually the FP ghost and anti-ghost fields can be identified as
connection fields in the superspace.  
Furthermore, BRST transformation is rewritten as
a geometrical condition, the horizontal condition.  Consequently,
the BRST transformation $\delta_B$ (resp. anti-BRST transformation
$\bar \delta_B$) of the field variable coincides with the 
derivative
${\partial \over \partial \theta}$ 
(resp. ${\partial \over \partial \bar \theta}$ ) in the direction
$\theta$ (resp. $\bar \theta$).  
Taking into account that the differentiation is equivalent to the 
integration for the Grassmannian variable,  we can write the MAG TFT
in a manifestly supersymmetric covariant form, 
\begin{eqnarray}
   S_{TFT} = \int d^D x \int d\theta \int d\bar \theta 
   {\cal O}(x,\theta,\bar \theta) ,
\end{eqnarray}
where ${\cal O}(x)$ is extended to the superfield
${\cal O}(X)={\cal O}(x,\theta,\bar \theta)$ 
and ${\cal O}(x,\theta,\bar \theta)$ has OSp(D/2) invariant form.
This implies the existence of the hidden supersymmetry in MAG TFT
which is an origin of the dimensional reduction shown in the next
section.  

\subsection{Choice of MAG}

In the previous section, we considered the simplest MAG condition,
(\ref{dMAG0}) which leads to
\begin{eqnarray}
  {\cal L}_{GF} &=& 
  \phi^a F^a[A,a] + {\alpha \over 2} (\phi^a)^2
  + i \bar C^a D^\mu{}^{ab}[a]  D_\mu^{bc}[a] C^c
  \nonumber\\
  &&  
- i  \bar C^a [A_\mu^a A^\mu{}^b - A_\mu^c A^\mu{}^c
\delta^{ab} ] C^b 
 + i \bar C^a \epsilon^{ab3} F^b[A,a]  C^3 .
    \label{BRSTGF2}
\end{eqnarray}
Note that we can take more general form for $G_{gf}$
\cite{HN93,KondoI},
\begin{eqnarray}
  G_{gf} &=& \sum_{\pm}
  \bar C^{\mp} (F^{\pm}[A,a] + {\alpha \over 2} \phi^{\pm})
  +  \zeta C^3 \bar C^+ \bar C^-  
  +  \eta \sum_{\pm} (\pm) 
  \bar C^3 \bar C^{\pm} C^{\mp} .
\label{G2}
\end{eqnarray}
\par
In what follows, we choose a specific form 
\begin{eqnarray}
  G_{gf}' = \sum_{\pm}
  \bar C^{\mp} (F^{\pm}[A,a] - \phi^{\pm})
  - 2 C^3 \bar C^+ \bar C^- ,
\label{dMAG2}
\end{eqnarray}
which corresponds in (\ref{G2}) to 
\begin{eqnarray}
 \alpha =-2, \quad \zeta =-2, \quad \eta=0 .
\label{coeff}
\end{eqnarray}
Then the gauge fixing part 
${\cal L}_{GF}=-i \delta_B G_{gf}$ has the additional
contribution,
\begin{eqnarray}
 {\cal L}_{GF}'
 &=& {\cal L}_{GF} - \zeta \sum_{\pm} (\pm) C^3 \bar C^{\mp}
\phi^{\pm} 
 - \zeta \bar C^+ \bar C^- C^+ C^-  
\nonumber\\
 &=& {\cal L}_{GF} 
 - \zeta \sum_{a,b} i  \epsilon^{ab3} C^3 \bar C^{a} \phi^b
  - \zeta \bar C^+ \bar C^- C^+ C^- .
\end{eqnarray}
The four-ghosts interaction term is generated. This is a general
feature of non-linear gauge-fixing.
\footnote{
Such a term is necessary to
renormalize the YM theory in MAG, since MAG is a non-linear
gauge-fixing.
This is reflected to a fact that the U(1) invariant four-ghosts
interaction term
$\bar C^+ \bar C^- C^+ C^-$ is produced through the
expansion of $\ln \det Q$ (see Ref.~\cite{KondoI}), 
\begin{eqnarray}
 (\bar C^a  C^b - \bar C^c C^c \delta^{ab})
(\bar C^b  C^a - \bar C^d C^d \delta^{ba})
= - 2 \bar C^1 C^1 \bar C^2 C^2
= - 2 \bar C^+ \bar C^- C^+ C^-  .
\end{eqnarray}
}
Separating  the $\phi^a$ dependent terms and integrating out the
field $\phi^a$, we obtain
\begin{eqnarray}
 S_{GF} &=&  \int d^Dx \Big[
 - {1 \over 2\alpha}(F^a[A,a]+J_\phi^a
 + \zeta i\epsilon^{ab3} C^3 \bar C^{b})^2
  + i \bar C^a D^\mu{}^{ab}[a]  D_\mu^{bc}[a] C^c
  \nonumber\\ &&  \quad \quad
- i  \bar C^a (A_\mu^a A^\mu{}^b - A_\mu^c A^\mu{}^c \delta^{ab} )
C^b 
 + i \bar C^a \epsilon^{ab3} F^b[A,a]  C^3 
  \nonumber\\ &&   \quad \quad
  - \zeta \bar C^+ \bar C^- C^+ C^- 
   +  A_\mu^a J^\mu{}^a \Big] ,
\end{eqnarray}
where we have included the source term,
$J_\phi^a \phi^a + J_\mu^a A_\mu^a$.
Thus the action is summarized as
\begin{eqnarray}
 S_{GF} &=&  \int d^Dx \Big[
  {1 \over 2g^2} A_\mu^a  Q_{\mu\nu}^{ab} A_\nu^b  
 + i \bar C^a D^\mu{}^{ac}[a]  D_\mu^{cb}[a] C^b
 \nonumber\\&& \quad \quad
+ A_\mu^a  \left( G_\mu^a + 
{1 \over \alpha}  D^\mu{}^{ab}[a]  J_\phi^b
 +  J^\mu{}^a \right) 
 \nonumber\\&& \quad \quad \quad
 + {1 \over 2\alpha}(\zeta \epsilon^{ab3} C^3 \bar C^{b})^2
 - \zeta \bar C^+ \bar C^- C^+ C^-
 \nonumber\\&& \quad \quad \quad
 - {i\zeta  \over \alpha} J_\phi^b \epsilon^{ab3}C^3 \bar C^a
 - {1 \over 2\alpha} (J_\phi^a)^2 
 \Big] ,
\label{actionGF}
   \\
     Q_{\mu\nu}^{ab} &:=&   
 - 2i g^2 \xi (\bar C^a  C^b - \bar C^c C^c \delta^{ab})
\delta_{\mu\nu}
 + {1 \over \alpha}  D_\mu[a]^{ac} D_\nu[a]^{cb} ,
\label{defQ}
\\
 G_\mu^c  &:=& i \left( {\zeta \over \alpha}  - 1 \right)
 D_\mu[a]^{cb}(\epsilon^{ab3}C^3 \bar C^a) ,
    \label{defG}
\end{eqnarray}
where 
$G_\mu^c(x) \equiv 0$ for the choice of (\ref{coeff}).
 
\par
An advantage of the choice (\ref{dMAG2}) is that $G_{gf}'$ is written
as the anti-BRST exact form,
\begin{eqnarray}
G_{gf}'
= \bar \delta_B \left( {1 \over 2}(A_\mu^a)^2
+ i C^a \bar C^a \right)
= \bar \delta_B \left(
A_\mu^+ A_\mu^- + i \sum_{\pm} C^{\pm} \bar C^{\mp}
\right),
\end{eqnarray}
where $\bar \delta_B$ is the nilpotent anti-BRST transformation
\cite{antiBRST},
\begin{eqnarray}
   \bar \delta_B {\cal A}_\mu(x)  &=&  {\cal D}_\mu \bar {\cal C}(x) 
   := \partial_\mu \bar {\cal C}(x) - i [{\cal A}_\mu(x), \bar {\cal
C}(x)],
    \nonumber\\
   \bar \delta_B  {\cal C}(x)  &=&   i \bar \phi(x)  ,
    \nonumber\\
   \bar \delta_B \bar {\cal C}(x)  &=& 
 i{1 \over 2} [\bar {\cal C}(x), \bar {\cal C}(x)] ,
    \nonumber\\
   \bar \delta_B \bar \phi(x)  &=&  0 ,
    \nonumber\\
  \bar \delta_B \psi(x)  &=&  i  \bar {\cal C}(x) \psi(x) ,
    \nonumber\\
\phi(x) + \bar \phi(x) &=& [{\cal C}(x), \bar {\cal C}(x)] ,
    \label{aBRST}
\end{eqnarray}
where $\bar \phi$ is defined in the last equation.
The BRST and anti-BRST transformations have the following properties,
\footnote{
The operation $\delta_B$ or $\bar \delta_B$ on the product of two
quantities is given by
\begin{eqnarray}
  \delta(XY) = (\delta X) Y \mp X \delta Y, \quad \delta = \delta_B,
\bar \delta_B ,
\end{eqnarray}
where $+(-)$ sign is taken for a bosonic (fermionic) quantity $X$.
}
\begin{eqnarray}
 (\delta_B)^2 = 0, \quad (\bar \delta_B)^2 = 0, \quad
 \{ \delta_B,  \bar \delta_B \} 
 := \delta_B  \bar \delta_B + \bar \delta_B  \delta_B = 0 .
\end{eqnarray}
Hence, we obtain
\begin{eqnarray}
{\cal L}_{GF} 
= i \delta_B \bar \delta_B \left( {1 \over 2}(A_\mu^a)^2
+ i C^a \bar C^a \right)
= i \delta_B \bar \delta_B \left(
A_\mu^+ A_\mu^- + i \sum_{\pm} C^{\pm} \bar C^{\mp}
\right),
\end{eqnarray}
which is invariant under the  BRST and anti-BRST transformations,
\begin{eqnarray}
 \delta_B {\cal L}_{GF} = 0 = \bar \delta_B {\cal L}_{GF} .
\end{eqnarray}
Thus the MAG TFT action can be written as
\begin{eqnarray}
 S_{TFT} 
&=& \int d^Dx \ i \delta_B \bar \delta_B \left( {1 \over
2}(\Omega_\mu^a(x))^2 + i C^a(x) \bar C^a(x) \right)
\\
&=& \int d^Dx  \ i \delta_B \bar \delta_B \left(
\Omega_\mu^+(x) \Omega_\mu^-(x) + i \sum_{\pm} C^{\pm}(x) \bar
C^{\mp}(x)
\right) 
\\
&=& \int d^Dx \ i \delta_B \bar \delta_B {\rm tr}_{\cal G - \cal H} 
\left( {1 \over 2}(\Omega_\mu(x))^2 + i {\cal C}(x) \bar
{\cal C}(x) \right) .
\end{eqnarray}
\par
For our choice of MAG, we find for (\ref{cond}) 
\begin{eqnarray}
  {\cal M}_\mu^A[U] &:=& \delta_B \bar \delta_B  
  [(UT^A U^\dagger)^a\Omega_\mu^a] ,
  \nonumber\\
  {\cal K}^{AB}[U] &:=& \delta_B \bar \delta_B  
  [(UT^A U^\dagger)^a(UT^B U^\dagger)^a] ,
\end{eqnarray}
where we have used
\begin{eqnarray}
 \delta_B {\cal V}_\mu(x) = 0 = \bar \delta_B {\cal V}_\mu(x) .
\end{eqnarray}
The BRST and anti-BRST transformations for $U$ are
\begin{eqnarray}
    \delta_B U(x)  =  i {\cal C}(x) U(x),  \quad
   \bar \delta_B  U(x)  =   i \bar {\cal C}(x) U(x)  .
\label{UBRST}
\end{eqnarray}
This reproduces the usual BRST and anti-BRST transformations of the
gauge field 
$\Omega_\mu:=i U\partial_\mu U^\dagger$.

\par
\subsection{Superspace formulation}

Now we explain the superspace formulation based on
\cite{PS79,MNTW83,MNT82,Cardy83,KLF84,BT81}.
We introduce a (D+2)-dimensional superspace ${\cal M}$ 
with coordinates 
\begin{eqnarray}
 X^M:=(x^\mu, \theta, \bar \theta) \in {\cal M},
 \quad x \in R^{D},
\end{eqnarray}
where $x^\mu$
denotes the coordinate of the $D$-dimensional Euclidean
space, $\theta$ and
$\bar
\theta$ are anti-Hermitian Grassmann numbers satisfying 
\begin{eqnarray}
 \theta^2 = 0, \quad \bar \theta^2 = 0, \quad 
\{ \theta, \bar \theta \} 
 := \theta \bar \theta +  \bar  \theta \theta = 0, 
 \quad
 \theta^\dagger = - \theta, 
 \quad
 \bar \theta^\dagger = - \bar \theta .
\end{eqnarray}
\par
We define the inner product of two vectors by introducing the
superspace (covariant) metric tensor $\eta_{MN}$ with components,
\begin{eqnarray}
 \eta_{\mu\nu} = \delta_{\mu\nu} , 
 \quad
 \eta_{\theta\bar\theta} = - \eta_{\bar\theta\theta} = - 2/\gamma,
 \quad
 {\rm others} = 0 .
\end{eqnarray}
The contravariant metric tensor is defined by 
$\eta^{MN}\eta_{NL} = \delta^M_L$.
Note that $\eta_{MN}$ is not symmetric. We introduce the covariant
supervector,
\begin{eqnarray}
  X_M = \eta_{MN} X^N
\end{eqnarray}
and the quadratic form
\begin{eqnarray}
  X^M X_M = X^M \eta_{MN} X^N = x^2 + (4/\gamma) \bar \theta \theta .
\end{eqnarray}
Note that $X^M X_M$ and $X_M X^M$ are different, because the metric
tensor is not symmetric,
\begin{eqnarray}
  X^M X_M \not= X_M X^M = \eta_{MN} X^N X^M =  x^2 - (4/\gamma) \bar
\theta \theta .
\end{eqnarray}
Integrations over $\bar \theta$ and $\theta$ are defined by
\begin{eqnarray}
 \int d \theta = \int d\bar \theta = 0,
 \quad
 \int d \theta \ \theta = \int d\bar \theta  \ \bar \theta = i,
\end{eqnarray}
or
\begin{eqnarray}
 \int d \theta \  d\bar \theta  
 \pmatrix{1 \cr \theta \cr \bar \theta \cr  \theta \bar \theta }
 = \pmatrix{ 0 \cr  0 \cr 0 \cr 1 } .
\end{eqnarray}
\par
Supersymmetry transformations are simply rotations in the
superspace leaving invariant the quadratic form,
\begin{eqnarray}
 \eta_{MN}X_1^M X_2^N = x_1^\mu x_2^\mu 
 + (2/\gamma) (\bar \theta_1 \theta_2 - \theta_1 \bar \theta_2) . 
\end{eqnarray}
This corresponds to the orthosymplectic supergroup $OSp(D/2)$.
It contains the rotation in $R^D$, i.e., the $D$-dimensional
orthogonal group $O(D)$ which leaves $x^2$ invariant and the
symplectic group $OSp(2)$ of transformations leaving
$\theta\bar\theta$ invariant.  In addition, 
$OSp(D/2)$ includes transformations that mix the commuting and
anticommuting variables,  
\begin{eqnarray}
 x^\mu &\rightarrow& x^\mu{}' 
 := x^\mu + 2 \bar a^\mu \xi \theta + 2 a^\mu \xi \bar \theta,
\nonumber\\
 \theta &\rightarrow& \theta' := \theta + \gamma a^\mu x_\mu \xi, 
\nonumber\\
 \bar \theta &\rightarrow& \bar \theta'  := \bar \theta -
\gamma \bar a^\mu x_\mu \xi,
\end{eqnarray}
where $a$, $\bar a$ are arbitrary $D$-vectors and $\xi$ is an
anticommuting c-number ($\xi^2 = \{ \xi, \theta \} = \{\xi, \bar
\theta\} = 0$). We call this transformation $\tau(a,\bar a)$.
\par
Any object 
$A^M=(A^\mu,A^{\theta},A^{\bar \theta})$
which transforms like the supercoordinate under $OSp(D/2)$ is
defined to be a (contravariant) supervector. If 
$A_1^M$ and $A_2^M$ are two
such supervectors, then the inner product 
\begin{eqnarray}
  A_1^M A_2{}_M = A_2^M A_1{}_M = A_1^\mu A_2{}_\mu 
  + (2/\gamma) (A_1^{\bar \theta} A_2^\theta 
  + A_2^{\bar \theta} A_1^{\theta}) ,
\end{eqnarray}
is invariant under superrotations. 
We define the the partial derivatives to be covariant supervectors in
superspace,
\begin{eqnarray}
   \partial_M := ({\partial \over \partial x^\mu}, {\partial \over
\partial \theta},
   {\partial \over \partial \bar \theta})
   := (\partial_\mu, \partial_\theta, \partial_{\bar \theta}) .
\end{eqnarray}
Then the superLaplacian defined by
\begin{eqnarray}
   \partial^M \partial_M := \Delta_{SS}
   =   \partial^\mu \partial_\mu 
   + \gamma \partial_{\bar \theta} \partial_\theta,
\end{eqnarray}
is an invariant.
\par
Introducing a grading $p(M)$ for each coordinate $X^M$ by
\begin{eqnarray}
  p(\mu)=0, \quad p(\theta)=p(\bar \theta)=1,
\end{eqnarray}
the coordinates obey the graded commutation relations,
\begin{eqnarray}
  X^M X^N - (-1)^{p(M)p(N)} X^N X^M = 0 .
\end{eqnarray}
 Similarly, objects $F_{MN}$ which transform like
$A_1^M A_2^N$ are defined to be (contravariant) supertensors and 
$F^M_M := F^{MN} \eta_{MN}$ is invariant.
The metric tensor defined above is a supertensor.  The metric has
another invariant called the supertrace in addition to the trace
$\eta^{MN}\eta_{MN}$,
\begin{eqnarray}
  {\rm str}(\eta) = (-1)^{p(M)} \eta^M_M .
\end{eqnarray}

\par
We introduce the superfield $\Phi(x,\theta, \bar \theta)$ by
\begin{eqnarray}
 \Phi(x,\theta, \bar \theta) 
 &=& \Phi_0(x)  + \theta \bar  \Phi_1(x) 
+ \bar \theta \Phi_2(x)
 + \bar \theta \theta \Phi_3(x)
 \nonumber\\ 
 &=& \Phi_0(x)  + \theta \partial_{\theta} \Phi_0(x) 
+ \bar \theta \partial_{\bar\theta} \Phi_0(x)
 + \bar \theta \theta \partial_{\theta}\partial_{\bar \theta}
\Phi_0(x) ,
\label{superf}
\end{eqnarray}
where $\Phi_i$ are complex-valued functions,
$\Phi_i: R^D \rightarrow C (i=0,1,2,3)$. It should be noted that all
component fields $\Phi_i$ transform according to the same
representation of $O(D)$.  Hence, in this formulation of superspace,
supersymmetry transformations mix fields obeying different
statistics, but with identical spin.

\par
For any superfield $\Phi$, the supertransformation $\tau$ acts as 
\begin{eqnarray}
  (\tau(a, \bar a)\Phi)(x,\theta,\bar\theta) &=&
\Phi(x,\theta,\bar\theta)
  + [\gamma a^\mu x_\mu \Phi_1(x) - \gamma \bar a^\mu x_\mu \Phi_2(x)
]\xi
  \nonumber\\&&
  + [-2 \partial_\mu \Phi_0(x) \bar a^\mu + \gamma \bar a^\mu x_\mu
\Phi_3(x)] \theta \xi
  \nonumber\\&&
+ [-2 \partial_\mu \Phi_0(x) a^\mu + \gamma a^\mu x_\mu \Phi_3(x)]
\bar \theta \xi
  \nonumber\\&&
+ 2[\partial_\mu \Phi_1(x) a^\mu - \partial_\mu \Phi_2(x) \bar
a^\mu] \bar \theta \theta \xi .
\end{eqnarray}
If the superfield $\Phi$ is invariant by $\tau$ for all 
$a,\bar a \in R^D$, the term with $\xi$ of the RHS of this equation
must be zero for all $a,\bar a \in R^D$.  Hence, 
\begin{eqnarray}
 \Phi_1(x) \equiv 0 \equiv \Phi_2(x), \quad 
 {2 \over \gamma} \partial_\mu \Phi_0(x) = x_\mu \Phi_3(x) .
\end{eqnarray}
This implies that $\Phi_0(x)$ is a function only of
$x^2:=x^\mu x^\mu$.  Then we can write $\Phi_0(x)=f(x^2)$ for a
function 
$f: [0, \infty) \rightarrow C$ and
$\Phi_3(x) = {4 \over \gamma} f'(x^2)$ 
\par
Therefore, if the superfield ${\cal O}(X)$ is
supersymmetric, then there exists a function 
$f: [0, \infty) \rightarrow C$ such that
\begin{eqnarray}
 {\cal O}(x,\theta, \bar \theta) 
 = f(x^2) + (4/\gamma) \bar \theta \theta f'(x^2) 
 = f(x^2 + (4/\gamma) \bar \theta \theta) .
 \label{ss}
\end{eqnarray}

\subsection{Geometric meaning of BRST transformation in superspace}
\par
We define the connection one-form (superspace vector potential) ${\cal
A}(X)$ and its curvature (superspace field strength)
${\cal F}(X)$ in the superspace, 
$X^M := (x^\mu, \theta, \bar \theta) \in {\cal M}$,
\begin{eqnarray}
   {\cal A}(X) &:=& {\cal A}_M(X) dX^M
   = {\cal A}_\mu(x, \theta, \bar \theta) dx^\mu
   + {\cal C}(x, \theta, \bar \theta) d\theta
   + \bar {\cal C}(x, \theta, \bar \theta) d\bar \theta ,
   \nonumber\\
   {\cal F}(X) &:=& \tilde d {\cal A}(X) 
   + {1 \over 2} [{\cal A}(X), {\cal A}(X)]
   = - {1 \over 2} {\cal F}_{NM}(X) dX^M dX^N ,   
   \nonumber\\
   {\cal A}_M(X) &:=& {\cal A}_M^A(X) T^A,
   \nonumber\\
   dX^M &:=& (dx^\mu, d \theta, d\bar \theta ),
\end{eqnarray}
where $\tilde d$ is the exterior differential in the superspace,
\begin{eqnarray}
   \tilde d  :=  d + \delta + \bar \delta
   := {\partial \over \partial x^\mu} dx^\mu
   + {\partial \over \partial \theta} d \theta
   + {\partial \over \partial \bar \theta} d \bar \theta .
\end{eqnarray}
These definitions are compatible when
\begin{eqnarray}
   dx^M dx^N &=& - (-1)^{p(M)p(N)} dx^N dx^M, 
   \nonumber\\
   (x^M, \partial_M) dx^N &=& (-1)^{p(M)p(N)} dx^N (x^M,\partial_M) .
\end{eqnarray}
The supergauge transformation is given by
\begin{eqnarray}
   {\cal A}(X)  \rightarrow {\cal A}'(X) &:=& 
 U(X){\cal A}(X)U^\dagger(X) + i U^\dagger(X) dx^M \partial_M U(X),
\nonumber\\
 U(X) &:=& \exp [ i\omega^A(X) T^A ] .
\end{eqnarray}
\par
In what follows, we show that the superfield 
${\cal A}_\mu(X)$, ${\cal C}(X)$, $\bar {\cal C}(X)$
are  respectively identified with a generalization of 
${\cal A}_\mu(x)$, ${\cal C}(x)$, $\bar {\cal C}(x)$ into the
superspace. 
First, we require that 
\begin{eqnarray}
  {\cal A}_\mu(x, 0, 0) 
  = {\cal A}_\mu(x) ,
\quad
  {\cal C}(x, 0, 0) 
  = {\cal C}(x)  ,
\quad
  \bar {\cal C}(x, 0, 0) 
  = \bar {\cal C}(x)   ,
\end{eqnarray}
and impose the {\it horizontal condition} \cite{BT81}  for any $M$, 
\begin{eqnarray}
  {\cal F}_{M\theta}(X) = {\cal F}_{M\bar \theta}(X) = 0 ,
\end{eqnarray}
which is equivalent to set
\begin{eqnarray}
   {\cal F}(X)  
   =  {1 \over 2} {\cal F}_{\mu\nu}(X) dx^\mu dx^\nu .
\end{eqnarray}
By solving the horizontal condition, the dependence of the
superfield $A_M(x, \theta, \bar \theta)$ on $\theta, \bar \theta$
is determined as follows.
The horizontal condition is rewritten as
\begin{eqnarray}
&&  (d+\delta+\bar \delta)({\cal A}^{1}+{\cal C}^{1}+\bar {\cal
C}^{1})
  + {1 \over 2} [{\cal A}^{1}+{\cal C}^{1}+\bar {\cal C}^{1}, 
  {\cal A}^{1}+{\cal C}^{1}+\bar {\cal C}^{1}]
  \nonumber\\
   &=&  d {\cal A}^{1} + {1 \over 2} [{\cal A}^{1}, {\cal A}^{1}] .
\label{hcond}
\end{eqnarray}
where we have defined the one-form,
\begin{eqnarray}
{\cal A}^{1} := {\cal A}_\mu(x, \theta, \bar \theta) dx^\mu, \quad 
{\cal C}^{1} := {\cal C}(x, \theta, \bar \theta) d\theta, \quad
\bar {\cal C}^{1} := \bar {\cal C}(x, \theta, \bar \theta) d\bar
\theta .
\end{eqnarray}
By comparing both sides of the equation (\ref{hcond}), we obtain
\begin{eqnarray}
   \partial_\theta {\cal A}_\mu(X)  &=& 
   \partial_\mu {\cal C}(X) - i [{\cal A}_\mu(X), {\cal C}(X)],
    \nonumber\\
   \partial_\theta {\cal C}(X)  &=&  i{1 \over 2} [{\cal C}(X), {\cal
C}(X)] ,
    \nonumber\\
   \partial_{\bar \theta} {\cal A}_\mu(X)  &=& 
   \partial_\mu \bar {\cal C}(X) - i [{\cal A}_\mu(X), \bar {\cal
C}(X)],
    \nonumber\\
   \partial_{\bar \theta} \bar {\cal C}(X)  &=& 
 i{1 \over 2} [\bar {\cal C}(X), \bar {\cal C}(X)] ,
    \nonumber\\
   \partial_\theta \bar {\cal C}(X) + \partial_{\bar \theta}
 {\cal C}(X) &=& - \{ {\cal C}(X), \bar {\cal C}(X) \} ,
    \label{B}
\end{eqnarray} 
where we have used that $d\theta d\theta\not=0$ and $d\theta, \theta$
anticommute with
${\cal C}$. For the components which can not be determined by the
horizontal condition alone, we use the following identification,
\begin{eqnarray}
 \partial_\theta \bar {\cal C}(x,0,0) := i \phi(x),
 \quad
 \partial_{\bar \theta} {\cal C}(x,0,0) := i \bar \phi(x) .
\end{eqnarray}
This corresponds to ${\cal F}_{\theta \bar \theta}=0$ and gives
\begin{eqnarray}
   i \phi(x) + i \bar \phi(x) + \{ {\cal C}(x), \bar {\cal C}(x) \}
   = 0 .
\end{eqnarray}
\par
From these results, it turns out that the  derivatives in the
direction of $\theta, \bar \theta$ give respectively the BRST and the
anti-BRST transformations,
\begin{eqnarray}
 {\partial \over \partial \theta} = \delta_B,
 \quad
 {\partial \over \partial \bar \theta} = \bar \delta_B ,
\end{eqnarray}
where we define the derivative as the left derivative.
This implies that the BRST and anti-BRST charges, $Q_B, \bar Q_B$ are
the generators of the translations in the variables $\theta, 
\bar \theta$. 
\par
Thus the superfields are determined as
\begin{eqnarray}
  {\cal A}_\mu(x, \theta, \bar \theta) 
  &=& {\cal A}_\mu(x) + \theta {\cal D}_\mu {\cal C}(x)
  + \bar \theta {\cal D}_\mu \bar {\cal C}(x)
  + \bar \theta \theta (i  {\cal D}_\mu \phi(x) + \{  {\cal D}_\mu
{\cal C}(x), \bar {\cal C}(x) \}) ,
\nonumber\\
  {\cal C}(x, \theta, \bar \theta) 
  &=& {\cal C}(x) + \theta (- {1 \over 2} [{\cal C}, {\cal C}](x))
  + \bar \theta i \bar \phi(x)
  + \bar \theta \theta [i \bar \phi(x), {\cal C}(x) ] ,
\nonumber\\
  \bar {\cal C}(x, \theta, \bar \theta) 
  &=& \bar {\cal C}(x) + \theta i \phi(x)
  + \bar \theta  (- {1 \over 2} [\bar {\cal C}, \bar {\cal C}](x)) 
  + \bar \theta \theta [-i \phi(x), \bar {\cal C}(x) ] .
\end{eqnarray}
The non-vanishing components of ${\cal F}_{\mu\nu}$ have
\begin{eqnarray}
  {\cal F}_{\mu\nu}(x, \theta, \bar \theta) 
  &=& {\cal F}_{\mu\nu}(x) 
  + \theta[{\cal F}_{\mu\nu}(x), \bar {\cal C}(x)]
  + \bar \theta[{\cal F}_{\mu\nu}(x), {\cal C}(x)]
  \nonumber\\&&
  + \bar \theta \theta (i[{\cal F}_{\mu\nu}(x), \phi(x)]
  + [[{\cal F}_{\mu\nu}(x), {\cal C}(x)],\bar {\cal C}(x)]) .
\end{eqnarray}
\par
For the matter field $\varphi(x)$, we define the superfield
$\varphi(X)$ and its covariant derivative by
\begin{eqnarray}
   {\cal \varphi}(X) &:=& \varphi(x) + \theta \varphi_1(x)
 + \bar \theta \varphi_2(x) 
 + \bar \theta \theta \varphi_3(x),
   \\
   \tilde {\cal D}[{\cal A}] \varphi(X) 
   &:=& (\tilde d + {\cal A}(X)) \varphi(X) .
\end{eqnarray}
The horizontal condition for the matter field is
\begin{eqnarray}
   {\cal D}_M \varphi(X) dX^M = {\cal D}_\mu \varphi(X) dx^\mu ,
\end{eqnarray}
which implies
\begin{eqnarray}
{\cal D}_\theta \varphi(X) = 0 ={\cal D}_{\bar \theta} \varphi(X) .
\end{eqnarray}
From this, we have for example
\begin{eqnarray}
 \varphi_1(x,0,0) = \partial_\theta \varphi(x,0,0) 
  = - {\cal A}_\theta(X)
\varphi(X)|_{\theta=\bar\theta=0}
 = - {\cal C}(x) \varphi(x) = \delta_B \varphi(x) .
\end{eqnarray}
\par
Accordingly, all the field variables obey the relation,
\begin{eqnarray}
  \Phi(x, \theta, \bar \theta) = \Phi(x) + \theta (\delta_B \Phi(x))
  + \bar \theta (\bar \delta_B \Phi(x)) + \bar \theta \theta 
  (\bar \delta_B \delta_B  \Phi(x)) .
\end{eqnarray}
\par
Let $\Phi_1(X)$ and $\Phi_2(X)$ be two superfields corresponding to
$\phi_1(x)$ and $\phi_2(x)$, respectively.  It is easy to show that
the following formula holds,
\begin{eqnarray}
 \Phi_1(x,\theta, \bar \theta) \Phi_2(x,\theta, \bar \theta)
 &=&  \phi_1(x)\phi_2(x) 
 + \theta \delta_B (\phi_1(x)\phi_2(x))
 + \bar \theta \bar \delta_B (\phi_1(x)\phi_2(x))
 \nonumber\\&&
 + \bar \theta \theta \bar \delta_B \delta_B (\phi_1(x)\phi_2(x)) .
\end{eqnarray}
Thus, for any (elementary or composite) field
${\cal O}(x)$, we can define the corresponding superfield 
${\cal O}(x,\theta, \bar \theta)$ using BRST and anti-BRST
transformation by
\begin{eqnarray}
 {\cal O}(x,\theta, \bar \theta) = {\cal O}(x) + \theta
\delta_B {\cal O}(x)
 + \bar \theta \bar \delta_B {\cal O}(x) 
 + \bar \theta \theta \bar \delta_B \delta_B {\cal O}(x) .
\end{eqnarray}
In the superspace ${\cal M}$, the BRST and anti-BRST transformations
correspond to the translation of $\theta$ and $\bar \theta$
coordinates, respectively. 

\par
For the Grassmann number, the integration 
$\int d\theta$ (resp. $\int d\bar \theta$) is
equivalent to the differentiation 
${d \over d\theta}$ (resp. ${d \over d\bar \theta}$).
Hence  
the BRS
$\delta_B$ and anti-BRST $\delta_B$) transformation has the
following correspondence,
\begin{eqnarray}
  \delta_B  \leftrightarrow 
{d \over d\theta} \leftrightarrow \int d\theta , \quad
  \bar \delta_B  \leftrightarrow 
 {d \over d\bar \theta} \leftrightarrow \int d\bar \theta .
\end{eqnarray}
This implies
\begin{eqnarray}
 \int d\theta d\bar \theta {\cal O}(x,\theta, \bar \theta) 
 =  - {\cal O}_3(x)
 =  - {\partial \over \partial \theta}
 {\partial \over \partial \bar \theta} {\cal O}(x,\theta, \bar
\theta) 
 = - \bar \delta_B \delta_B  {\cal O}(x)  
 =  \delta_B \bar \delta_B {\cal O}(x) .
\end{eqnarray}
Therefore, if the Lagrangian (density) of the form 
$\delta_B \bar \delta_B {\cal O}(x)$ is given for an operator 
${\cal O}$, the operator ${\cal O}$ can be extended into the
superfield 
${\cal O}(x,\theta, \bar\theta)$ in the superspace,
\begin{eqnarray}
\int d^D x \delta_B \bar \delta_B {\cal O}(x) 
= \int d^D x \int d\theta d\bar \theta {\cal O}(x,\theta, \bar
\theta) .
\end{eqnarray}

\subsection{MAG TFT as a supersymmetric theory}
\par
Especially, the operator 
\begin{eqnarray}
 {\cal O}(x) := - i {\rm tr}_{{\cal G} \setminus {\cal H}} \left( {1
\over 2}({\cal A}_\mu(x))^2 + i {\cal C}(x) \bar {\cal C}(x) \right)
,
\end{eqnarray}
has the corresponding superfield given by
\begin{eqnarray}
 {\cal O}(X) := {-i \over 2} {\rm tr}_{{\cal G} \setminus {\cal H}} \left(
 ({\cal A}_\mu(X))^2 + 2i {\cal C}(X) \bar {\cal C}(X) \right) ,
\end{eqnarray}
where we have chosen
\begin{eqnarray}
 {1 \over \gamma} := {i \over 2} .
\end{eqnarray}
The superfield ${\cal O}(X)$ is written in $OSp(D/2)$ invariant form,
\begin{eqnarray}
 {\cal O}(X) 
 = {-i \over 2} {\rm tr}_{{\cal G}\setminus{\cal H}} \left( 
\eta_{NM} {\cal A}^M(X) {\cal A}^N(X) \right) .
\end{eqnarray}
Thus the action of MAG TFT can be written in the manifestly
superspace covariant form,
\begin{eqnarray}
 S_{TFT} 
= \int d^D x \int d\theta d\bar \theta {-i \over 2}
{\rm tr}_{{\cal G} \setminus {\cal H}}
 (\eta_{NM} \Omega^M(x,\theta, \bar \theta) 
 \Omega^N(x,\theta, \bar \theta)) .
 \label{Saction}
\end{eqnarray}

%\newpage
\section{Dimensional reduction of Topological field theory}
\setcounter{equation}{0}

After giving a basic knowledge for the dimensional reduction of
Parisi and Sourlas in the supersymmetric model, we apply this
mechanism to the MAG TFT.  
We show that the $D$-dimensional MAG TFT is reduced to the
$(D-2)$-dimensional coset $G/H$ non-linear $\sigma$ model (NLSM). 
This implies that a class of correlation functions in the
$D$-dimensional MAG TFT can be calculated in the equivalent
$(D-2)$-dimensional coset NLSM.

\subsection{Parisi and Sourlas dimensional reduction}
\par
Now we split the $D$-dimensional Euclidean space into two subsets,
\begin{eqnarray}
 x = (z, \hat x) \in R^D, \quad z \in R^{D-2}, \quad \hat x \in R^2
. 
\end{eqnarray}
The relation (\ref{ss}) holds for any $D$. Hence, for supersymmetric
operator
${\cal O}(X)$, we obtain
\begin{eqnarray}
 {\cal O}(x,\theta, \bar \theta) 
 = f(z, \hat x^2 + (4/\gamma) \bar \theta \theta) 
 \equiv f(z, \hat x^2) + {4 \over \gamma} \bar \theta \theta {d \over
d\hat x^2}f(z,\hat x^2) .
\end{eqnarray}
Therefore, for supersymmetric model, we find
\footnote{
An alternative derivation is as follows.
By integration by parts, we find for $D>2$
$$
 \int d^D x f'(x^2) = S_D \int_0^\infty r^{D-1} dr f'(r^2)
 = - S_D {D-2 \over 2} \int_0^\infty dr^2 (r^2)^{D/2-2} f(r^2)
 = - \pi \int d^{D-2}x f(x^2) ,
$$
where $S_D=2\pi^{D/2}/\Gamma(D/2)$ is the area of the unit sphere in
$D$-dimensional space.
}
\begin{eqnarray}
 S_{GF} 
&=& \int d^D x \int d\theta \int d\bar \theta 
{\cal O}(x,\theta, \bar \theta) 
\nonumber\\
&=&  \int d^{D-2}z \int d^2\hat x  \int d\theta \int d\bar \theta 
\ {4 \over \gamma} \bar \theta \theta {d \over
d\hat x^2}f(z,\hat x^2)   
\nonumber\\
&=& -  {4 \over \gamma} \int d^{D-2}z \int d^2\hat x  {d \over d\hat
x^2} f(z,\hat x^2)
\nonumber\\
&=& -  {4 \over \gamma} \int d^{D-2}z  \int_0^\infty  \pi dr^2 {d
\over dr^2}f(z,r^2)  
\nonumber\\
&=&  {4\pi \over \gamma} \int d^{D-2}z  f(z,0)  
\nonumber\\
&=&  {4\pi \over \gamma} \int d^{D-2}z  {\cal O}_0((z,0),0,0)  ,
\label{equiva}
\end{eqnarray}
where we have assumed 
$f(z,\infty) \equiv {\cal O}_0((z,\infty),0,0) =0$ and used the
notation of (\ref{superf}).
\par
This shows the dimensional reduction by two units.  The
supersymmetric $D$-dimensional model is equivalent to a purely
bosonic model in $D-2$ dimensions.  This fact was first discovered
 by Parisi and Sourlas (PS) \cite{PS79}.
\par
The correlation function in supersymmetric theory are generated by
the partition function in the presence of external sources,
\begin{eqnarray}
 Z_{SUSY}[{\cal J}] := \int [d\Phi]
 \exp \left\{ - \int d^Dx d\theta d\bar \theta [{\cal L}_{SUSY}[\Phi]
 - \Phi(x,\theta, \bar \theta) {\cal J}(x,\theta, \bar \theta) ]
\right\} ,
\end{eqnarray}
where we write all the fields by $\Phi$ collectively for the
supersymmetric Lagrangian ${\cal L}_{SUSY}[\Phi]$. Restricting the
source to a $D-2$ dimensional subspace, 
\begin{eqnarray}
  {\cal J}(x,\theta, \bar \theta) 
  = J(z) \delta^2(\hat x) \delta(\theta) \delta(\bar \theta) ,
\end{eqnarray}
and taking the derivatives of $Z_{SS}[J]$ with respect to $J(z)$, we
obtain the correlation functions of the superspace
theory which are restricted to the $D-2$ dimensional subspace. 
These are identical to the  correlation functions of the
corresponding $D-2$ dimensional quantum theory,
\begin{eqnarray}
 Z_{SUSY}[{\cal J}] = Z_{D-2}[J] ,
\end{eqnarray}
where $Z_{D-2}[J]$ is the generating functional for $D-2$
dimensional theory,
\begin{eqnarray}
  Z_{D-2}[J] 
 := \int [d\Phi_0]
 \exp \left\{ - \int d^{D-2}z  
\left[{4\pi \over \gamma}{\cal L}_{0}[\Phi_0]
 - \Phi_0(z) J(z) \right] \right\} .
\end{eqnarray}
\par
When the PS dimensional reduction occurs, the three-way
equivalence is known among (1) a field theory in a superspace of $D$
commuting and 2 anticommuting dimensions,
(2) the corresponding $D-2$ dimensional quantum field theory,
(3) the $D$-dimensional classical stochastic theory, namely,
the stochastic average of the $D$-dimensional classical theory in
the presence of random external sources.
The final point is not yet made clear in this paper.

\subsection{Dimensional reduction of TFT to NLSM}

The action (\ref{Saction}) of TFT is manifestly invariant by all
supertransformations. 
Therefore, the $D$-dimensional MAG TFT is dimensionally reduced to
the
$(D-2)$-dimensional model in the sense of Parisi and Sourlas.  
From (\ref{Saction}) and (\ref{equiva}), the
equivalent $(D-2)$-dimensional theory is given by
\begin{eqnarray}
 S_{NLSM} 
&=&   2\pi   \int d^{D-2}z  \ 
  {\rm tr}_{{\cal G}\setminus{\cal H}} 
[{1 \over 2}\delta_{\mu\nu}\Omega^\mu(z)
\Omega^\nu(z)], 
  \quad
   \Omega_\mu(z) := {i \over g} U(z) \partial_\mu U^\dagger(z) 
\\
&=& - {\pi  \over g^2} \int d^{D-2}z \  
  {\rm tr}_{{\cal G}\setminus{\cal H}}  [ 
   U(z) \partial^\mu U^\dagger(z) U(z) \partial^\mu U^\dagger(z)] 
\nonumber\\
&=& {\pi \over g^2} \int d^{D-2}z \  
  {\rm tr}_{{\cal G}\setminus{\cal H}}  
  [ \partial^\mu U(z)  \partial^\mu U^\dagger(z)] .
\end{eqnarray}
Thus the $D$-dimensional MAG TFT is reduced to the
$(D-2)$-dimensional $G/H$ non-linear $\sigma$ model (NLSM) whose
partition function is given by
\begin{eqnarray}
  Z_{NLSM} := \int [dU] \exp \{ - S_{NLSM}[U] \} ,
\end{eqnarray}
where we have dropped the ghost contribution $iC(z)\bar C(z)$.
The  correlation functions of the $D$-dimensional TFT  
coincide with the same correlation function calculated in the
 equivalent  $(D-2)$-dimensional NLSM if the arguments $x_i$ are
located on the $(D-2)$-dimensional subspace,
\begin{eqnarray}
  \langle \prod_i {\cal F}_i(x_i) \rangle_{G MAG TFT_D}
  = \langle \prod_i {\cal F}_i(x_i) \rangle_{G/H NLSM_{D-2}}  
  \quad {\rm if} \quad  x_i \in R^{D-2} .
\end{eqnarray}
 
\par
The propagator of NLSM$_{D-2}$ in momentum representation is obtained
by taking
 $\hat p=p_\theta=p_{\bar\theta}=0$  in the supersymmetric quantity,
\begin{eqnarray}
&& {1 \over 2\pi i } \int d^D x d\theta d\bar \theta 
\ e^{i p_\mu x_\mu - p_{\bar\theta}\theta + p_\theta \bar \theta}  
  \eta_{NM} \langle 
  \Omega_M^a(x,\theta,\bar\theta)   \Omega_N^b(0,0,0)
\rangle_{TFT_D} |_{\hat p=p_\theta=p_{\bar\theta}=0}
\nonumber\\
 &=& \int d^{D-2} z \ e^{i p_k \cdot z_k} 
 \delta_{ij}
  \langle \Omega_i^a(z)  \Omega_j^b(0) \rangle_{NLSM_{D-2}} 
\nonumber\\
 &=&  {g^2 \over \pi}  \delta_{ab} \delta_{ij}
 \left\{ [1+u(p_k^2)]{p_i p_j \over p_k^2}
 - u(p_k^2) \delta_{ij} \right\} , (p_i,p_j,p_k \in R^{D-2}) .
 \label{iden1}
 \end{eqnarray}
 From $OSp(D/2)$ invariance, we have
\begin{eqnarray}
&& {1 \over 2\pi i } \int d^D x d\theta d\bar \theta 
\ e^{i p_\mu x_\mu - p_{\bar\theta}\theta + p_\theta \bar \theta}  
  \langle 
  \Omega_M^a(x,\theta,\bar\theta)  \Omega_N^b(0,0,0)
\rangle_{TDT_D}  
\nonumber\\
 &=&  {g^2 \over \pi} \delta_{ab} 
 \left\{ [1+u(p_L^2)]{p_M p_N \over p_L^2}
 - u(p_L^2) \delta_{MN} \right\} ,  
\label{iden2}
 \end{eqnarray}
where
\begin{eqnarray}
  p_L^2 := p_\mu^2 + 2i p_{\bar\theta} p_\theta 
 = p_k^2 + \hat p^2 +  2i p_{\bar\theta} p_\theta  .
 \end{eqnarray}
By setting $M=\mu,  N=\nu$ and differentiating both sides of
(\ref{iden2})  by 
$\partial^2/\partial p_\theta \partial p_{\bar \theta}$, 
we obtain the propagator in $D$-dimensional TFT,
\begin{eqnarray}
   {1 \over 2\pi i } \int d^D x  \ e^{i p_\mu x_\mu}  
  \langle 
  \Omega_\mu^a(x)   \Omega_\nu^b(0)
\rangle_D  
=  {g^2 \over \pi} \delta_{ab} \left\{ v(p) \delta_{\mu\nu}
 + (\delta_{\mu\nu} - p_\mu p_\nu) v'(p^2)  \right\} ,
 \label{iden3}
 \end{eqnarray}
where
\begin{eqnarray}
 v(p^2) &:=& {1+u(p^2) \over p^2} , \quad p^2 := p_\mu^2 .
 \end{eqnarray}
We compare (\ref{iden3}) with (\ref{iden1}) following \cite{HK85}.
If the particle spectrum has a mass gap in $D-2$ dimensions
  (\ref{iden1}), then the function $v(p^2)$ is
analytic around $p^2=0$ and hence there is no massless particle at
all in the channel $A_\mu=\Omega_\mu$ in $D$ dimensions
(\ref{iden3}).
\par
The dimensional reduction says the equivalence of
the correlation functions at special coordinates, 
$\hat x=\theta=\bar \theta=0$ or
$\hat p = p_\theta=p_{\bar\theta}=0$.
It should be remarked that the spectra of particles in the channel
$U(x)$ differ  between $D$- and $D-2$-dimensional models. 
It is worthwhile to remark that the PS dimensional reduction implies
neither the equivalence of the state vector spaces nor the
equivalence of the S matrices between the original model and the
dimensionally reduced model.
\par
The existence of mass gap in two-dimensional O(3) NLSM has been shown
in
\cite{PW83,HMN90}.  Therefore, the off-diagonal
gluons $A_\mu^a=\Omega_\mu^a (a=1,2)$ in four-dimensional SU(2) MAG
TFT have a non-zero mass,
$m_A\not=0$. Although this was assumed in the previous study of APEGT
of YM theory
\cite{KondoI}, it was supported by Monte Carlo simulation
\cite{ASu97}. 
  If we restrict the YM theory to the TFT part, the
existence of non-zero gluon mass has been just proven.  This will
hold also in the full YM theory, since the perturbation is not
sufficient to diminish this mass to yield the massless gluons.

%\newpage
\section{Non-linear $\sigma$ model, instanton and monopole}
\setcounter{equation}{0}

In the previous section, we have shown that, thanks to the
dimensional reduction, the calculation of correlation functions in
TFT$_D$ is reduced to that in the NLSM$_{D-2}$.  
In what follows, we restrict our considerations to the SU(2) YM
theory. 
In this section we study  the correspondence between O(3) NLSM$_2$
and SU(2) MAG TFT$_4$, especially focusing on the topological
non-trivial configurations. It is well known that the
two-dimensional O(3) NLSM has instanton solutions.  We find that the
instanton in two dimensions corresponds to the magnetic monopole in
four dimensions. This correspondence is utilized to prove quark
confinement in the next section.

\subsection{NLSM from TFT}

For concreteness, we consider the case of $G=SU(2)$. 
The case of $G=SU(N), N >2$ will be separately discussed in the next
section.
\par
First of all, we define
\begin{eqnarray}
&& R_\mu(x) := i U^\dagger(x) \partial_\mu U(x) 
= R_\mu^A(x) T^A
 \nonumber\\
 &=& {-1 \over 2}  \pmatrix{ 
 \partial_\mu \varphi(x)+ \cos \theta(x) \partial_\mu \chi(x) &
 - e^{-i\chi(x)}[i \partial_\mu \theta(x) 
  - \sin \theta(x) \partial_\mu \chi(x)
\cr
 e^{+i\chi(x)}[i \partial_\mu \theta(x) 
  + \sin \theta(x) \partial_\mu \chi(x)]
 &
-[\partial_\mu \varphi(x)+ \cos \theta(x) \partial_\mu \chi(x)] } ,
\end{eqnarray}
and
\begin{eqnarray}
&& L_\mu(x) := i U(x) \partial_\mu U(x)^\dagger 
= L_\mu^A(x) T^A
 \nonumber\\
 &=& {1 \over 2}  \pmatrix{ 
 \partial_\mu \chi(x)+ \cos \theta(x) \partial_\mu \varphi(x) &
 - e^{+i\chi(x)}[i \partial_\mu \theta(x) 
  + \sin \theta(x) \partial_\mu \varphi(x)]
\cr
 e^{-i\chi(x)}[i \partial_\mu \theta(x) 
  - \sin \theta(x) \partial_\mu \varphi(x)]
 &
-[\partial_\mu \chi(x)+ \cos \theta(x) \partial_\mu \varphi(x)] }  ,
\end{eqnarray}
where we have used the Euler angles 
$\theta, \varphi, \chi$ and the fundamental
representation,
\begin{eqnarray}
  T^A = {1 \over 2} \sigma^A, \quad
  \sigma^1 := \pmatrix{0 & 1 \cr 1 & 0}, \quad 
  \sigma^2 := \pmatrix{0 & -i \cr i & 0}, \quad 
  \sigma^3 := \pmatrix{1 & 0 \cr 0 & -1} . 
\end{eqnarray}
Note that $R_\mu$ and $L_\mu$ are Hermitian, 
$R_\mu^\dagger = R_\mu$, $L_\mu^\dagger = L_\mu$.
\par
For later purposes, it is convenient to write various quantities in
terms of Euler angle variables,
\begin{eqnarray}
L_\mu^{\pm}(x) &:=& {1 \over \sqrt{2}}(L_\mu^1 \pm i L_\mu^2)
= {\pm i \over \sqrt{2}}
e^{\mp i\chi(x)}[ \partial_\mu \theta(x) 
  \mp i \sin \theta(x) \partial_\mu \varphi(x)] ,
\\
  L_\mu^3(x) 
&=& \partial_\mu \chi(x)+ \cos \theta(x) \partial_\mu \varphi(x) ,
\end{eqnarray}
and
\begin{eqnarray}
L_\mu^1(x) L_\mu^1(x) + L_\mu^2(x) L_\mu^2(x)
&=& 2 L_\mu^{+}(x) L_\mu^{-}(x)  
\nonumber\\
&=&  \partial_\mu \theta(x) \partial_\mu \theta(x) 
  + \sin^2 \theta(x) 
  \partial_\mu \varphi(x) \partial_\mu \varphi(x)  ,
\\
  L_\mu^3(x) L_\mu^3(x) 
&=& [\partial_\mu \chi(x)+ \cos \theta(x) \partial_\mu \varphi(x)]^2
.
\end{eqnarray}

\par
The O(3) NLSM is defined by introducing
a three-dimensional unit vector ${\bf n}(x)$ on each point of
space-time,
${\bf n}: R^{d} \rightarrow S^2$ ($d:=D-2$),
\begin{eqnarray}
 {\bf n}(x) 
  := \pmatrix{n^1(x) \cr n^2(x) \cr n^3(x)}
:= \pmatrix{\sin \theta(x) \cos \varphi(x) \cr
 \sin \theta(x) \sin \varphi(x) \cr
\cos \theta(x)  } .
\end{eqnarray}
The direction of the unit vector in internal space is specified by
two angles
$\theta(x), \varphi(x)$ at each point $x \in R^{d}$.
Note that
\begin{eqnarray}
{\bf n}(x) \cdot {\bf n}(x) &:=& \sum_{A=1}^{3} n^A(x) n^A(x) = 1,
\label{const}
\\
  {\bf n}(x) \cdot  \partial_\mu {\bf n}(x) &=& 0 .
\end{eqnarray}
\par
Using
\begin{eqnarray}
 \partial_\mu {\bf n}(x) 
 := \pmatrix{\cos \theta(x) \cos \varphi(x) \partial_\mu \theta(x)
 - \sin \theta(x) \sin \varphi(x) \partial_\mu \varphi(x) \cr
 \cos \theta(x) \sin \varphi(x) \partial_\mu \theta(x)
 + \sin \theta(x) \cos \varphi(x) \partial_\mu \varphi(x) \cr
 - \sin \theta(x) \partial_\mu \theta(x) } ,
\end{eqnarray}
we find 
\footnote{
It is easy to see that we can write an alternative form for the
action,
\begin{eqnarray}
 ({\bf n} \times \partial_\mu {\bf n}) \cdot
 ({\bf n} \times \partial_\mu {\bf n})
 = \partial_\mu {\bf n} \cdot \partial_\mu {\bf n} ,
\end{eqnarray}
where the explicit form is written as
\begin{eqnarray}
 {\bf n}(x) \times \partial_\mu {\bf n}(x) 
 := \pmatrix{-\sin \varphi(x) \partial_\mu \theta(x)
 - \sin \theta(x) \cos \theta(x) \cos \varphi(x) \partial_\mu
\varphi(x)
\cr
 \cos \varphi(x) \partial_\mu \theta(x)
 - \sin \theta(x) \cos \theta(x) \sin \varphi(x) \partial_\mu
\varphi(x)
\cr
  \sin^2 \theta(x) \partial_\mu \varphi(x) } .
\end{eqnarray}
}
\begin{eqnarray}
 {1 \over 2}[(\Omega_\mu^1(x))^2+ (\Omega_\mu^2(x))^2]
&=&  {1 \over 2g^2} 
 [(L_\mu^1(x))^2+ (L_\mu^2(x))^2]
\nonumber\\
&=&  {1 \over 2g^2} 
\partial_\mu {\bf n}(x)  \cdot \partial_\mu {\bf n}(x)  
\\
&=&  {1 \over 2g^2}[ (\partial_\mu \theta(x))^2
  + \sin^2 \theta(x) (\partial_\mu \varphi(x))^2] .
  \label{NLSMaction0}
\end{eqnarray}
Following the argument in the previous section, we conclude that the
$SU(2)/U(1)$ MAG TFT in $D$ dimensions (TFT$_D$) is "equivalent" to
O(3) NLSM in $D-2$ dimensions (NLSM$_{D-2}$)
with the action,
\begin{eqnarray}
 S_{NLSM} = \int d^{D-2} x  {\pi \over 2g^2} 
\partial_\mu {\bf n}(x)  \cdot \partial_\mu {\bf n}(x) .
\label{action}
\end{eqnarray}
Both the action (\ref{action}) and the constraint (\ref{const}) are
invariant under global
$O(3)$ rotation in internal space.
The vector ${\bf n}$ is related to $U$  through the  adjoint orbit
parameterization (see e.g. \cite{KTT97} for more rigorous
mathematical presentation) as
\begin{eqnarray}
 n^A(x)T^A =  U^\dagger(x) T^3 U(x), \quad 
 n^A(x) = {\rm tr}[U(x)T^A U^\dagger(x) T^3] \quad (A=1,2,3) .
 \label{aop}
\end{eqnarray}
The residual U(1) invariance corresponds to a rotation about the
vector ${\bf n}$.  In other words, ${\bf n}$ is a U(1)
gauge-invariant quantity and the NLSM is a theory written in terms
of gauge invariant quantity alone. In fact, under the transformation 
$U \rightarrow e^{i \theta T^3}U$, $n^A$ is invariant. Then the
$U(1)$ part in the Haar measure is factored out.  This can be seen
as follows.
\par
In general, the action of NLSM  is determined as follows.
The infinitesimal distance in the group manifold 
$SU(2)/U(1) \cong S^2$ is
given by
\begin{eqnarray}
 ds^2 = g_{ab}(\Phi) d\Phi^a d\Phi^b 
 = R^2[(d\theta)^2 + \sin^2 \theta (d\varphi)^2] . 
\end{eqnarray}
This implies that the metric $g_{ab}$ and its determinant $g$ are
given by
\begin{eqnarray}
  g_{\theta\theta} = R^2,   \quad
  g_{\varphi\varphi} = R^2 \sin^2 \theta,   \quad
  g = \det(g_{ab}) = R^4 \sin^2 \theta . 
\end{eqnarray}
Hence the corresponding action of NLSM is given by
\begin{eqnarray}
  S = \int d^dx  g_{ab}(\Phi(x)) \partial_\mu \Phi^a(x) \partial_\mu
\Phi^b(x) ,
\end{eqnarray}
where coordinates $x^\mu, \mu=1, \cdots, d$ span a $d$-dimensional
flat space-time and the fields $\Phi^a (a=1,2)$ are
coordinates in
two-dimensional Riemann manifold ${\cal M}$ called the target
space.  The symmetric matrix $g_{ab}(\Phi)$ is the corresponding
metric tensor. Indeed, this action (Lagrangian) agrees with
(\ref{NLSMaction0}) for 
$\Phi^a=(\theta, \varphi)$.
Consequently, the integration measure is given by
\begin{eqnarray}
   d\mu(\Phi)  := \prod_{x\in R^d} \sqrt{g(\Phi(x))} d\Phi^1  
d\Phi^2
  = \prod_{x\in R^d} R^2 \sin \theta(x) d\theta(x) d\varphi(x) .
\end{eqnarray}
This is the area element of two-dimensional sphere of radius $R$.
Thus the partition function is defined by
\begin{eqnarray}
 Z_{NLSM} &:=& \int [d\mu({\bf n})]
\prod_{x\in R^d} \delta({\bf n}(x) \cdot {\bf n}(x) - 1) \exp
(-S_{NLSM}) ,
\\
 d\mu({\bf n}) &=& \prod_{x\in R^d} \sin \theta(x) d\theta(x)
d\varphi(x) .
\end{eqnarray}
\par
The constraint (\ref{const}) is removed by introducing the Lagrange
multiplier field $\lambda(x)$ as
\begin{eqnarray}
 S_{NLSM} = \int d^{D-2} x  \left[ {\pi \over 2g^2} 
\partial_\mu {\bf n}(x)  \cdot \partial_\mu {\bf n}(x) 
+ \lambda(x)({\bf n}(x) \cdot {\bf n}(x)-1) \right] .
\end{eqnarray}
 For this action, the field equation is
\begin{eqnarray}
 \partial_\mu \partial_\mu {\bf n}(x)   
+ \lambda(x) {\bf n}(x) = 0 .
\end{eqnarray}
Using the constraint and this field equation, we see
\begin{eqnarray}
  \lambda(x) = \lambda(x)  {\bf n}(x) \cdot {\bf n}(x)
  = - {\bf n}(x) \cdot \partial_\mu \partial_\mu {\bf n}(x) .
\end{eqnarray}
Therefore $\lambda$ is eliminated from the field equation, 
\begin{eqnarray}
 \partial_\mu \partial_\mu {\bf n}(x)   
- ({\bf n}(x) \cdot \partial_\mu \partial_\mu {\bf n}(x))
{\bf n}(x) = 0 .
\label{feq}
\end{eqnarray}

\subsection{Instanton solution}
 Instantons are solutions of field equations with non-zero but finite
action.  For this, the field ${\bf n}(x)$ must satisfy 
\begin{eqnarray}
\partial_\mu {\bf n}(x) \rightarrow 0 \quad (r \rightarrow \infty) ,
\end{eqnarray}
namely, ${\bf n}(x)$ approach the same value ${\bf n}^{(0)}$ at
infinity where ${\bf n}^{(0)}$ is any unit vector in internal space, 
${\bf n}^{(0)} \cdot {\bf n}^{(0)} =1$.
\par
It is important to remark that the coset
$SU(2)/U(1)$ is isomorphic to the two-dimensional surface $S^2$
($S^n \cong SO(n+1)/SO(n)$),
\begin{eqnarray}
 SU(2)/U(1) \cong S^2 := S^2_{int} .
\end{eqnarray}
Moreover, by one-point compactification (i.e., adding a point of
infinity) two-dimensional plane can be converted into the
two-dimensional sphere,
\begin{eqnarray}
 R^2 \cup \{\infty\} \cong S^2 = S^2_{phy} .
\end{eqnarray}
This implies that any finite action configuration
${\bf n}(x)$ is just a mapping from $S^2_{phy}$ to $S^2_{int}$.  The
mapping can be classified by the homotopy theory.
The O(3) NLSM$_2$ has instanton and anti-instanton
solutions, because the homotopy group is non-trivial,
\begin{eqnarray}
 \Pi_2(SU(2)/U(1)) = \Pi_2(S^2) = Z .
\end{eqnarray}
The instanton (topological soliton) is characterized by the
integer-valued topological charge $Q$. 
This is seen as follows.
\par
The mathematical identity
\begin{eqnarray}
\partial_\mu {\bf n}  \cdot \partial_\mu {\bf n}
= {1 \over 2} 
(\partial_\mu {\bf n} \pm \epsilon_{\mu\rho} {\bf n} \times
\partial_\rho {\bf n}) \cdot 
(\partial_\mu {\bf n} \pm \epsilon_{\mu\sigma} {\bf n} \times
\partial_\sigma {\bf n})
\pm   \epsilon_{\mu\nu} {\bf n} \cdot 
(\partial_\mu {\bf n}  \times  \partial_\nu {\bf n}) 
\end{eqnarray}
implies
\begin{eqnarray}
\partial_\mu {\bf n}  \cdot \partial_\mu {\bf n}
\ge \pm   \epsilon_{\mu\nu} {\bf n} \cdot 
(\partial_\mu {\bf n}  \times  \partial_\nu {\bf n}) .
\label{ineq1}
\end{eqnarray}
Hence the action has a lower bound,
\begin{eqnarray}
 S_{NLSM} := \int d^{D-2} x  {\pi \over 2g^2} 
\partial_\mu {\bf n}(x)  \cdot \partial_\mu {\bf n}(x) 
\ge  S_Q := {4\pi^2 \over g^2} |Q| ,
\end{eqnarray}
where $Q$ is the Pontryagin index (winding number) defined by
\begin{eqnarray}
 Q := {1 \over 8\pi} \int d^2 x 
 \epsilon_{\mu\nu} {\bf n} \cdot (\partial_\mu {\bf n}  .
\times \partial_\nu {\bf n}) .
\end{eqnarray}
The Euclidean action $S_{NLSM}$ of NLSM is minimized when the
inequality (\ref{ineq1}) is saturated.  This happen if and only if
\begin{eqnarray}
 \partial_\mu {\bf n} = \pm \epsilon_{\mu\nu} {\bf n} \times
\partial_\nu {\bf n} .
\label{ieq}
\end{eqnarray}
Any field configuration that satisfies (\ref{ieq}) as well as the
constraint (\ref{const}) will minimize the action 
and therefore
automatically satisfies the extremum condition given by the field
equation (\ref{feq}).
The converse is not necessarily true. 
Note that (\ref{ieq}) is a first-order differential equation and
easier to solve than the field equation (\ref{feq}) which is a
second-order differential equation.
\par
Now we proceed to construct the topological charge.
\begin{eqnarray}
 \partial_\mu {\bf n}   \times \partial_\nu {\bf n} 
 = \sin \theta (\partial_\mu \theta \partial_\nu \varphi
 - \partial_\mu \varphi \partial_\nu \theta) {\bf n}
 = \sin \theta {\partial(\theta, \varphi) 
 \over \partial(x^\mu,x^\nu)} {\bf n} ,
\end{eqnarray}
where 
${\partial(\theta, \varphi) \over \partial(x^\mu,x^\nu)}$ 
is the Jacobian of the transformation from coordinates
$(x^\mu,x^\nu)$ on
$S^2_{phy}$ to $S^2_{int}$ parameterized
by $(\theta, \varphi)$ where $\mu, \nu$ are any pair from $1, \cdots
, D$.  Using
\begin{eqnarray}
 {\bf n} \cdot (\partial_\mu {\bf n}   \times \partial_\nu {\bf n}) 
 = \sin \theta (\partial_\mu \theta \partial_\nu \varphi
 - \partial_\mu \varphi \partial_\nu \theta) 
 = \sin \theta {\partial(\theta, \varphi) 
 \over \partial(x^\mu,x^\nu)}  ,
\end{eqnarray}
it is easy to see that $Q$ is an integer, since
\begin{eqnarray}
 Q &:=& {1 \over 8\pi} \oint_{S^2_{phy}} d^2 x 
  \epsilon_{\mu\nu} \sin \theta {\partial(\theta, \varphi) 
 \over \partial(x^\mu,x^\nu)}
\nonumber\\
  &=& {1 \over 4\pi} \oint_{S^2_{phy}} d\sigma_{\mu\nu} 
  \sin \theta {\partial(\theta, \varphi) 
 \over \partial(x^\mu,x^\nu)}
 \nonumber\\
&=&  {1 \over 4\pi} \int_{S^2_{int}} \sin \theta d\theta  d\varphi  ,
\end{eqnarray}
where $S^2_{int}$ is a surface of a unit sphere with area $4\pi$. 
Hence
$Q$ gives a number of times the internal sphere $S^2_{int}$ is
wrapped by a mapping from the physical space $S^2_{phys}$ to the
space of fields
$S^2_{int}$.

\par
The instanton equation (\ref{ieq}) can be rewritten as
\begin{eqnarray}
 \partial_1 n = \mp i (n \partial_2 n_3 - n_3 \partial_2 n),
 \quad 
 \partial_2 n = \pm i (n \partial_1 n_3 - n_3 \partial_1 n),
 \quad 
 n := n_1 + i n_2 .
\end{eqnarray}
By change of variables (stereographic projection from north pole),
\begin{eqnarray}
 w_1(x) := {n_1(x) \over 1-n_3(x)},
 \quad 
 w_2(x) := {n_2(x) \over 1-n_3(x)},
\end{eqnarray}
the instanton equation reads
\begin{eqnarray}
 \partial_1 w = \mp i  \partial_2 w ,
 \quad 
 w := w_1 + i w_2 .
\end{eqnarray}
This is equivalent to the Cauchy-Riemann equation,
\begin{eqnarray}
 {\partial w_1(z) \over \partial x_1} = \pm
 {\partial w_2 \over \partial x_2} ,
 \quad 
 {\partial w_1(z) \over \partial x_2} = \mp 
 {\partial w_2(z) \over \partial x_1} , 
 \quad z := x_1 + i x_2 .
\end{eqnarray}
For the upper (resp. lower) signs, $w$ is an analytic function of
$z^*$ (resp. $z$).  Any analytic function $w(z), w(z^*)$ is a
solution of instanton equation and also of the field equation. Note
that $w$ is not an entire function and allows isolated poles in
$w(z)$, while cuts are prohibited by the single-valuedness of
$n_a(x)$.  The divergence $w \rightarrow \infty$ corresponds to
$n_3=1$, i.e., the north pole in $S^2_{int}$.
The Euler angles are related to the new variables as
\begin{eqnarray}
 w_1  := \tan {\theta \over 2} \cos \varphi,
 \quad 
 w_2  := \tan {\theta \over 2} \sin \varphi,
 \quad
 w = {n_1 + i n_2 \over 1-n_3} = e^{i\varphi} \cot {\theta \over 2} ,
\end{eqnarray}
corresponding to the stereographic projection from the north
pole. 
\footnote{
The stereographic projection from the south pole is
\begin{eqnarray}
 w_1  := \cot {\theta \over 2} \cos \varphi,
 \quad 
 w_2  := \cot {\theta \over 2} \sin \varphi,
 \quad
 w = {n_1 + i n_2 \over 1+n_3} = e^{i\varphi} \tan {\theta \over 2} .
\end{eqnarray}
}
\par
By using the new variables, we obtain the expressions for 
the topological charge
\begin{eqnarray}
  Q = {i \over 2\pi} \int_{S^2} {dw dw^* \over (1+w w^*)^2}
  = {i \over 2\pi} \int_{S^2} {dx_1 dx_2 \over (1+|w|^2)^2} \left(
  {\partial w \over \partial x_1}{\partial w^* \over \partial x_2}
-   {\partial w \over \partial x_2}{\partial w^* \over \partial x_1}
\right) ,
\end{eqnarray}
and an action
\begin{eqnarray}
\int d^2 x   {1 \over 2} \partial_\mu {\bf n} \cdot  \partial_\mu
{\bf n}  
= \int_{S^2} {dx_1 dx_2 \over (1+|w|^2)^2}
\left(
  {\partial w \over \partial x_1}{\partial w^* \over \partial x_1}
+   {\partial w \over \partial x_2}{\partial w^* \over \partial x_2}
\right) .
\end{eqnarray}
\par
A typical instanton solution with topological charge $Q=n$ is given
by
\begin{eqnarray}
  w(z) = [(z-z_0)/\rho]^n ,
\end{eqnarray}
where the constants $\rho$ and $z_0$ is regarded as the size and
location of the instanton solution.  The theory has the
translational and scale invariance
($x \rightarrow x-a$ and $x \rightarrow \rho x$ respectively),
since the solution exists for arbitrary
$\rho$ and
$z_0$, but neither the action nor the topological charge depend on
these constant.  The parameters $\rho, z_0$ are called collective
coordinates.

\subsection{One instanton solution}
\par
The one instanton solution at the origin $z_0=0$,
\begin{eqnarray}
  w(z) = z/\rho ,
  \label{oneinst}
\end{eqnarray}
implies the solution for the $O(3)$ vector,
\begin{eqnarray}
  n_1 = {2\rho x_1 \over |z|^2+\rho^2}, \quad
  n_2 = {2\rho x_2 \over |z|^2+\rho^2}, \quad
  n_3 = {|z|^2-\rho^2 \over |z|^2+\rho^2}, \quad
  |z|^2 := x_1^2 + x_2^2 .
  \label{oneinstn}
\end{eqnarray}
This solution is regarded as representing a monopole or a
projection of the four-dimensional instanton onto the
two-dimensional plane in the following sense. First, we observe that
the field of an instanton at infinity   points in the positive $3$
direction ${\bf n}^{(0)}$ while the field at the origin points in
the opposite direction, 
\begin{eqnarray}
  |z| &=& 0 \rightarrow {\bf n} = (0,0,-1) \equiv - {\bf n}^{(0)},
  \nonumber\\
  |z| &=& \rho \rightarrow {\bf n} = (x_1/\rho,x_2/\rho,0),
  \nonumber\\
  |z| &=& \infty \rightarrow {\bf n} = (0,0,1) \equiv {\bf n}^{(0)}.
\end{eqnarray}
If we identify the plane with the sphere $S^2$ by stereographic
projection from north pole, the north (resp. south) pole of $S^2$
corresponds to the infinity point (resp. the origin) and equator to
the circle
$|z|=\rho$.  Therefore, one instanton solution (\ref{oneinst}) looks
like a magnetic monopole (or a sea urchin).  The winding
number
$Q$ of this configuration is determined by the area of the sphere
divided by
$4\pi$, i.e., $Q=1$.  Thus one instanton has winding number $+1$ (One
anti-instanton  has $Q=-1$.). 
Equivalently, this denotes the magnetic charge $g_m=1$.
\par
Alternative interpretation is possible as follows.
The configuration (\ref{oneinstn}) leads to
\begin{eqnarray}
  {\bf n}(z) \cdot (\partial_i {\bf n}(z) \times \partial_j {\bf
n}(z))
  = -  \epsilon_{ij} {4\rho^2 \over (|z|^2+\rho^2)^2} .
  \label{2dinst0}
\end{eqnarray}
This should be compared with the four-dimensional instanton solution
in the non-singular gauge,
\begin{eqnarray}
 {\cal A}_\mu^A(x) =  \eta_{A\mu\nu} {2x_\nu \over x^2+\rho^2} ,
 \quad 
{\cal F}_{\mu\nu}^A(x) = - \eta_{A\mu\nu} 
 {4\rho^2 \over (x^2+\rho^2)^2} ,
 \quad x^2 = x_1^2+ \cdots + x_4^2 ,
\end{eqnarray}
which implies
\begin{eqnarray}
 {\cal A}_i^3(z) =  \epsilon_{ij} {2x_j \over |z|^2+\rho^2} ,
\quad 
{\cal F}_{ij}^3(z) =  - \epsilon_{ij} 
 {4\rho^2 \over (|z|^2+\rho^2)^2} ,
 \label{4dinst}
\end{eqnarray}
where we have used $\eta_{3ij}=\epsilon_{3ij}=\epsilon_{ij}$.
Therefore, the instanton solution (\ref{2dinst0}) in 2 dimensions 
is equal to the projection of the field strength ${\cal F}_{12}^3$
of the four-dimensional instanton solution (in the non-singular
gauge) onto two-dimensional plane. 
So it is expected that there is an interplay between the instanton
and the monopole in 4 dimensions. However, this does not imply that
the four-dimensional instanton configuration play the dominant role
in the confinement. The degrees of freedom responsible for the
confinement is the magnetic monopole which has complete
correspondence with the two-dimensional instantons as shown in the
next subsection furthermore.

\subsection{Instanton and magnetic monopole}

\par
By dimensional reduction, we can convert the calculation of
correlation functions in MAG TFT$_D$ into
that in NLSM$_{D-2}$, if all the arguments sit on the $D-2$
dimensional subspace. 
Euler angle expression yields
\begin{eqnarray}
 {\bf n} \cdot (\partial_\mu {\bf n}   \times \partial_\nu {\bf n}) 
=   C_{\mu\nu}[\Omega]
= \sin \theta (\partial_\mu \theta \partial_\nu \varphi
 - \partial_\mu \varphi \partial_\nu \theta) .
\label{corre}
\end{eqnarray}
Hence we obtain an alternative expression for the
winding number,
\begin{eqnarray}
 Q :=   {1 \over 8\pi} \int d^2 z \  \epsilon_{\mu\nu}
C_{\mu\nu}[\Omega]
 = {1 \over 4\pi} \int d^2 \sigma_{\mu\nu}
C_{\mu\nu}[\Omega] .
\end{eqnarray}
We can define the topological charge density by
\begin{eqnarray}
 \epsilon_{\mu\nu} C_{\mu\nu}[\Omega]
 = 
 \epsilon_{\mu\nu} {\bf n} \cdot (\partial_\mu {\bf n}   \times
\partial_\nu {\bf n})   .
\end{eqnarray}
 From (\ref{dec}), (\ref{magc1}) and (\ref{corre}), if we restrict
$\mu, \nu$ to the two dimensions, the monopole contribution in
four dimension corresponds to the instanton contribution in
two dimension. 
However, the monopole current defined by the divergence of the dual
field strength ${}^*f_{\mu\nu}$ can not be calculated in the
dimensionally reduced model, since all the derivatives is not
necessarily contained in two-dimensional space. 
However, if the four-dimensional diagonal field strength
$f_{\mu\nu}^\Omega$ is self-dual, 
\begin{eqnarray}
 {}^*f_{\mu\nu}^\Omega = f_{\mu\nu}^\Omega , 
\end{eqnarray}
the monopole charge in
4 dimensions completely agrees with the winding number (instanton
charge) in 2 dimensions
\begin{eqnarray}
  g_{m} = Q    . 
\end{eqnarray}
Intimate relationship between the magnetic monopole and instantons
may be a reflection of this observation.
Intuitively speaking, the magnetic monople and anti-monople currents
piercing the surface of the (planar) Wilson loop corresponds to the
instanton and anti-instanton in the dimensionally reduced
two-dimensional world.  In order to derive the area law of the Wilson
loop, the currents must piece the surface uniformly.  In this sense,
the monopole current condensation must occur in 4 dimensions.
\par
The dimensional reduction of TFT implies the self-duality at the
level of correlation function, 
\begin{eqnarray}
 \langle {\cal F}_{\mu\nu} \rangle_{TFT_4} 
 =    {1 \over 2}
\epsilon_{\mu\nu\rho\sigma} 
\langle {\cal F}_{\rho\sigma} \rangle_{TFT_4} ,
\end{eqnarray}
since both sides coincide with the same correlation function in the
dimensionally reduced two-dimensional model.

\par
If we define 
\begin{eqnarray}
 h_\mu &:=& - {1 \over g}  {1 \over 1 \mp n^3}
( {\bf n} \times \partial_\mu {\bf n})^3  
=  - {1 \over g}   {1 \over 1 \mp n^3}
\epsilon^{ab3}  n^a \partial_\mu n^b ,
\\
 a_\mu &:=& {\cal A}_\mu^A n^A + h_\mu, 
\end{eqnarray}
then we obtain the field strength,
\begin{eqnarray}
 f_{\mu\nu} := \partial_\mu a_\nu -  \partial_\nu a_\mu
 = \partial_\mu ({\cal A}_\nu^A n^A) -  \partial_\nu ({\cal A}_\mu^A
n^A)
-  {1 \over g} \epsilon^{ABC} n^A \partial_\mu n^B \partial_\nu n^C .
\end{eqnarray}
This field strength is regular everywhere and does not contain the
Dirac string.  This is nothing but the field strength of 't
Hooft-Polyakov monopole, since $n^A$ is obtained from $T^3$
by gauge rotation (\ref{aop}), 
$n^A(x)T^A =  U^\dagger(x) T^3 U(x)$. The Euler angle
expression 
\begin{eqnarray}
 h_\mu &:=& - {1 \over g}  {\sin^2\theta \partial_\mu \varphi \over 1
\mp \cos \theta} ,
\end{eqnarray}
is constructed from the instanton (vortex) solution in two dimensions
by the stereographic projection.
\par
The expression for the instanton charge in two-dimension is
equivalent to the magnetic charge in 4 dimensions, because
\begin{eqnarray}
 K_\mu &:=&  {1 \over 8\pi} \epsilon_{\mu\nu\rho\sigma}
\epsilon^{ABC}
\partial_\nu  n^A \partial_\rho n^B \partial_\sigma n^C 
 \nonumber\\
 &=& {1 \over 8\pi} \epsilon_{\mu\nu\rho\sigma} \epsilon^{ABC}
\partial_\nu (n^A \partial_\rho n^B \partial_\sigma n^C) 
 \nonumber\\
&=& {1 \over 8\pi} \epsilon_{\mu\nu\rho\sigma}  
\partial_\nu [{\bf n} \cdot (\partial_\rho {\bf n}   \times
\partial_\sigma {\bf n})] ,
\end{eqnarray}
we have
\begin{eqnarray}
 g_m := \int d^3x K_0 
&:=& {1 \over 8\pi} \int d^3x \epsilon_{ijk}  
\partial_i 
[{\bf n} \cdot (\partial_j {\bf n} \times \partial_k {\bf n})] 
 \nonumber\\
 &=& {1 \over 8\pi} \int d^2\sigma_i  \epsilon_{ijk} 
[{\bf n} \cdot (\partial_j {\bf n} \times \partial_k {\bf n})] 
 \nonumber\\
 &=& {1 \over 8\pi} \int_{S^2} d^2x  \epsilon_{jk} 
[{\bf n} \cdot (\partial_j {\bf n} \times \partial_k {\bf n})] 
  = Q .
\end{eqnarray}
The magnetic current is topologically conserved 
$\partial_\mu K_\mu = 0$ without equation of motion.
\par
We can also define the three-dimensional topological current,
\begin{eqnarray}
 J_\mu = {1 \over 8\pi} \epsilon_{\mu\rho\sigma} \epsilon^{ABC}
 (n^A \partial_\rho n^B \partial_\sigma n^C) 
= {1 \over 8\pi} \epsilon_{\mu\rho\sigma}  
  [{\bf n} \cdot (\partial_\rho {\bf n}   \times
\partial_\sigma {\bf n})] .
\end{eqnarray}
Then $Q$ is obtained from
\begin{eqnarray}
 Q &:=& \int d^2x J_0 .
\end{eqnarray}
This is related to Hopf invariant and Chern-Simons
theory \cite{Nakahara}.  The details will be presented in a
forthcomming paper.

%\newpage
\section{Wilson loop and linear potential}
\setcounter{equation}{0}

First of all, in order to see explicitly that the dimensionally
reduced two-dimensional NLSM has
$U(1)$ gauge invariance (corresponding to the residual $H$
symmetry), we study the  $CP^1$ formulation of O(3) NLSM.
The $CP^1$ formulation shows gauge structure more clearly than the
$O(3)$ NLSM and helps us to see analogy of NLSM with the 1+1
dimensional Abelian Higgs model, i.e., GL model.
The $CP^{N-1}$ model can have instanton solution for any $N$,
whereas the O(N) NLSM does not have for $N>3$. 
For $SU(N)$ YM theory in MAG, the dimensionally
reduced $SU(N)/U(1)^{N-1}$ NLSM has instanton solutions for any
$N$. The instanton solution of O(3) NLSM$_2$ is identified as a
vortex solution.  
\par
Next, we give relationship among the $CP^1$ model, O(3) NLSM and
TFT. It turns out that the calculation of the Wilson loop in
four-dimensional TFT is reduced to that in the two-dimensional $CP^1$
model owing to the dimensional reduction. 
\par
In subsection 6.3, we show that summing up the contribution of
instanton and anti-instanton configurations to the Wilson loop in
NLSM$_2$ or $CP^1$ model leads to quark confinement in 
four-dimensional TFT and YM theory in the sense of area law of the
Wilson loop.  We emphasize that the coset
$G/H$ is quite important for the existence of instanton and that the
coset structure is a consequence of the MAG together with the
dimensional reduction. We find that the magnetic monopole in
four dimensions corresponds to instanton (or vortex) in two
dimensions.
\par
Finally, we discuss some extensions of the proof of quark confinement
for the general gauge group and in higher dimensional cases.

\subsection{$CP^{N-1}$ model and instanton solution}
\par
\par
The $CP^{N-1}$ model is described by the $N$ complex scalar field 
$\phi_a(x) (a=1, \cdots, N)$ and the action of $d=D-2$
dimensional $CP^{N-1}$ model is given by
\begin{eqnarray}
 S_{CP}[\phi] = {\beta \over 2} \int d^{d}x \left[ 
 \partial_\mu \phi^*(x) \cdot \partial_\mu \phi(x)
 + (\phi^*(x) \cdot \partial_\mu \phi(x))(\phi^*(x) \cdot
\partial_\mu \phi(x)) \right] ,
\label{CPN}
\end{eqnarray}
where there is a constraint.
\begin{eqnarray}
 \phi^*(x) \cdot  \phi(x) := \sum_{a=1}^{N} 
 \phi_a^*(x) \phi_a(x) = 1 .
 \label{con}
\end{eqnarray}
\par
By introducing an auxiliary vector field $V_\mu$, $CP^{N-1}$ model
can be equivalently rewritten as
\begin{eqnarray}
 S_{CP} = {\beta \over 2} \int d^{d}x \left[ 
 \partial_\mu \phi^*(x) \cdot \partial_\mu \phi(x)
 + V_\mu^2(x) - 2 V_\mu(x) (i\phi^*(x) \cdot \partial_\mu \phi(x))
\right]  .
 \label{CPNb}
\end{eqnarray} 
In fact, integrating out $V_\mu$ field in (\ref{CPNb}) recovers
(\ref{CPN}). Here $V_\mu$ corresponds to the composite operator,
\begin{eqnarray}
 V_\mu(x) = i\phi^*(x) \cdot \partial_\mu \phi(x) .
 \label{defV}
\end{eqnarray}
This is real and 
$\phi^*(x) \cdot \partial_\mu \phi(x)$ is pure
imaginary, since from the constraint,
\begin{eqnarray}
 \phi^*(x) \cdot  \partial \phi(x) 
 + \partial \phi^*(x) \cdot  \phi(x) 
 = 2 Re(\phi^* \cdot \partial \phi(x)) = 0 .
\end{eqnarray}
Then, using the constraint (\ref{con}), the $CP^{N-1}$ model can be
further rewritten 
\begin{eqnarray}
 S_{CP}[\phi,V] = {\beta \over 2} \int d^{d}x  
(D_\mu[V] \phi^*(x)) \cdot (D_\mu[V] \phi(x))  ,
 \\
 D_\mu[V] \phi(x) := (\partial_\mu + iV_\mu) \phi(x) .
\end{eqnarray} 
The partition
function is defined by
\begin{eqnarray}
 Z_{CP} := \int [dV_\mu][d\phi][d\phi^*] 
 \prod_{x\in R^d} \delta(\phi(x) \cdot \phi(x)-1)
 \exp (-S_{CP}[\phi, V]) .
\end{eqnarray}
Here $D_\mu[V]$ is actually interpreted as the covariant derivative,
because the Lagrangian is invariant under the U(1) gauge
transformation,
\begin{eqnarray}
 \phi_a(x) &\rightarrow& \phi_a(x)' := \phi_a(x) e^{i \Lambda(x)},
 \nonumber\\
 V_\mu(x) &\rightarrow& V_\mu(x)' := V_\mu(x) - \partial_\mu
\Lambda(x),
\end{eqnarray}
where $\Lambda$ is independent of the index $a$ and 
\begin{eqnarray}
 D_\mu \phi_a(x) &\rightarrow& (D_\mu \phi_a(x)) e^{i \Lambda(x)} .
\end{eqnarray}
By this property, this model is called the $CP^{N-1}$ model
(target space is the complex projective space). Note that
\begin{eqnarray}
  CP^{N-1} \cong U(N)/U(1)/U(N-1) \cong SU(N)/U(N-1) .
\end{eqnarray}
The $CP^{N-1}$ model has global $SU(N)$ symmetry and a $U(1)$
subgroup of this $SU(N)$ is a local gauge symmetry. 
Hence the $CP^{N-1}$ model is U(1) gauge theory for any $N$.
However, 
$V_\mu$ is an auxiliary vector field and does not represent
independent degrees of freedom, since the kinetic term is absent. 
Apart from this fact, the $CP^{N-1}$ model is similar to the Abelian
Higgs model or scalar quantum electrodynamics (QED).
It is known that the kinetic term of $V_\mu$ is generated through
radiative correction, see \cite{Coleman85}.
\par
The constraint is included in the action by introducing the Lagrange
multiplier field $\lambda$ as
\begin{eqnarray}
 S_{CP} = {\beta \over 2} \int d^{d}x \left[ 
(D_\mu[V] \phi^*(x)) \cdot (D_\mu[V] \phi(x)) 
+ \lambda(x) (\phi^*(x) \cdot \phi(x) - 1) \right],
\end{eqnarray} 
The field equation is 
\begin{eqnarray}
  D_\mu[V] D_\mu[V] \phi(x) + \lambda(x)  \phi(x)  = 0 .
\end{eqnarray} 
The multiplier field is eliminated using
\begin{eqnarray}
  \lambda(x) = \lambda(x) \phi^*(x) \cdot \phi(x) 
  = -  \phi^*(x) \cdot D_\mu[V] D_\mu[V] \phi(x) ,
\end{eqnarray} 
to yield
\begin{eqnarray}
  D_\mu[V] D_\mu[V] \phi(x) 
  -  (\phi^*(x) \cdot D_\mu[V] D_\mu[V] \phi(x)) \phi(x)   = 0 .
\end{eqnarray} 

\par
Instantons are finite action solutions of field equations.
Finiteness of the action requires the boundary condition, 
\begin{eqnarray}
 D_\mu \phi_a := \partial_\mu \phi_a + i V_\mu \phi_a \rightarrow 0 
 \quad {\rm as} \quad r := |{\bf x}| \rightarrow \infty .
\end{eqnarray}
Separating $\phi_a$ into the modulus and the angular part,
\begin{eqnarray}
 \phi_a(x) := |\phi_a(x)| e^{i \Theta_a(x)},
\end{eqnarray}
the boundary condition yields
\begin{eqnarray}
  V_\mu = i {\partial_\mu \phi_a \over \phi_a} 
 = i {\partial_\mu |\phi_a| \over |\phi_a|} - \partial_\mu \Theta_a .
\end{eqnarray}
Here $V_\mu$ must be real and independent of $a$.  Hence, 
$\partial_\mu |\phi_a|=0$ and $\partial_\mu \Theta_a$ is independent
of $a$. This means $ |\phi_a|=\phi_0$ for a fixed complex vector
with 
$(\phi_0)^* \cdot \phi_0=1$ and
$\Theta_a = \Theta(\varphi)$ for a common phase angle
$\Theta(\varphi)$ which can depend on $\varphi$ parameterizing a
circle, $S^1_{phy}$.  Consequently, the boundary condition is given
by
\begin{eqnarray}
 \phi_a(x) \rightarrow  \phi_0 e^{i \Theta(\varphi)},
 \quad
 V_\mu  \rightarrow  - \partial_\mu \Theta(\varphi),
\end{eqnarray}
where the allowed values of the phase $\Theta$ form a circle
$S^1_{int}$.
The mapping $\Theta$ from $S^1$ to $S^1$ is
characterized by an  winding number 
\begin{eqnarray}
 Q := {1 \over 2\pi} \int_{S^1_{phy}} d\varphi {d \Theta \over
d\varphi} ,
\end{eqnarray}
which has an integral value corresponding to the fact,
\begin{eqnarray}
 \Pi_1(S^1) = Z  .
\end{eqnarray}
Though the global $SU(N)$ rotations can continuously change the value
of
$\phi_0$, this freedom does not introduce further homotopy
classification.
\par
The winding number can be rewritten in terms of $V_\mu$ as follows.
From (\ref{defV}), 
\begin{eqnarray}
 V_\varphi 
 = {i \over r} \phi^* \cdot {\partial \phi \over \partial \varphi}
 \rightarrow - {1 \over r} {d\Theta \over \partial \varphi} .
\end{eqnarray}
This leads to
\begin{eqnarray}
 Q = - {1 \over 2\pi} \int_{S^1_{phy}} d\varphi \ r V_\varphi
 = - {1 \over 2\pi} \int_{S^1_{phy}} d\ell \cdot V 
 = - {1 \over 2\pi} \int d^2x \epsilon_{\mu\nu} \partial_\mu V_\nu ,
 \label{chargeV}
\end{eqnarray}
where the integrand is a pure divergence. Using the constraint,
we can show that this is rewritten as
\begin{eqnarray}
 Q =  \int d^2x \epsilon_{\mu\nu} (D_\mu \phi)^* (D_\nu \phi) .
\end{eqnarray}
\par
 From an identity
\begin{eqnarray}
 (D_\mu \phi) \cdot (D_\mu \phi) = {1 \over 2} 
 (D_\mu \phi \pm i \epsilon_{\mu\nu} D_\nu \phi)^* \cdot 
 (D_\mu \phi \pm i \epsilon_{\mu\nu} D_\nu \phi)
 \mp i \epsilon_{\mu\nu} (D_\mu \phi)^* (D_\nu \phi) ,
\end{eqnarray}
we obtain
\begin{eqnarray}
  (D_\mu \phi) \cdot (D_\mu \phi) \ge
 \mp i \epsilon_{\mu\nu} (D_\mu \phi)^* (D_\nu \phi) .
\end{eqnarray}
Hence the lower bound of the action is obtained,
\begin{eqnarray}
 S_{CP} \ge {\pi^2 \over g^2} |Q| := S_Q.
\end{eqnarray}
The action has the minimum value when the inequality is saturated,
\begin{eqnarray}
 D_\mu \phi_a = \pm i \epsilon_{\mu\nu} D_\nu \phi_a  .
 \label{ieq2}
\end{eqnarray}
This is a self-duality equation which is analogous to the
self-duality equation of YM theory.
This equation is first order (partial differential) equation and
easier to solve than the field equation.  Solution of this equation
automatically satisfies the field equation, but the converse is not
necessarily true.
\par
To solve (\ref{ieq2}), we introduce the gauge invariant field,
\begin{eqnarray}
 \omega_a(x) := \phi_a(x)/\phi_1(x) \quad (a=1, \cdots, N) .
\end{eqnarray}
The covariant derivative is eliminated by substituting 
$\phi_a(x)=\omega_a(x) \phi_1(x)$ into (\ref{ieq2}),
\begin{eqnarray}
 \partial_\mu \omega_a(x) = \pm i \epsilon_{\mu\nu} \partial_\nu
\omega_a  .
 \label{ieq3}
\end{eqnarray}
This is nothing but the Cauchy-Riemann  equation. 
For the minus (resp. plus) sign, each $\omega_a$ is an analytic
function of $z:=x_1+ix_2$ (resp. $z^*=x_1-ix_2$).
\par
The expression for $V_\mu$ in terms of $\omega$ is
\begin{eqnarray}
 V_\mu &=& {i \over 2|\omega|^2}(\omega^* \cdot \partial_\mu \omega
 - \omega \cdot \partial_\mu \omega^*)
 = {i \over 2}(\hat \omega^* \cdot \partial_\mu \hat \omega
 - \hat \omega \cdot \partial_\mu \hat \omega^*),
 \nonumber\\
 && \hat \omega := \omega/|\omega| , \quad
 |\omega|:= (\omega^* \cdot \omega)^{1/2} = |\phi_1|^{-1} .
\end{eqnarray}
Taking into account the Cauchy-Riemann relations, we obtain
\begin{eqnarray}
 V_\mu = \pm \epsilon_{\mu\nu} {\omega^* \cdot \partial_\nu
\omega + \omega \cdot \partial_\nu \omega^* \over 2|\omega|^2}
= \pm \epsilon_{\mu\nu}  \partial_\nu \ln |\omega| .
\label{V}
\end{eqnarray}
The topological charge is expressed as
\begin{eqnarray}
 Q  = - {1 \over 2\pi} \int d^2x \epsilon_{\mu\nu} \partial_\mu
V_\nu  
= \pm {1 \over 4\pi} \int d^2x \partial_\mu \partial_\mu  
\ln |\omega|^2 .
\end{eqnarray}
\par
An example of one instanton solution is given by
\begin{eqnarray}
 \omega(z) = u + [(z-z_0)/\rho] v,
 \label{CPsol}
\end{eqnarray}
where $u,v$ are any pair of orthonormal complex vectors satisfying 
\begin{eqnarray}
 u_1 = v_1 = 1, \quad
 u^* \cdot u = v^* \cdot v = 1, \quad u^* \cdot v = 0.
\end{eqnarray}
Here the constant $\rho, z_0$ represent the size and location
(in the $z$ plane) of instanton.
Reflecting the scale and translational invariance of the action, we
can choose arbitrary values for $\rho, z_0$.
The solution (\ref{CPsol}) is inverted to become
\begin{eqnarray}
 \phi_a(z) = {\rho u_a + (z-z_0) v_a \over
(\rho^2+|z-z_0|^2)^{1/2}} .
\label{CPsol2}
\end{eqnarray}
As $z \rightarrow \infty$, this solution satisfies the boundary
condition with a phase angle $\Theta(\varphi)=\varphi$,
\begin{eqnarray}
 \phi_a(z)  \rightarrow (z/|z|) u = e^{i \varphi} u .
\end{eqnarray}
Hence this solution leads to $Q=1$, the single instanton.  The
anti-instanton is obtained by replacing $z$ by $z^*$,

\par
Using the solution (\ref{CPsol}), the vector potential (\ref{V}) and
its field strength reads
\begin{eqnarray}
 V_\mu &=& \pm \epsilon_{\mu\nu} {x_\nu \over |x|^2+\rho^2} ,
 \quad |x|^2 = x_1^2 + x_2^2 ,
 \\
 V_{\mu\nu} &:=& \partial_\mu V_\nu - \partial_\nu V_\mu 
 = \mp \epsilon_{\mu\nu} 
 {2\rho^2 \over (|x|^2+\rho^2)^2} .
 \label{2dinst}
\end{eqnarray}
Note that $V_\mu$ tends to a pure U(1) gauge field configuration at
infinity,
\begin{eqnarray}
 V_\mu \rightarrow \pm \epsilon_{\mu\nu} {x_\nu \over |x|^2} 
 = \partial_\mu \Theta, \quad
 \Theta := \arctan {x_2 \over x_1} .
\end{eqnarray}
Hence $V_\mu$ denotes the vortex with a center at $x=0$.
This is consistent with (\ref{chargeV}).  
This implies that the magnetic field of the magnetic current induces
the (quantized) current around it on the plane perpendicular to the
magnetic field. This is regarded as the dual of the usual Ampere
law  
where the electric current induces the magnetic field around it,
\begin{eqnarray}
 I = {1 \over 2\pi} \oint_C V 
 = {1 \over 2\pi} \oint_C d \Theta = n , \quad
 V := V_\mu dx^\mu .
\end{eqnarray}
In two dimensions the dual of the vector is again the vector.
The two descriptions are dual to each other.
\par
The solution (\ref{2dinst}) should be compared with the
four-dimensional instanton solution in the non-singular gauge
(\ref{4dinst}). The instanton solution (\ref{2dinst}) in two
dimensions are regarded as the projection of the four-dimensional
counterparts (\ref{4dinst}) on the two-dimensional plane.  
However, this does not imply that the four-dimensional
instanton configuration play the dominant role in the confinement.
The degrees of freedom responsible for the confinement is the
magnetic monopole which has complete correspondence with the
two-dimensional instantons.  This has been shown in sections 5.2 and
5.3.

\subsection{$CP^1$ model, $O(3)$ NLSM and TFT}
\par
The $CP^1$ model is locally isomorphic to the $O(3)$ NLSM with the
identification, 
\begin{eqnarray}
 n^A(x) := {1 \over 2} 
 \phi_a^*(x) (\sigma^A)_{ab}  \phi_b(x)  \quad (a,b = 1,2) ,
\end{eqnarray}
or
\begin{eqnarray}
 n^1 =  Re (\phi_1^* \phi_2), \quad n^2 =  Im(\phi_1^* \phi_2),
 \quad n^3 = {1 \over 2}(|\phi_1|^2 - |\phi_2|^2) .
 \label{corresp}
\end{eqnarray}
Actually, the constraint is satisfied,
$n^A n^A = (|\phi_1|^2 + |\phi_2|^2)^2 =1$. 
Hence $CP^1$ model has three independent parameters, whereas
O(3) vector ${\bf n}$ has two.  One of three parameters in $CP^1$
model is unobservable, since a global change of the phase does not
lead to any observable effect.  In fact, ${\bf n}$ is invariant
under the
$U(1)$ gauge transformation.  It is possible to show
that the Lagrangian (\ref{CPN}) for $N=2$ reduces to O(3) NLSM.
The $CP^{N-1}$ has instantons for arbitrary
$N \ge 2$, while $O(N)$ NLSM does not have them for $N>3$.
The map from $CP^1$ model to $O(3)$ NLSM is identified with a Hopf
map
$H: S^3 \rightarrow S^2$ where $S^3$ denotes the unit three-sphere
embedded in $R^4$ by
$|\phi_1|^2 + |\phi_2|^2 =1$.  In the language of
mathematics,
$S^3$ is a $U(1)$ bundle over $S^2$, see e.g. \cite{Nakahara}.

\par
The field variables of $CP^1$ model is written in terms of Euler
angles,
\begin{eqnarray}
 \phi_1 = \sqrt{2S} \exp [{i \over 2}(\varphi+\chi)] 
 \cos {\theta \over 2} , \quad
 \phi_2 = \sqrt{2S} \exp [-{i \over 2}(\varphi-\chi)] 
 \sin {\theta \over 2} ,
 \label{defspin}
\end{eqnarray}
which satisfies the constraint
$\phi_a^* \phi_a = 2S$.
Indeed, substitution of (\ref{defspin}) into (\ref{corresp}) leads to
\begin{eqnarray}
 n^1 &=&  2 {\rm Re} (\phi_1 \phi_2^*) = 2S \sin \theta \cos
\varphi, 
 \nonumber\\
 n^2 &=& 2 {\rm Im}(\phi_1 \phi_2^*) = 2S \sin \theta \sin \varphi,
 \nonumber\\ 
 n^3 &=&  |\phi_1|^2 - |\phi_2|^2  
 = 2S \cos \theta .
 \label{corresp2}
\end{eqnarray}
The is nothing but the Schwinger-Wigner representation of spin
$S$ operator in terms of two Bose creation and annihilation
operators $\phi_a^\dagger, \phi_a$.  In the path integral formalism,
they are not operators, but $c$-numbers.
\par
Substituting (\ref{defspin}) into (\ref{defV}) yields 
\begin{eqnarray}
 V_\mu(x) = i\phi^*(x) \cdot \partial_\mu \phi(x) 
 = - S [\partial_\mu \chi + \cos \theta \partial_\mu \varphi]
 = - S L_\mu^3 .
 \label{defV2}
\end{eqnarray}
Hence, the vector field $V_\mu$ is equivalent to $\Omega_\mu^3$ when
$\mu$ is restricted to $\mu=1,\cdots,d$. 
Furthermore, 
\begin{eqnarray}
  \partial_\mu \phi^*(x) \cdot \partial_\mu \phi(x)
  = {S \over 2}[(L_\mu^1)^2+(L_\mu^2)^2+(L_\mu^3)^2] .
\end{eqnarray} 
Owing to the dimensional reduction, the $D$-dimensional SU(2) MAG
TFT is equivalent to the $d (=D-2)$-dimensional $CP^1$ model,
\begin{eqnarray}
 S_{CP^1} = {\beta \over 2} \int d^{d}x \left[ 
 (L_\mu^1(x))^2+(L_\mu^2(x))^2 \right]  ,
\quad \beta := {\pi \over g^2}.
\end{eqnarray} 
Consequently, when the Wilson loop has the support on the
$(D-2)$-dimensional subspace $R^d \subset R^D$, 
then  the diagonal Wilson loop in $D$-dimensional SU(2) MAG TFT 
\begin{eqnarray}
 W_C[a^\Omega] &:=& \exp \left( i q \oint_C a_\mu^\Omega(z) dz^\mu
\right) , \quad z \in R^d
\\
\quad  a_\mu^\Omega(x) &:=& {\rm tr}[T^3 \Omega_\mu(x)] = L_\mu^3(x),
\end{eqnarray}
corresponds to the Wilson loop in $d(=D-2)$-dimensional $CP^1$ model,
\begin{eqnarray}
 W_C[V]  := \exp \left( i q \oint_C V_\mu(z) dz^\mu \right) 
=  \exp \left( {i \over 2} q \int_S V_{\mu\nu}(z)
d\sigma^{\mu\nu} \right)  .
\end{eqnarray}

\subsection{Area law for the diagonal Wilson loop}

\par
Now we evaluate the Wilson loop expectation value to obtain the
static potential for two widely separated charges
$\pm q$ (in a $\theta$ vacuum). 
We define the diagonal Wilson loop operator \cite{Polyakov77} for a
closed loop
$C$ by
\begin{eqnarray}
 W_C[a^U] := \exp \left( i q \oint_C a_\mu^U(x) dx^\mu \right) ,
\quad  a_\mu^U(x) := {\rm tr}[T^3 {\cal A}_\mu^U(x)] . 
\end{eqnarray}
According to  the Stokes theorem, this is equal to
\begin{eqnarray}
 W_C[a^U] = \exp \left( {i \over 2} q \int_S f_{\mu\nu}^U(x)
d\sigma^{\mu\nu} \right)  ,
\end{eqnarray}
for any surface $S$ with a boundary $C$.
We restrict the loop $C$ to be planar.
Otherwise, we can not receive any benefit of dimensional reduction to
calculate the Wilson loop expectation. 
In what follows we calculate the contribution from
$\Omega_\mu$, namely, topological contribution alone. 
Then the dimensional reduction implies
\begin{eqnarray}
  \langle W_C[a^\Omega] \rangle_{MAG TFT_4}
  = \langle W_C[a^\Omega] \rangle_{O(3) NLSM_{2}} ,
\quad  a_\mu^\Omega(x) := {\rm tr}[T^3 \Omega_\mu(x)] .
\end{eqnarray}
Following the procedure in section 2, we regard other contributions
as perturbative deformation
$W[U;J^\mu,0,0]$, see (\ref{pert}). 

\par
According to section 5.3 (or 6.2) for $G=SU(2)$,  the Wilson loop in
two-dimensional O(3) NLSM is rewritten as
\begin{eqnarray}
 W_C[a^\Omega] = \exp \left( i{2\pi q \over g}  \int_S  
d^2 x {1 \over 8\pi}
\epsilon_{\mu\nu} {\bf n} \cdot (\partial_\mu {\bf n}  
\times \partial_\nu {\bf n}) \right) .
\label{nlsmWl}
\end{eqnarray}
Note that the integrand is the density of instanton number as shown
in the previous section. This implies that the Wilson loop 
$W_C[a^\Omega]$ (\ref{nlsmWl}) counts the number of instanton -
anti-instanton  (or vortex--anti-vortex in $CP^1$ formulation)
existing in the area
$S$ bounded by the loop $C$ in O(3) NLSM.
The Wilson loop expectation value is written as
\begin{eqnarray}
 \langle W_C[a^\Omega]  \rangle_{O(3) NLSM_{2}} = 
 { \int d\mu({\bf n}) \delta({\bf n}\cdot{\bf n}-1) 
 e^{-S_{NLSM}+i \theta Q} W_C[a^\Omega]
 \over \int d\mu({\bf n}) \delta({\bf n}\cdot{\bf n}-1)
 e^{-S_{NLSM}+i \theta Q}  } 
 =: {I_2^\theta \over I_1^\theta} ,
 \label{wle}
\end{eqnarray}
where we have included the topological term $i\theta Q$.
\footnote{
  Note that the non-zero $\theta$ is not essential to show the area
law of the Wilson loop in the following.
We can put $\theta=0$ in the final results, (\ref{wler}) and
(\ref{pot}).
 }
Inclusion of topological term
$i \theta Q$ in the action is equivalent to consider the $\theta$
vacuum defined by
\begin{eqnarray}
  |\theta \rangle := \sum_{n=-\infty}^{+\infty} e^{in\theta} 
  |n \rangle .
\end{eqnarray}
The action with a topological angle $\theta$ is written as
\begin{eqnarray}
 S_{NLSM}^\theta = S_{NLSM} -  i \theta Q
 =  (n_{+}+n_{-})S_1 - i \theta (n_{+}-n_{-}) ,
 \quad S_1(g) = {4\pi^2 \over g^2} .
\end{eqnarray}
We regard (\ref{wle}) as the average of the instanton number
$Q$ inside
$S$ over all the instanton--anti-instanton ensembles generated from
the action of NLSM.

\par
In the following, we use the dilute instanton-gas approximation as a
technique to calculate (\ref{wle}). 
 This method is well known, see e.g. chapter 11 of Rajaraman
\cite{Rajaraman89} or Chapter 7 of Coleman \cite{Coleman85}.
(We will give the Wilson loop calculation based on other methods
elsewhere.)  We first classify the configurations of the field,
${\bf n}$, that contribute to the tunneling amplitude of instantons
$\langle n|e^{-HT} | 0 \rangle$
according to the number of well-separated instantons
$n_{+}$ and anti-instantons $n_{-}$ such that $Q=n=n_{+}-n_{-}$.  
Then we sum over all configurations with $n_+$ instantons and $n_-$
anti-instantons, all widely separated.  
In  the dilute-gas approximation, the
calculation of tunneling amplitude is reduced to that of a single
instanton (resp. anti-instanton) contribution 
$n \rightarrow n+1$ (resp. $n \rightarrow n-1$).
The term with
$n_{+}=1, n_{-}=0$ (or $n_{+}=0, n_{-}=1$)  is given by
\begin{eqnarray}
\langle n=\pm 1|e^{-HT} | 0 \rangle
&=& \int d\mu(\rho) \int d^2 x \exp (-S_1(g)) \exp (\pm i\theta)
\nonumber\\
&=& B L_1 L_2 \exp (-S_1(g)) \exp (\pm i\theta) . 
\end{eqnarray}
Here $T=L_1$ or $L_2$ and the prefactor $BL_1 L_2$ comes from
integration of the collective coordinates, the size and position of
the instanton,
\begin{eqnarray}
 \int d\mu(\rho) \int d^2 x  =  B L_1 L_2  , \quad
B \sim O(m_A^2) ,
\end{eqnarray}
where $L_1 L_2$ is the (finite but large) volume of two-dimensional
space and
$B$ is a normalization constant of order $m_A^2$, because  instanton
size is proportional to the inverse mass $m_A^{-1}$ of off-diagonal
gluons. In order to know the precise form of $B$, we must determine
the measure
$\mu(\rho)$ for the collective coordinate
$\rho$, see 
\cite{Jevicki77,Forster77,BL79,FFS79,BL81}.
\par
In the dilute-gas approximation, the denominator $I_1^\theta$ is
calculated as
\begin{eqnarray}
I_1^\theta :=  \langle \theta|e^{-HT} |\theta \rangle 
  &=& \sum_{n_{+},n_{-}=0}^{\infty} {(BL_1 L_2)^{n_{+}+n_{-}} \over
n_{+}!n_{-}!}  
   \exp \left[-(n_{+}+n_{-})S_1(g) 
  + i \theta (n_{+}-n_{-}) \right]
  \nonumber\\
  &=& \sum_{n_{+},n_{-}=0}^{\infty}  
  {1 \over n_{+}!} (BL_1 L_2 e^{-S_1(g) + i\theta})^{n_{+}} 
  {1 \over n_{-}!} (BL_1 L_2 e^{-S_1(g) - i\theta})^{n_{-}} 
  \nonumber\\
  &=&  \exp [ BL_1 L_2 e^{-S_1(g) + i\theta} 
  + BL_1 L_2 e^{-S_1(g) - i\theta}  ]
  \nonumber\\
  &=& \exp \left[2 (BL_1 L_2) \cos \theta e^{-S_1(g)} \right] ,
\end{eqnarray}
where there is no constraint on the integers $n_+$ or $n_-$, since
we are summing over all $Q =n_+ - n_-$. 
The sum is precisely the grand partition function for a classical
perfect gas (i.e. non-interacting particles)
\footnote{
By the fermionization method, the non-interacting instanton and
anti-instanton system can be rewritten as the free massive fermion
models with two flavors.
From this viewpoint, including the interactions between instantons
and anti-instantons is equivalent to introducing the
four-fermion interaction of Thirring type 
\cite{BL81}. 
By bosonization, the interacting fermionic model is converted into
the sine-Gordon-like bosonic model \cite{BL81}.
The Wilson loop calculation from this point of view will be given in
a forthcomming paper.
}
 containing two species of particles with equal chemical potential
$e^{-S_1(g)}$ and volume measured in units of
$B$.   The energy (action) for a configuration with $n_{+}$ and
$n_{-}$ members of each species is $(n_{+}+n_{-})S_1(g)$ while the
entropy of the configuration is 
$\ln [(BL_1 L_2)^{n_{+}+n_{-}}/n_{+}!n_{-}!]$.
\par
The configuration of instanton and anti-instanton is not an
exact solution to the equation of motion.  
However, the dominant term is given by the configuration for which
the free energy (energy  minus entropy) is smallest. 
For large coupling the action of a given field configuration
decreases like $g^{-2}$ while the entropy which is obtained as the
log of the volume of function space occupied by the configuration is
less sensitive to $g$.  Thus for moderate or strong coupling the
entropy of a field configuration can be more important than its
action.  The exact multi-instanton solutions are of essentially no
relevance in constructing the vacuum state because they have so
little entropy.  In fact, the sum over all terms with either $n_{+}$
or $n_{-}$ equal to zero is exponentially small compared to the
complete sum for large $T$ \cite{CDG78}.
When $g$ is small the instanton gas is extremely dilute.  For larger
$g$ instantons and anti-instantons come closer together.
\par
When $\theta=0$, the most dominant term in this sum is given for
large
$T$ at
\begin{eqnarray}
 n_{+} = n_{-} = BL_1 L_2 e^{-S_1(g)} ,
\end{eqnarray}
and as $T\rightarrow \infty$ the entire sum comes essentially from
this term alone. 
The important lessons learned from \cite{CDG78} are
(i) the dominant term contains both instantons and anti-instantons
and cannot be computed by a strict saddle-point method that relies
on exact solutions to the (Euclidean) equation of motion.
(ii) the dominant term is not the one for which the classical action
$\exp[-S]$ is minimum.

\par
 
The calculation of the numerator $I_2^\theta$ reduces to the
construction of a system in a
$\theta$ vacuum outside the loop and that in a $\theta + 2\pi q$
vacuum inside the loop. 
Let $A(C)$ be the area enclosed by the loop $C$.
In the dilute-gas approximation, the numerator is
\begin{eqnarray}
I_2^\theta  
  &=& \sum_{n_{+}^{in},n_{-}^{in}=0}^{\infty}
{(B A(C))^{n_{+}^{in}+n_{-}^{in}}
\over n_{+}^{in}!n_{-}^{in}!}  
   e^{-(n_{+}^{in}+n_{-}^{in})S_1(g)
   + i (\theta + {2\pi q \over g})(n_{+}^{in}-n_{-}^{in}) }
  \nonumber\\ 
  && \times \sum_{n_{+}^{out},n_{-}^{out}=0}^{\infty}
{(B(L_1 L_2-A(C)))^{n_{+}^{out}+n_{-}^{out}}
\over n_{+}^{out}!n_{-}^{out}!}  
   e^{-(n_{+}^{out}+n_{-}^{out})S_1(g) 
  + i \theta (n_{+}^{out}-n_{-}^{out}) } 
  \nonumber\\
  &=& \exp \left\{2 B \left[ A(C)\cos \left(\theta + {2\pi q \over
g}\right) 
  + (L_1 L_2-A(C))\cos \theta \right]
e^{-S_1(g)} \right\} . 
\end{eqnarray}
Here we decomposed the sum inside the Wilson loop and outside it.
The decomposition $n_{\pm}=n_{\pm}^{in}+n_{\pm}^{out}$ is meaningful
only when the loop $C$ is sufficiently large and the instanton size
is negligible  compared with the size of the loop $C$ so that the
overlapping of the instanton and anti-instanton with the loop is
neglected (This is equivalent to neglecting the perimeter decay part
of the Wilson loop). Then we can write
\begin{eqnarray}
 W_C[a^\Omega] = \exp \left[{2\pi q \over g}
i(n_{+}^{in}-n_{-}^{in}) \right].
\end{eqnarray}
Finally we notice that the volume dependence disappears in the ratio 
$I_2^\theta/I_1^\theta$.
The above derivation is very similar to the two-dimensional
Abelian Higgs model, see \cite{RU78}.

\par
In the vacuum with the topological angle
$\theta$, therefore, the Wilson loop expectation value has
\begin{eqnarray}
 \langle W_C[a^\Omega]  \rangle =  \exp \left\{ - 2B e^{-S_1} 
 \left[ \cos \theta - \cos \left(\theta + {2\pi q \over g}\right)
\right]A(C) \right\} .
\label{wler}
\end{eqnarray}
The Wilson loop integral exhibits area law.  
If we take the rectangular Wilson loop, the static quark potential
is derived.
If $q/g$ is an integer, the potential vanishes because
the vacuum is periodic in $\theta$ with period $2\pi$.
The integral charge is screened by the formation of neutral bound
states. When $q$ is not an integral multiples of an elementary
charge $g$, the static quark potential $V(R)$ is given by the linear
potential with string tension
$\sigma$,
\begin{eqnarray}
 V(R) = \sigma R ,
 \quad
 \sigma =   2B e^{-S_1} \left[ \cos \theta - \cos 
\left(\theta + {2\pi q \over g}\right) \right]  ,
\label{pot}
\end{eqnarray}
where $B \sim m_A^2$ and $S_1= 4\pi^2/g^2$ is the action for
one instanton. It should be remarked that the confining potential is
very much a non-perturbative quantum effect caused by instantons,
because the linear potential has a factor $e^{-S_1/\hbar}$ (if we
had retained
$\hbar$ dependence) which is exponentially small in
$\hbar$  and vanishes as $\hbar \rightarrow 0$.  This is a
crucial difference between the linear potential (\ref{pot}) and  the
linear Coulomb potential in two dimensions.  

\par
On the other hand, the four-dimensional 
Coulomb potential is calculated by perturbation theory
\cite{Kogut83} (see \cite{KondoIV}),
\begin{eqnarray}
   V(R) =  - {C_2 \over 4\pi} {g^2 \over R} + {\rm constant} .
\end{eqnarray}
Therefore, we arrive at the conclusion that the total static
quark potential in four-dimensional YM theory is given by
\begin{eqnarray}
   V(R) = \sigma R - {C_2 \over 4\pi} {g^2 \over R} 
 + {\rm constant} .
\end{eqnarray}

\par
The two-dimensional O(N+1) NLSM is asymptotic free and the $\beta$
function \cite{Polyakov75b} is given by
\begin{eqnarray}
  \beta(g) := \mu {dg(\mu) \over d\mu} = - {N-1 \over 8\pi^2} g^3 
  + O(g^5) ,
  \label{betaNLSM}
\end{eqnarray}
where $g$ is the renormalized coupling constant and $\mu$ the
renormalization scale (mass) parameter.  
By the dimensional transmutation as in QCD, the mass and  the
"string tension" of NLSM  should be given by \cite{HMN90} 
\begin{eqnarray}
  m \sim \Lambda  \exp \left( - \int^g {dg \over \beta(g)}
\right) , \quad
  \sigma \sim \Lambda^2 \exp \left( -2 \int^g {dg \over \beta(g)}
\right) .
\end{eqnarray}
For the $\beta$ function (\ref{betaNLSM}), this implies for $N=2$
\begin{eqnarray}
  \sigma \sim \Lambda^2 \exp \left( - {4\pi^2 \over g^2} \right) ,
\end{eqnarray}
in agreement essentially with the above result (\ref{pot}).
In this case, the scale $\Lambda$ of the theory is given by the
off-diagonal gluon mass $m_A$. This result does not agree with the
four-dimensional SU(N) YM theory in which
\begin{eqnarray}
  \beta(g)  =  - {b_0 \over 16\pi^2} g^3 + O(g^5) , 
\quad  b_0 = {11N \over 3} > 0 ,
  \label{betaYM}
\end{eqnarray}
because we have taken into account only the MAG TFT part of YM
theory and neglected an additional contribution coming from the
perturbative part (Note that the correspondence of SU(N) YM theory
to O(N+1) NLSM is meaningful only for $N=2$).  
By integrating out the off-diagonal gluons $A_\mu^{\pm}$ in MAG TFT
(\ref{actionGF}),
we can obtain the APEGT of MAG TFT, as performed for YM theory in the
previous paper \cite{KondoI}.
The APEGT of MAG TFT is given by the $H=U(1)$ gauge theory with the
running coupling $g(\mu)$ governed by $\beta$ function
(\ref{betaNLSM}), 
\begin{eqnarray}
  S_{APEGT} = \int d^4 x \left[ - {1 \over 4g^2(\mu)} 
  f_{\mu\nu} f^{\mu\nu} \right] ,
\end{eqnarray}
where ghost interactions and higher derivative terms are neglected.
\par
The naive instanton calculus given above can be improved by including
the correction around the instanton solutions following the works
\cite{BL79,FFS79,BL81,Silvestrov90}.
Although we have identified the two-dimensional space with the
sphere in the above, instanton solutions exist also for 
the torus \cite{RR83,MS94}
and the cylinder \cite{Snippe94}.  
However the torus only admits multi-instantons with topological
charge two or more (no single-instanton solution).

\subsection{Importance of coset $G/H$}
\par
In our approach, it is important to choose the coset $G/H$ so that 
$\Pi_1(G/H)\not=0$, because for any compact connected Lie group $G$, 
\begin{eqnarray}
 \Pi_2(G) = 0 ,
\end{eqnarray}
where the two-dimensional NLSM model fails to contain the instanton.
The MAG naturally leads to such a coset G/H NLSM. This is a reason
why the PGM based on $G$ can not contain non-trivial topological
structure and dynamical degrees of freedom except for unphysical
gauge modes, although the authors \cite{Hata84,HK85} tried to
include the physical modes as perturbation of PGM.  It is
interesting to clarify the relationship between the Wilson criterion
of quark confinement and color confinement criterion by Kugo and
Ojima
\cite{KO79} and Nishijima \cite{Nishijima94}.  This issue is deserved
for  future investigations.

\subsection{Generalization to $SU(N)$}
\par
The above consideration can be generalized to more general case,
$G=SU(N)$.  Using
\begin{eqnarray}
  \Pi_1(SU(N)) = 0 ,
\end{eqnarray}
we obtain
\begin{eqnarray}
 \Pi_2(SU(N)/U(1)^{N-1}) = \Pi_1(U(1)^{N-1}) = Z^{N-1} .
 \label{math}
\end{eqnarray}
This formula guarantees the existence of the instanton and
anti-instanton  solution in the 
$SU(N)/U(1)^{N-1}$ NLSM$_2$ model obtained from $SU(N)$ MAG TFT$_4$
by dimensional reduction.  Therefore, the whole strategy adopted in
this paper to prove the quark confinement will be valid for $SU(N)$
gauge theory in 4 dimensions. 
The origin of instantons in the dimensionally reduced model is the
monopole in the original model, as suggested by the mathematical
formula (\ref{math}). 
\par
In order to study the case $N=3$ in more detail, it will
be efficient to perform the $1/N$ expansion to the
$SU(N)/U(1)^{N-1}$ NLSM$_2$ model.
\par

\subsection{Higher-dimensional cases}

Our strategy of proving quark confinement in $D$
dimensions is based on the existence of instanton solutions in the
dimensionally reduced $(D-2)$-dimensional NLSM.
This can be generalized to arbitrary dimension, $D>4$.   
Remember the mathematical formula for the Homotopy
group,
\begin{eqnarray}
 \Pi_{n}(SU(2)/U(1)) = \Pi_{n}(S^2)  \ (n:=D-2>2) ,
\end{eqnarray}
and
\begin{eqnarray}
  \Pi_{3}(S^2) &=& Z  \ (D=5), \nonumber\\
  \Pi_{4}(S^2) &=& Z_2  \ (D=6), \nonumber\\
  \Pi_{5}(S^2) &=& Z_2  \ (D=7) , \cdots .
\end{eqnarray}
This opens  a possibility of proving quark confinement based on
instanton and anti-instantons even for $D>4$ dimensions.

\subsection{Exact results in two dimensions}

The classical O(3) NLSM in 1+1 dimensions is characterized by an
infinite number of conserved quantities and by B\"acklund
transformations for generating solutions.
The quantized O(3) NLSM is
asymptotically free and the conserved quantities exist free of
anomalies
\cite{Polyakov75b}. An exact factorised S-matrix has been
constructed using the existence of the infinite conserved quantities
\cite{ZZ79}.
\par
It is known \cite{Witten84} that the $\sigma$ model 
\begin{eqnarray}
 S =  {1 \over 4\lambda^2} \int d^2 x {\rm tr}(\partial_\mu U^{-1}
\partial^\mu U) + k \Gamma(U)  ,
\label{wzw}
\end{eqnarray}
with a Wess-Zumino (WZ) term,
\begin{eqnarray}
 \Gamma(U) := {1 \over 24\pi} \int d^3 x
\epsilon^{\alpha\beta\gamma} {\rm tr}[L_\alpha L_\beta L_\gamma],
\quad L_\mu := U^{-1} \partial_\mu U ,
\end{eqnarray}
becomes massless and possesses an infrared stable fixed point when
\begin{eqnarray}
 \lambda^2 = {4\pi \over k} \quad (k=1,2,\cdots) .
\end{eqnarray}
At these special values of $k$, the model (\ref{wzw}) is called the
level $k$ Wess-Zumino-Novikov-Witten (WZNW) model.
The familiar $\sigma$ model
corresponds to
$k=0$ case where the theory is asymptotically free and massive.
WZNW model is invariant under the conformal transformation and
with respect to infinite-dimensional current (Kac-Moody)
algebra.  
\par
The $\sigma$ model with arbitrary coupling $\lambda$ can
be solved exactly by means of the Bethe ansatz technique
\cite{PW83}.  However, the computation of correlation function
remain beyond the powers of the Bethe ansatz method.  Though the
conformal field theory approach \cite{BPZ84} is restricted to the
fixed-point case, but it provides much more detailed information
about the theory including the correlation functions \cite{KZ84}. 
We can calculate exactly all correlation functions in rational
conformal field theories  which include the WZNW and minimal models
as subsets.  The off-critical theory can be considered as
perturbation of conformal theories by a suitable relevant field. The
perturbed field theory is called a deformation and corresponds to
the renormalization group trajectory starting from the corresponding
fixed point.  The integrable deformation \cite{FOZ93} among all
possible deformations gives the integrable perturbed field theory
and  the factorized scattering theory.
\par
The NLSM with a topological angle $\theta$ is integrable at two
particular points $\theta=0$ and $\theta=\pi$
\cite{Polyakov77,FZ91,HMN90}.
At $\theta=0$ the correlation length is finite and all the
excitations are massive.  The spectrum consists of a single O(3)
triplet of massive particles with a non-perturbatively generated
mass 
$m \sim r_0^{-1} e^{-2\pi/g^2}$.
On the other hand, at
$\theta=\pi$ the scale invariant behavior is observed in the IR
limit, infinite correlation length.  The large-distance asymptotics
is described by the $SU(2)
\times SU(2)$ WZNW theory at level $k=1$.  So the NLSM at
$\theta=\pi$ can be considered as an interpolating trajectory ending
up at the IR fixed point characterized by level 1 CFT.

\par
The three-dimensional Chern-Simons gauge theory is a topological
field theory in the sense that the integrand of the action is a
total derivative and it is generally covariant without any metric
tensor.  If we quantize CS theory and take a time slice, one
dimension is lost, and the theory becomes a two-dimensional
conformal field theory. The correlation function in CS theory are
purely topological invariants and the correlation functions over
Wilson lines gives invariant knot polynomial \cite{Witten89}.  The
knot theory can describe all known rational conformal field theories.
\par
All the exact results in two dimensions  mentioned above will be
utilized to understand  more quantitatively the quark confinement in
four-dimensional QCD  by dimensional reduction.

%\newpage
\section{Discussion}

In this paper we have considered one of the most important problems
in modern particle physics: quark confinement in four-dimensional
QCD. 
In order to prove quark confinement in QCD, we have suggested to use
a TQFT which is extracted from the YM theory in the MAG. This TQFT
describes the dynamics of magnetic monopole and anti-monopole in YM
theory in MAG.     We have proposed a reformulation of QCD in which
QCD can be considered as a perturbative deformation of the TQFT. In
other words, in this reformulation the non-perturbative
dynamics of QCD is saturated by the TQFT we proposed, as far as the
issue of quark confinement is concerned. 
Needless to say, additional non-perturbative dynamics
responsible for quark confinement could possibly come from the
self-interaction among the gluon fields reflecting the non-Abelian
nature of the gauge group. 
However, additional non-perturbative contributions to quark
confinement are expected to be rather small, if any. This claim is
strongly supported by the recent numerical simulations
\cite{SY90,review} of lattice gauge theory with the maximal Abelian
gauge fixing, since the magnetic monopole dominance as well as the
Abelian dominance in low-energy physics of QCD has been observed in
this gauge  for various quantities including the string tension. 
\par
The idea of reformulating the gauge theory as a deformation of a
TQFT  works also for the Abelian gauge theory
\cite{KondoIII}. In the Abelian case, on the other hand, there is no
self-interaction for the gauge field.  Hence, using the similar
reformulation of Abelian gauge theory, we can prove the existence 
of the quark (fractional charge) confinement phase in the strong
coupling region of the four-dimensional QED
\cite{KondoIII} without worrying about the additional
non-perturbative effect.  This result implies the existence of
non-Gaussian fixed point in QED.
\par
In this reformulation, the dimensional reduction occurs as a result
of the supersymmetry hidden in the TQFT.  Hence the calculation of
the Wilson loop in four-dimensional QCD is reduced to that in
two-dimensional NLSM.  It should be remarked that this equivalence
between TQFT$_4$ and   NLSM$_2$ is exact.
\par
In this paper we have used the instanton calculus to calculate the
Wilson loop in two dimensions. We have shown that the area law of the
Wilson loop is derived from the naive instanton calculus, i.e.,
dilute instanton-gas approximation. 
The improvement of the instanton calculus can be performed along the
lines shown in \cite{BL79,FFS79,BL81}.  
 The two-dimensional instanton (resp. anti-instanton) is considered
as the intersection of the magnetic monopole (resp. anti-monopole)
current with the two-dimensional space (plane). 
In other words, the area law behaviour of the Wilson loop average
is understood in a simple geometrical manner, i.e., as summing
up the higher linking numbers between loop and surface. 
This implies that the quark confinement in QCD is caused by
condensation of magnetic monopole and anti-monopole (currents),
together with the previous result \cite{KondoI}. Therefore, these
results support  the scenario of quark confinement proposed by
Nambu, 't Hooft and Mandelstam, i.e. dual superconductor picture of
QCD vacuum.
\par
Note that we have used the instanton calculus merely to see the
correspondence between the two-dimensional instanton and
four-dimensional magnetic monopole (current), we need not to use
the instanton calculus for exactly calculating the Wilson loop in
two-dimensional NLSM.  We can use other methods too, e.g.
fermionization \cite{KondoIII}.   There is some hope to perform the
calculation exactly, since the two-dimensional O(3) NLSM is exactly
soluble
\cite{HMN90,Wiegmann85}.
 \par
Our formulation is also able to estimate the perturbative
correction around the non-perturbative (topologically non-trivial)
background  without ad hoc assumption.  As an example, the
calculation of static potential  is given in
\cite{KondoIV} where the perturbative Coulomb potential is
reproduced in addition to the linear potential part coming from the
TQFT. 
The relationship between the full non-Abelian Wilson loop and the
diagonal Abelian Wilson loop can be given based on the non-Abelian
Stokes theorem \cite{KondoIV}.
Consequently, the paper \cite{KondoIV} completes (together with
the results of this paper) the proof of area decay of the full
non-Abelian Wilson loop {\it within} the reformulation of
four-dimensional QCD as a perturbative deformation of the TQFT.

\par
The advantage of this reformulation is that one can in principle
check whether this reformulation is reliable or not, since the
calculations of the Wilson loop (and therefore string tension) are
reduced to calculations in a two-dimensional NLSM. 
In fact, one can check by direct numerical simulation whether the
string tension obtained from the  diagonal Wilson loop in
two-dimensional NLSM saturates that of full non-Abelian Wilson loop
in four-dimensional QCD, as proposed in \cite{Kondo98}.
This is nothing but the test of Abelian dominance and magnetic
monopole dominance through the dimensionally reduced two-dimensional
model.  Such simulations will prove or disprove the validity of the
reformulation of QCD proposed in this paper.
\par

\section*{Acknowledgments}
 I would like to thank Yoshio Kikukawa and Ryu Sasaki for helpful
discussions in the early stage of this work.  
I am also grateful to Mauro Zeni for sending many comments and
remarks on the first version of this paper. This work is supported in
part by the Grant-in-Aid for Scientific Research from the Ministry of
Education, Science and Culture.


\newpage
\baselineskip 10pt
\begin{thebibliography}{999}
\bibitem{GWP73}
  D.J. Gross and F. Wilczek,
  Ultraviolet behavior of non-abelian gauge theories,
  Phys. Rev. Lett. 30, 1343-1345 (1973).
  \\
  H.D. Politzer,
  Reliable perturbative results for strong interactions?,
  Phys. Rev. Lett. 30, 1346-1349 (1973).
%  \\
%  D.J. Gross and F. Wilczek,
%  Asymptotically free gauge theories, I.,
%  Phys. Rev. D 8, 3633-3652 (1973),
%  Asymptotically free gauge theories, II.,
%  Phys. Rev. D 9, 980-993 (1974).

\bibitem{KSW87}
  A. Kronfeld, G. Schierholz and U.-J. Wiese,
  Topology and dynamics of the confinement mechanism,
  Nucl. Phys. B 293, 461-478 (1987).
%\bibitem{KLSW87}
  A. Kronfeld, M. Laursen, G. Schierholz and U.-J. Wiese,
  Monopole condensation and color confinement,
  Phys. Lett. B 198, 516-520 (1987).
  
\bibitem{review}
  T. Suzuki,
  Monopole condensation in lattice SU(2) QCD,
  hep-lat/9506016.
\\ 
  A. Di Giacomo, 
  Monopole condensation and color confinement,
  hep-lat/9802008;  
  Mechanisms for color confinement,
  hep-th/9603029.
\\
  M.I. Polikarpov,
  Recent results on the abelian projection of lattice
gluodynamics,
  hep-lat/9609020.
%  \\
  M.N. Chernodub and M.I. Polikarpov,
  Abelian projections and monopoles,
  hep-th/9710205.

\bibitem{NO73}
  H.B. Nielsen and P. Olesen,
  Vortex-line models for dual strings,
  Nucl. Phys. B 61, 45-61 (1973).
  
\bibitem{tHooft74}
  G. 't Hooft,
  Magnetic monopoles in unified gauge theories,
  Nucl. Phys. B 79, 276-284 (1974).
  
\bibitem{BPST75}
  A.A. Belavin, A.M. Polyakov, A.S. Schwartz and Yu.S. Tyupkin,
  Pseudoparticle solutions of the Yang-Mills equations,
  Phys. Lett. B 59, 85-87 (1975).
  
\bibitem{Shifman94}
  M. Shifman,
  {\it Instantons in gauge theories}
  (World Scientific, 1994).

\bibitem{Polyakov77}
  A.M. Polyakov,
  Compact gauge fields and the infrared catastrophe,
  Phys. Lett. B 59, 82-84 (1975).
%\bibitem{Polyakov77}
  A.M. Polyakov,
  Quark confinement and topology of gauge theories,
  Nucl. Phys. B 120, 429-458 (1977).

\bibitem{SW94}
  N. Seiberg and E. Witten,
  Electric-magnetic duality, monopole condensation, and
confinement in N=2 supersymmetric Yang-Mills theory,
hep-th/9407087,
Nucl. Phys. B 426, 19-52 (1994).
\\
 Monopoles, duality and chiral symmetry breaking in N=2
supersymmetric QCD,
hep-th/9408099,
Nucl. Phys. B 431, 484-550 (1994).

\bibitem{Witten}
  E. Witten, 
  Topological quantum field theory,
  Commun. Math. Phys. 117, 353-386 (1988).
  \\
  Topological sigma model,
  Commun. Math. Phys. 118, 411-449 (1988).
  
\bibitem{Schwarz}
  A.S. Schwarz, 
  The partition function of a degenerate quadratic functional and
the Ray-Singer invariants,
  Lett. Math. Phys. 2, 247-252 (1978).
  
\bibitem{TQFT}
  D. Birmingham, M. Blau, M. Rakowski and G. Thompson,
  Topological field theory,
  Phys. Report 209, 129-340 (1991).
  
\bibitem{AN93}
  H. Abe and N. Nakanishi,
  How to solve the covariant operator formalism of gauge theories
and quantum gravity in the Heisenberg picture. III two-dimensional
nonabelian BF theory,
  Prog. Theor. Phys. 89, 501-522 (1993).

\bibitem{Izawa93}
  K.-I. Izawa,
  Another perturbative expansion in nonabelian gauge theory,
  hep-th/9309150,
  Prog. Theor. Phys. 90, 911-916 (1993).
  
%BF-YM

\bibitem{BFYM}
  M. Martellini and M. Zeni, 
  Feynman rules and $\beta$-function for the BF Yang-Mills
theory,
  hep-th/9702035,
  Phys. Lett. B 401, 62-68 (1997).
\\
  F. Fucito, M. Martellini and M. Zeni,
  The BF formalism for QCD and quark confinement,
  hep-th/9605018,
  Nucl. Phys. B 496, 259-284 (1997).
\\
  A.S. Cattaneo, P. Cotta-Ramusino, F. Fucito, M.
Martellini, M. Rinaldi, A. Tanzini and M. Zeni,
  Four-dimensional Yang-Mills theory as a deformation of
topological BF theory,
  hep-th/9705123.


%%%%%
\bibitem{QR97}
  M. Quandt and H. Reinhardt,
  Field strength formulation of SU(2) Yang-Mills theory in
the maximal abelian gauge: perturbation theory,
  hep-th/9707185.

\bibitem{KondoI}
  K.-I. Kondo,
  Abelian-projected effective gauge theory of QCD with asymptotic
freedom and quark confinement,
%Chiba Univ. Preprint, CHIBA-EP-99,
  hep-th/9709109 (revised), Phys. Rev. D 57, 7467-7487 (1998).
\\
  K.-I. Kondo, Talk given at YKIS'97: Non-perturbative QCD, 2-12
December 1997, Kyoto, Chiba Univ. Preprint, CHIBA-EP-104,
hep-th/9803063,
  Prog. Theor. Phys. Supplement, No. 130, in press.
 
\bibitem{antiBRST}
  G. Curci and R. Ferrari,
  Slavnov transformations and supersymmetry,
  Phys. Lett. B 63, 91-94 (1976).
  \\
  I. Ojima,
  Another BRS transformation,
  Prog. Theor. Phys. 64, 625-638 (1980).
  
\bibitem{Nambu74}
  Y. Nambu,
  Strings, monopoles, and gauge fields,
  Phys. Rev. D 10, 4262-4268 (1974).

\bibitem{tHooft81}
  G. 't Hooft,
  Topology of the gauge condition and new confinement
phases in non-Abelian gauge theories,
  Nucl. Phys. B 190 [FS3], 455-478 (1981).
  
\bibitem{Mandelstam76}
  S. Mandelstam,
  Vortices and quark confinement in non-abelian gauge
theories, 
  Phys. Report, 23, 245-249 (1976).
  
\bibitem{Wilson} 
  K.G. Wilson and J. Kogut,
  The renormalization group and the $\epsilon$ expansion,
  Phys. Rep. 12 C, 75-200 (1974).
\\
  K.G. Wilson,
  {\it The renormalization group and critical phenomena},
  Rev. Mod. Phys. 55, 583-600 (1983).
  
\bibitem{EI82}
  Z.F. Ezawa and A. Iwazaki,
  Abelian dominance and quark confinement in Yang-Mills
theories,
  Phys. Rev. D 25, 2681-2689 (1982).
  \\
  Abelian dominance and quark confinement in Yang-Mills
theories,
  II. Oblique confinement and $\eta'$ mass,
  Phys. Rev. D 26, 631-647 (1982).

\bibitem{SY90}
  T. Suzuki and I. Yotsuyanagi,
  Possible evidence of abelian dominance in quark
confinement,
  Phys. Rev. D 42, 4257-4260 (1990).
  \\
  S. Hioki, S. Kitahara, S. Kiura, Y. Matsubara, O.
Miyamura, S. Ohno and T. Suzuki,
  Abelian dominance in SU(2) color confinement,
  Phys. Lett. B 272, 326-332 (1991).
  \\
  Monopole distribution in momentum space in SU(2) lattice
gauge theory,
  Phys. Lett. B 285, 343-346 (1992).

\bibitem{ASu97}
  K. Amemiya and H. Suganuma,
  Gluon propagator in maximally abelian gauge and abelian dominance
for long-range interaction, hep-lat/9712028.

\bibitem{Wilson74}
  K. Wilson, 
  Confinement of quarks,
  Phys. Rev. D 10, 2445-2459 (1974).
  
\bibitem{KondoIV}
  K.-I. Kondo, 
  Abelian magnetic monopole dominance in quark confinement,
	 Chiba Univ. Preprint, CHIBA-EP-106, 
  hep-th/9805153, Phys. Rev. D, to appear.
  
%%%%%SUSY, BRS

\bibitem{PS79}  
  G. Parisi and N. Sourlas,
  Random magnetic fields, supersymmetry, and negative dimensions,
  Phys. Rev. Lett. 43, 744-745 (1979).
  
\bibitem{MNTW83}
  B. McClain, A. Niemi, C. Taylor and L.C.R. Wijewardhana,
  Superspace, dimensional reduction, and stochastic quantization,
  Nucl. Phys. B 217, 430-460 (1983).
  \\
  Superspace, negative dimensions, and quantum field theories,
  Phys. Rev. Lett. 49, 252-255 (1982).
  
\bibitem{MNT82}
  B. McClain, A. Niemi and C. Taylor,
  Stochastic quantization of gauge theories,
  Ann. Phys. 140, 232-246 (1982).

\bibitem{Cardy83}
  J. Cardy,
  Nonperturbative effects in a scalar supersymmetric theory,
  Phys. Lett. B 125, 470-472 (1983).
  
\bibitem{KLF84}
  A. Klein, L.J. Landau and J.  Fernando Perez,
  Supersymmetry and Parisi-Sourlas dimensional reduction: a rigorous
proof,
  Commun. Math. Phys. 94, 459-482 (1984).
  \\
  A. Klein and J. Fernando Perez,
  Supersymmetry and dimensional reduction: a non-perturbative proof,
  Phys. Lett. B 125, 473-475 (1983).

%%%%%
\bibitem{BT81}  
  L. Bonora and M. Tonin,
  Superfield formulation of extended BRS symmetry,
  Phys. Lett. B 98, 48-50 (1981).

\bibitem{Baulieu85}
  L. Baulieu,
  Perturbative gauge theories,
  Phys. Rept. 129, 1-74 (1985).
  
%%%%%exact result
\bibitem{Luscher78}
  M. L\"uscher,
  Quantum non-local charges and absence of particle production in
the two-dimensional non-linear $\sigma$ model,
  Nucl. Phys. B 135, 1-19 (1978).
  
\bibitem{BZL76}
  E. Brezin, J. Zinn-Justin and J.C. LeGuillou,
  Renormalization of the nonlinear $\sigma$ model in 2+$\epsilon$
dimensions,
  Phys. Rev. D 14, 2615-2621 (1976).
  
\bibitem{ZZ79}
  A.B. Zamolodchikov and Al. B. Zamolodchikov,
  Factorized S-matrices in two dimensions as the exact solutions of
certain relativistic quantum field theory models,
  Ann. Phys. 120, 253-291 (1979).

\bibitem{AAR91}
  E. Abdalla, M.C.B. Abdalla and K.D. Rothe,
  {\it Non-perturbative methods in 2 dimensional quantum
feild theory},
  (World Scientific, 1991).

\bibitem{FZ91}
  V. Fateev and Al. Zamolodchikov, 
  Integrable perturbations of $Z_N$ parafermion models and the O(3)
sigma model,
  Phys. Lett. B 271, 91-100 (1991).
\\
%\bibitem{YZ91}
  V.P. Yurov and Al.B. Zamolodchikov,
  Correlation functions of integrable 2D models of the relativistic
field theory; Ising model,
  Intern. J. Mod. Phys. A6, 3419-3440 (1991).
\\
%\bibitem{ZZ92}
  A. Zamolodchiov and Al. Zamolodchikov, 
  Massless factorized scattering and sigma models with topological
terms,
  Nucl. Phys. B 379, 602-623 (1992).

\bibitem{FOZ93}
  V.A. Fateev, E. Onofri and Al.B. Zamolodchikov,
  Integrable deformations of the O(3) sigma model. The sausage model,
  Nucl. Phys. B 406 [FS], 521-565 (1993).

\bibitem{BLZ94}
  V.V. Bazhanov, S.L. Lukyanov and A.B. Zamolodchikov,
  Integrable structure of conformal field theory, quantum KdV theory
and thermodynamic Bethe ansatz, hep-th/9412229.

\bibitem{HMN90}
  P. Hasenfratz, M. Maggiore and F. Nedermayer,
  The exact mass gap of the O(3) and O(4) non-linear $\sigma$ models,
  Phys. Lett. B 245, 522-528 (1990).
\\
  P. Hasenfratz and F. Nedermayer,
  The exact mass gap of the O(N) $\sigma$ model for arbitrary $N \ge
3$ in $d=2$,
  Phys. Lett. B 245, 529-532 (1990).
  
\bibitem{Wiegmann85}
  P.B. Wiegmann,
  Exact solution of O(3) nonlinear $\sigma$-model,
  Phys. Lett. B 152, 209-214 (1985).
  
\bibitem{Tsvelik88}
  A.M. Tsvelik,
  Exact solution of a model of one-dimensional fermions with SU(N)
$\times$ SU(N) symmetry,
  Sov. Phys. JETP 66, 754-760 (1988).

\bibitem{NT91}
  V. Nikos Nicopoulos and A.M. Tsvelik,
  (1+1)-dimensional O(3) nonlinear $\sigma$ model in a magnetic
field: magnetization and effective potential,
  Phys. Rev. B 44, 9385-9391 (1991).

\bibitem{PW83}
  A. Polyakov and P.B. Wiegmann,
  Theory of nonAbelian Goldstone bosons in two dimensions,
  Phys. Lett. B 131, 121-126 (1983).
\\
%\bibitem{Wiegmann84}
  P.B. Wiegmann,
  On the theory of nonAbelian Goldstone bosons in two dimensions;
exact solution of the SU(N) $\times$ SU(N) nonlinear $\sigma$ model,
  Phys. Lett. B 141, 217-222 (1984).
\\
%\bibitem{PW84}
  A.M. Polyakov and P.B. Wiegmann,
  Goldstone fields in two dimensions with multivalued action,
  Phys. Lett. B 141, 223-228 (1984).



%%%%%RG of NLSM
\bibitem{Polyakov75b}
  A.M. Polyakov,
  Interaction of Goldstone particles in two dimensions. 
Applications to ferromagnets and massive Yang-Mills fields,
  Phys. Lett. B 59, 79-81 (1975).

\bibitem{BHZ80}
  E. Brezin, S. Hikami and J. Zinn-Justin,
  Generalized non-linear $\sigma$-models with gauge invariance, 
  Nucl. Phys. B 165, 528-544 (1980).

%%%%%Non-linear sigma model
\bibitem{BP75}
  A.A. Belavin and A.M. Polyakov,
  Metastable states of two-dimensional isotropic ferromagnets,
  JETP Lett. 22, 245-248 (1975).
  
\bibitem{Perelomov87}
  A.M. Perelomov,
  Chiral models: geometrical aspects,
  Phys. Rept. 146, 135-213 (1987).

\bibitem{Woo77}
  G. Woo,
  Pseudoparticle configurations in two-dimensional ferromagnets,
  J. Math. Phys. 18, 1264-1266 (1977).
  
\bibitem{GP78}
  V.L. Golo and A.M. Perelomov,
  Solution of the duality equations for the two-dimensional
SU(N)-invariant chiral model,
  Phys. Lett. B 79, 112-113 (1978).
  
\bibitem{Perelomov78}
  A.M. Perelomov,
  Instantons and K\"ahler manifolds,
  Commun. Math. Phys. 63, 237-242 (1978).

%%%%%instanton 2D
\bibitem{CDG77}
  C.G. Callan, Jr., R. Dashen and D.J. Gross,
  Pseudoparticles and massless fermions in two dimensions,
  Phys. Rev. D 16, 2526-2534 (1977).

\bibitem{RU78}
  S. Raby and A. Ukawa,
  Instantons in (1+1)-dimensional Abelian gauge theories,
  Phys. Rev. D 18, 1154-1173 (1978).
  
\bibitem{CDG78}
  C.G. Callan, Jr., R. Dashen and D.J. Gross,
  Towards a theory of the strong interactions,
  Phys. Rev. D 17, 2717-2763 (1978).

\bibitem{Coleman85} 
  S. Coleman, 
  {\it Aspect of Symmetry}
  (Cambridge Univ. Press, New York, 1985) .

\bibitem{Rajaraman89}
  R. Rajaraman,
  {\it Solitons and Instantons}
  (North-Holland, Amsterdam, 1989).

%%%%%RG MK
\bibitem{MK75}
  A. Migdal,
  Recursion equations in gauge field theories,
  Zh. Eksp. Teor. Fiz. 69, 810, 1457 (1975), 
  Sov. Phys. JETP 42, 413, 743 (1975).
\\
  L.P. Kadanoff,
  Notes on Migdal's recursion formulas,
  Ann. Phys. 100, 359-394 (1976).
  
 
%%%%%confinement

\bibitem{HK85}
  H. Hata and T. Kugo,
  Color confinement, Becchi-Rouet-Stora symmetry, and negative
dimensions,
  Phys. Rev. D 32, 938-944 (1985).

\bibitem{Hata84}
  H. Hata,
  Asymptotic free chiral model in four dimensions,
  Phys. Lett. B 143, 171-174 (1984).
  
\bibitem{HN93}
  H. Hata and I. Niigata,
  Color confinement, abelian gauge and renormalization group
flow,
  Nucl. Phys. B 389, 133-152 (1993).

\bibitem{HT94}
  H. Hata and Y. Taniguchi,
  Finite temperature deconfining transition in the BRST formalism,
  hep-th/9405145,
  Prog. Theor. Phys. 93, 797-812 (1995).
  
\bibitem{HT95}
  H. Hata and Y. Taniguchi,
  Color confinement in perturbation theory from a topological model,
  hep-th/9502083,
  Prog. Theor. Phys. 94, 435-444 (1995).

\bibitem{Kugo95}
  T. Kugo,
  The universal renormalization factors $Z_1/Z_3$ and color
confinement condition in non-Abelian gauge theory,
hep-th/9511033.

\bibitem{KO79}
  T. Kugo and I. Ojima,
  Local covariant operator formalism of non-Abelian gauge theories
and quark confinement problem,
  Prog. Theor. Phys. Suppl. 66, 1-130 (1979).
  
\bibitem{Nishijima94}
  K. Nishijima,
  Confinement of quarks and gluons,
  Intern. J. Mod. Phys. A 9, 3799-3819 (1994).
%\bibitem{Nishijima95}
%  K. Nishijima,
  Confinement of quarks and gluons  II,
  Intern. J. Mod. Phys. A 10, 3155-3167 (1995).

\bibitem{Suzuki83}
  T. Suzuki,
  Color confinement and asymptotic freedom,
  Prog. Theor. Phys. 69, 1827-1830 (1983).
  \\
  T. Suzuki and K. Shimada,
  Confinement criteria and compact QED in (2+1)-dimensions,
  Prog. Theor. Phys. 69, 1537 (1983).
  
%%%%%
\bibitem{BLS76}
  W.A. Bardeen, B.W. Lee and R.E. Schrock,
  Phase transition in the nonlinear $\sigma$ model in a
(2+$\epsilon$)-dimensional continuum,
  Phys. Rev. D 14, 985-1005 (1976)
  
\bibitem{DLD78}
  A. D'adda, M. L\"uscher and P. Di Vecchia,
  A 1/n expandable series of non-linear $\sigma$ models with
instantons,
  Nucl. Phys. B 146, 63-76 (1978).

\bibitem{Eichenherr78}
  H. Eichenherr,
  SU(N) invariant non-linear $\sigma$ models,
  Nucl. Phys. B 146, 215-223 (1978).
  
\bibitem{Macfarlane79}
  A.J. Macfarlane,
  The SU(N+1) $\sigma$ model or $CP^N$ model as a non-linear
realization of SU(N+1) symmetry,
  Nucl. Phys. B 152, 145-152 (1979).

\bibitem{DDL79}
  A. D'adda, P. Di Vecchia and M. L\"uscher,
  Confinement and chiral symmetry breaking in $CP^{n-1}$ models with
quarks,
  Nucl. Phys. B 152, 125-144 (1979).
  
\bibitem{Witten79}
  E. Witten,
  Instantons, the quark model, and the 1/N expansion,
  Nucl. Phys. B 149, 285-320 (1979).

\bibitem{Jevicki79}
  A. Jevicki,
  Instantons and the 1/N expansion in nonlinear $\sigma$ models,
  Phys. Rev. D20, 3331-3335 (1979).

\bibitem{Affleck80}
  I. Affleck,
  Testing the instanton method,
  Phys. Lett. B 92, 149-152 (1980).
  \\
  I. Affleck,
  The role of instantons in scale-invariant gauge theories,
  Nucl. Phys. B 162, 461-477 (1980).
  (II). The short-distance limit,
  Nucl. Phys. B 171, 420-444 (1980).
  
\bibitem{Iwasaki81}
  Y. Iwasaki,
  Low temperature behaviors of classical O(3) Heisenberg model in
two dimensions,
  Prog. Theor. Phys. 66, 1089-1092 (1981).
  \\
  Correlation length of classical O(3) Heisenberg model in two
dimensions, 
  Prog. Theor. Phys. 66, 1093-1094 (1981).
  \\
  The structure of the vacuum. I: two-dimensional non-linear O(3)
$\sigma$  model,
  Prog. Theor. Phys. 68, 448-470 (1982).

\bibitem{AJ78}
  M.F. Atiyah and J.D. Jones,
  Topological aspects of Yang-Mills theory,
  Commun. Math. Phys. 61, 97-118 (1978).

%%%%%
\bibitem{BPZ84}
  A.A. Belavin, A.M. Polyakov and A.B. Zamolodchikov,
  Infinite conformal symmetry in two-dimensional quantum field
theory,
  Nucl. Phys. B 241, 333-380 (1984).
  
\bibitem{KZ84}
  V.G. Knizhnik and A.B. Zamolodchikov,
  Current algebra and Wess-Zumino model in two dimensions,
  Nucl. Phys. B 247, 83-103 (1984).
  
\bibitem{Witten84}
  E. Witten,
  Non-Abelian bosonization in two dimensions,
  Commun. Math. Phys. 92, 455-472 (1984).
  
\bibitem{BQ94}
  C.P. Burgess and F. Quevedo,
  Nonabelian bosonization as duality,
  hep-th/9403173, 
  Phys. Lett. B329, 457-462 (1994).
  
\bibitem{Affleck89}
  I. Affleck,
  Field theory methods and quantum critical phenomena,
 in Fields, Strings and Critical Phenomena, Les Houches, 1988 eds. by
E. Brezin and J. Zinn-Justin (Elsevier Science Publishers, Amsterdam,
1989).

\bibitem{Witten89}
  E. Witten,
  Quantum field theory and Jones polynomial,
  Commun. Math. Phys. 121, 351-399 (1989).
  
\bibitem{MT97}
  H. Miyazaki and I. Tsutsui,
  Quantum mechanically induced Wess-Zumino term in the principal
chiral model,
  hep-th/9706167.
  
\bibitem{KTT97}
  H. Kobayashi, I. Tsutsui and S. Tanimura,
  Quantum mechanically induced Hopf term in the O(3) nonlinear sigma
model, hep-th/9705183.

\bibitem{MW96}
  L.J. Mason and N.M.J. Woodhouse,
  {\it Integrability, self-duality and twistor theory}
  (Oxford Univ. Press, Oxford, 1996).
  

\bibitem{KondoIII}
  K.-I. Kondo,
  Existence of confinement phase in quantum electrodynamics,
  Chiba Univ. Preprint, CHIBA-EP-105,
  hep-th/9803133,
  Phys. Rev. D, in press.

\bibitem{Nakahara}
  M. Nakahara,
  {\it Geometry, Topology and Physics}
  (Institute of Physics, Bristol, 1990).
  
%%%%%NLSM
\bibitem{Jevicki77}
  A. Jevicki,
  Quantum fluctuations of pseudoparticles in the non-linear $\sigma$
model, 
  Nucl. Phys. B 127, 125-140 (1977).
  
\bibitem{Forster77}
  D. F\"orster,
  On the structure of instanton plasma in the two-dimensional O(3)
non-linear $\sigma$-model,
  Nucl. Phys. B 130, 38-60 (1977).
  
\bibitem{BL79}
  B. Berg and M. L\"uscher,
  Computation of quantum fluctuations around multi-instanton field
from exact Green's functions: the $CP^{n-1}$ case,
  Commun. Math. Phys. 69, 57-80 (1979).
  
\bibitem{FFS79}
  V.A. Fateev, I.V. Frolov and A.S. Schwarz,
  Quantum fluctuations of instantons in the non-linear $\sigma$
model,
  Nucl. Phys. B 154, 1-20 (1979).

\bibitem{BL81}
  A.P. Bukhvostov and L.N. Lipatov,
  Instanton--Anti-instanton interaction in the O(3) non-linear
$\sigma$ model and an exactly soluble fermion theory,
  Nucl. Phys. B 180 [FS2], 116-140 (1981).
 
\bibitem{Silvestrov90}
  P.G. Silvestrov,
  A new way of instanton--anti-instanton interaction description. 
The nonlinear O(3) sigma model example,
  Sov. J. Nucl. Phys. 51, 1121-1127 (1990).


\bibitem{Kogut83}
  J.B. Kogut,
  The lattice gauge theory approach to quantum chromodynamics,
  Rev. Mod. Phys. 55, 775-836 (1983).

\bibitem{RR83}
  J.-L. Richard and A. Rouet,
  The CP$_1$ model on the torus, contribution of instantons,
  Nucl. Phys. B 211, 447-464 (1983).
  \\
  P. Chiappetta and J.L. Richard,
  Numerical analysis of the "fugacity" factor in the periodic CP$_1$
model,
  Nucl. Phys. B 211, 465-470 (1983).
  
\bibitem{MS94}
  C. Michael and P.S. Spencer,
  Instanton size distribution in O(3),
  hep-lat/9404001,
  Phys. Rev. D 50, 7570-7577 (1994).
  
\bibitem{Snippe94}
  J. Snippe,
  Tunneling through sphalerons: the O(3) $\sigma$-model on a
cylinder,
  hep-th/9405129,
  Phys. Lett. B 335, 395-402 (1994).
  \\
  J. Snippe and P. van Baal,
  A new approach to instanton calculations in the O(3) nonlinear
$\sigma$-model,
  hep-lat/9411055,
  Nucl. Phys. Proc. Suppl. 42, 779-781 (1995).
  
\bibitem{Kondo98}
  K.-I. Kondo,
  A proof of quark confinement in QCD,	
%  Chiba Univ. Preprint, CHIBA-EP-107, 
  Talk given at the third International Conference
on Quark Confinement and the  Hadron Spectrum, 7-12 June
1998,  Jefferson Lab., Newport News, VA, USA.

\end{thebibliography}
\end{document}